\documentclass[a4paper,10pt]{scrartcl}

\usepackage{amsmath}
\usepackage{amsthm}
\usepackage{amssymb}
\usepackage[affil-it]{authblk}
\usepackage{bbm}
\usepackage{caption}
\usepackage{chngcntr}
\usepackage{color}
\usepackage{float}
\usepackage[margin=3cm]{geometry}
\usepackage{grffile}
\usepackage{lmodern}
\usepackage{mathtools}
\usepackage{natbib}
\usepackage[low-sup]{subdepth}
\usepackage{tabu}
\usepackage{tikz, caption, subcaption}
\usepackage{url}

\counterwithin{figure}{subsection}
\counterwithin{table}{subsection}
\newtheorem{defi}{Definition}[section]

\newtheorem{lem}[defi]{Lemma}
\newtheorem{pro}[defi]{Proposition}
\newtheorem{rmk}[defi]{Remark}

\newtheoremstyle{NoItalic}{}{\baselineskip}{}{}{\bfseries}{.}{.5em}{\thmname{#1}\thmnumber{ #2}\thmnote{ #3}}\theoremstyle{NoItalic}
\setcapindent{0pt}

\newcommand{\valuefn}{V}

\begin{document} 

\title{Modern tontine with bequest: innovation in pooled annuity products}
\author{Thomas Bernhardt\footnote{\texttt{T.Bernhardt@hw.ac.uk}} \hspace{0em} and Catherine Donnelly\footnote{\texttt{C.Donnelly@hw.ac.uk} (corresponding author)}}
\affil{{\normalfont{\texttt{https://risk-insight-lab.com/}}} \\ \vspace{0.5cm} Risk Insight Lab, 
\\Department of Actuarial Mathematics and Statistics, 
\\Heriot-Watt University, Edinburgh, Scotland EH14 4AS}
\date{\today}
\maketitle

\vspace{-4em}

\section*{\begin{center}Abstract\end{center}}
\begin{abstract}
\noindent We introduce a new pension product that offers retirees the opportunity for a lifelong income and a bequest for their estate.  Based on a tontine mechanism, the product divides pension savings between a tontine account and a bequest account.  The tontine account is given up to a tontine pool upon death while the bequest account value is paid to the retiree's estate.  The values of these two accounts are continuously re-balanced to the same proportion, which is the key feature of our new product.

Our main research question about the new product is what proportion of pension savings should a retiree allocate to the tontine account.  Under a power utility function, we show that more risk averse retirees allocate a fairly stable proportion of their pension savings to the tontine account, regardless of the strength of their bequest motive.   The proportion declines as the retiree becomes less risk averse for a while.   However, for the least risk averse retirees, a high proportion of their pension savings is optimally allocated to the tontine account.  This surprising result is explained by the least risk averse retirees seeking the potentially high value of the bequest account at very old ages.   
\\[1em]
\textbf{Keywords:} Longevity credit; uncertain lifespan; decumulation; maximizing expected utility.
\end{abstract}

\section{Introduction}
Longevity risk, the risk of living longer than expected, is one of the major financial issues for retirees.  It impacts on how much a retiree can withdraw as an income from their savings.  Withdraw too much too soon and they risk running out of money.  Withdraw too little and they risk having an unnecessarily low standard of living in retirement.  Mitigation against longevity risk should be an integrated part of a retirement plan.  This can be done through life annuities or modern tontines.  A tontine allows its participants to pool directly their longevity risk together rather than indirectly through an insurance company, as is done with life annuities.

We study how much of their pension savings a retiree should put into an innovative tontine structure.  We allow for the retiree to withdraw a sustainable income and their desire to leave a bequest.  Surprisingly, the formulation can be and, indeed, is different from that of a life annuity contract.  A tontine can be structured more flexibly while having regard to adverse selection effects.  Although our results are derived in a theoretical framework, the high-level conclusion is that tontines can form a significant part of retirement investment to the financial benefit of the retiree and their estate.

The purpose of a modern tontine is to pay a life-long income to its participants.  Those who die earlier will, in effect, subsidize the income of the survivors.  The benefits of a modern tontine are gained by all of its surviving participants.  This feature of modern tontines contrasts with the popular image of historical tontines in which it is only the last survivor who gains everything.  Indeed, modern tontines can be thought of as life annuity contracts without the attendant guarantees.

Yet participants in modern tontines can have a better financial outcome than can be achieved through life annuities.  The tontine participants avoid the expensive, decades-long guarantees inherent in conventional life annuities.  In doing so, they can achieve a higher expected income in return for accepting more risk in the level and duration of their tontine-derived income. 

Participants should also have a better financial outcome while alive than those in income drawdown, the latter being those who withdraw an income from their invested pension savings without any type of longevity insurance mitigation.   The tontine participants can follow the same investment strategy as they would do in income drawdown.  However, due to their participation in a tontine, they receive additional money, called longevity credits, as long as they survive.  This means that there is a good chance that they have an income for life and the sustainable level of that life-long income is higher than in income drawdown.  

Surprisingly, in our proposed tontine structure it is possible for retirees to leave a higher bequest than an individual who is in income drawdown, in addition to a lifelong income.  While it is not always true and depends on the age of death of the individual, it illustrates that tontines allow us to re-think what is possible for retirees.

In our tontine structure, savings are divided between two accounts called a tontine account and a bequest account.  We show that the proportion of pension savings that are allocated to a tontine account is close to a constant, regardless of the strength of the bequest motive for typical levels of risk aversion.  As the retiree becomes more risk-seeking, the conclusion is very sensitive to the model's parameter values and should be studied further in a more nuanced setting.  

The pension savings that are not put in the tontine account are allocated to a bequest account.  Both the bequest account and the tontine account are invested in the financial market.  However, only the tontine account attracts longevity credits from the tontine pooling, and only the tontine account value is given up if the retiree dies.  The value of the bequest account is paid to the retiree's estate upon their death.

The proportion of savings in the tontine account is maintained at a fixed proportion at all times.  Both accounts grow at the same investment return rate and both are reduced by the same rate due to consumption.  Thus some of the longevity credits paid into the tontine account must be instantaneously transferred to the bequest account, in order to maintain the fixed ratio of the two account values.

Our contribution to the literature is two-fold.  First, we introduce an entirely new contract based on mondern tontines that allows retirees to benefit from a potentially life-long income and to satisfy their desire to leave a bequest.  This contract may be very attractive; in a poll of around $180$ actuaries and actuarially-interested individuals, and after they had the contract explained to them, $50$\% of them chose the modern tontine with bequest as a vehicle for decumulation, whereas $25$\% of them chose a pure modern tontine (i.e. with no bequest feature) and the remaining respondents were equally split between a conventional life annuity and an income drawdown choice.

Second, we analyze our new contract and find some surprising results.  For example, the least risk averse retirees put more into the tontine account because they hope to live long enough for their estate to benefit from an increasingly large bequest account.  For more risk averse retirees, the proportion of their savings allocated to the tontine account is relatively stable regardless of the strength of their bequest motive.

We assume an idealized set-up where the tontine is a perfect pool and there is no loss of longevity credits to the wider tontine pool due to the transfers from the tontine account to the bequest account.  The applicability of this assumption in a finite pool is a future area of study.  Nonetheless, analyzing the contract in our idealized model allows us to gain insights into how different retiree's risk preferences affect their  consumption rate and allocation of pension savings to the tontine account. 

\section{Background}
The pooling of longevity risk in a modern tontine is the same principle that underlies life annuities.  The latter have been extensively studied in the economics literature particularly as retirees do not buy life annuities as much as predicted by economic models, a phenomenon called the annuity puzzle.  There is no consensus view on the causes of the annuity puzzle and there is possibly no single explanation for it.  Reasons proposed for its existence include: a desire to leave a bequest; annuities being perceived as expensive even if retirees' estimate of their own life expectancy concurs with the basis used to price the annuity; retirees under-estimating their life expectancy; and viewing the life annuity in an investment frame rather than in a consumption frame.  See, for example, \citet{Mitchelletal2011} for a discussion of the possible causes of the annuity puzzle.

As modern tontines are relatively new and their risk-sharing mechanisms are the foremost topic of papers on them, not many papers consider how to incorporate different tontine designs into the life-cycle (as exceptions, see for example \citealt{hanewaldetal2013, valdezetal2006}).  There are many more papers which analyze life annuities in a life-cycle setting which, due to their similarities with tontines, are pertinent to the integration of tontines into the life-cycle setting.  Here we review some of the papers on life annuities which have results of some relevance to our paper.

Some papers have focused on the optimal time to annuitize, using different objective functions to optimize \citep{Milevsky1998, MilevskyYoung2007, DiGiacintoVigna2012, GerrardHojgaardVigna2012}.  The optimal time depends on the would-be annuitant's attitude to risk, as encapsulated by the chosen objective function, their future distribution of death and the anticipated returns on financial assets.  It is likely to be at older ages, as this is when the ``mortality return'' from a life annuity contract starts becoming increasingly significant.  We choose a fixed age of $65$ years for our numerical illustrations, with the retiree immediately allocating at age $65$ a fixed proportion of their total pension savings to the tontine account.

Other papers have asked a similar question to us, i.e. what is the optimal balance between a bequest and a life annuity, where we substitute a bequest account and tontine account, respectively.  However, as the sum of a life annuity contract and a bequest is not the same as our new tontine product, the existing answers in the literature cannot answer our question.  \citet{BabbelMerrill2006} ask how much of your wealth should you use to buy a life annuity.  They allow for life annuities to be more expensive than the actuarially fair price due to real-world considerations such as costs, fees and profit margins.  Even when annuities are $40\%$ above their actuarially fair price, a risk-averse retiree should still annuitize half of their wealth according to their CRRA-utility model.  \citet{MilevskyYoung2007} determine an optimal annuitization strategy and allow the retiree a good deal of flexibility in the timing of their purchases.   Investors are free to annuitize their wealth as they want: using lump sums and continuous purchases.  They find that retirees should always hold some annuities.  Moreover, they should increase their annuity holdings as they become wealthier, which is called phased annuitization.  \citet{KingstonThorp2005} extend the results of \citet{MilevskyYoung2007} to HARA utility functions, while \citet{BlakeWrightZhang2014} also find that phased annuitization is optimal.  For our product, we ask a much simpler question: what fixed proportion of total pension savings should be allocated to the tontine account, with the remainder allocated to the bequest account.

\citet{HorneffMaurerStamos2008} have an extended version of the setting of \citet{MilevskyYoung2007}, which is closer to our setting but again, they use irreversible life annuity contracts.  Like us, they allow for a bequest motive.  They find that partial annuitisation is preferred when there is a bequest motive.  Further, when the retiree is forced to annuitize all or none of their wealth at a time of their choosing, then it is optimal to annuitize none of it when there is a bequest motive.  Eventual full annuitization is optimal when there is no bequest motive.    \citet{HainautDevolder2006} obtain fairly similar results when determining the optimal allocation at the retirement date.  In our setting, we find that putting all of the retiree's pension savings in the tontine account is optimal only for the least risk averse retirees.  As risk aversion increases, less than $100$\% is allocated to the tontine account.

While many papers assume that the investment return underlying a life annuity contract is the risk-free rate, \citet{Horneffetal2010} allow for variable payout annuities, also called investment-linked annuities, in their asset universe.  The investment return is random in these contracts.  We also allow for random investment returns.  In \citet{Horneffetal2010}, the investor has to decide how much to consume and how much to invest in a variable payout annuity, a risky stock and a risk-free bond. The solution is a pair consisting of an optimal withdrawal strategy and an optimal investment strategy.  The expectation of the sum of the investor's discounted utility of lifetime consumption and utility due to a bequest are maximized.  \citet{Horneffetal2010} find that the retiree never fully annuitizes exactly at the retirement date, even without any bequest motive.  Instead, full annuitization is completed at a later age.  In our setting, we constrain the proportion of pension savings allocated to the tontine account to be a constant at all times, and so we do not allow for a varying proportion, as effectively occurs in \citet{Horneffetal2010}.

\section{Modern tontines}
\subsection{A brief review of modern tontines}
Modern tontines allow retirees to make cost-efficient use of their savings in order to gain a higher income for longer, compared to someone who is in income drawdown.  Simultaneously, it allows them to mitigate their longevity risk which, in this context, we interpret as the risk of outliving their savings.

Modern tontines are a re-interpretation of the life annuity contract.  They mimic the fundamental mechanism underlying life annuities; the shorter-lived participants subsidize the longer-lived.  The subsidies happen throughout the participants' lifetimes; modern tontines are \emph{not} structured so that the ``last survivor'' takes all.  While tontines have a bad connotation in the popular imagination, the existence of the annuity puzzle and the challenges facing many retirees in today's world in converting a pension pot into an income suggest that they may have a strong role to play \citep{MilevskyBook2015}.

A key difference of a pure tontine from the life annuity is that there are no guarantees built into the tontine.   While longevity risk is mitigated for the tontine participant, it is not eliminated as it is for the purchaser of a life annuity contract.  Guarantees can be bought on tontines, but there are not a fundamental or necessary part of them.

The absence of guarantees means that a pure tontine is not for the very risk-averse.  However, the lack of guarantees also means that tontine participants can avoid paying for and being constrained by guarantees.  They can benefit from pooling longevity risk without having to buy the associated investment and longevity guarantees which are an integral part of conventional life annuities.  In exchange for bearing the volatility of the longevity risk, tontine participants may gain a higher return and consequently a higher income than a life annuitant.

A downside of pure tontines, just as in life annuities, is the absence of a bequest for those who wish to leave money after their death.  The value of funds allocated to a tontine by a retiree are shared among the tontine's participants.  This is the flip-side of the mitigation of the risk of outliving one's savings; the tontine participant must give up any money allocated to the tontine upon their death.  

Several tontine structures have been proposed recently, with the common goal of paying an income for life to the tontine members.  Some of them are set up to pay a specified income to their members, with the calculation of the income specified up front.  Others pay ``longevity credits'' to their members, namely the re-distribution of the funds of deceased members among the tontine participants, but do not specify how money should be withdrawn as an income by each participant in the tontine.

The tontines that explicitly allocate longevity credits can be classified as explicit tontines.  Each member has an individual account value, which accumulates with investment returns, any new money paid in and longevity credits from the tontine pooling.  Income is withdrawn by the member for their needs, although there are likely to be limitations on withdrawals to minimize adverse selection and moral hazard effects.

In contrast, implicit tontines promise to pay the participants an income for life, but longevity credits are not explicitly allocated to the participants.  For example, an implicit tontine is the group self-annuitization scheme of \citet{Piggottetal2005}, which has been further studied in \citet{hanewaldetal2013, qiaosherris2013, valdezetal2006}.  Similarly, \citet{MilevskySalisbury2015} propose an implicit tontine and extend it to heterogeneous groups in \citet{MilevskySalisbury2016}.  Real-world examples of an implicit tontine are some annuities issued by TIAA \citep{FormanSabin2015, MilevskyBook2015}.

Explicit tontines can still pay an income to their members and look like an implicit tontine.  However, they allocate explicitly longevity credits to the individual accounts of participants.  Thus an income paid to a participant is dependent on their own personal account value.  Different rules of calculating the longevity credits are proposed in \citet{donnellyetal2014, sabin2010, stamos2008}.
	
\citet{brautigametal2017} compare the tontine of \citet{donnellyetal2014} against that of \citet{MilevskySalisbury2016}, but are unable to fit the two structures into a single, over-arching structure.  Similarly, \citet{Donnelly2015} compares the scheme of \citet{donnellyetal2014} against the group self-annuitization scheme of \citet{Piggottetal2005}, and finds that the financial outcomes due to pooling longevity risk are similar if there are enough people in each scheme.

\subsection{A new tontine structure that incorporates bequests}
We study a structure that incorporates an idealized, explicit, modern tontine which is instantaneously actuarially fair.  It assumes that there is an infinite number of participants in the tontine such that the Law of Large Numbers holds with respect to the deaths occurring among the participants.  This ``perfect pool'' assumption enables us to determine analytical solutions to the studied problems.

The consequences of the structure are that:
\begin{itemize}
    \item participants are paid an income that is expected to last for their lifetime,
    \item participants are paid an income that is higher, on average, than if they were not in any longevity-pooling structure, and
    \item the estate of each participant receives a sum of money upon the participant's death.
\end{itemize}

We consider one of the participants in the tontine structure.  Like every other participant, the pension savings they bring into the structure is ring-fenced as their own savings.  They earn investment returns and longevity credits on their savings as long as they are alive, and withdraw income at a constant rate from their savings.

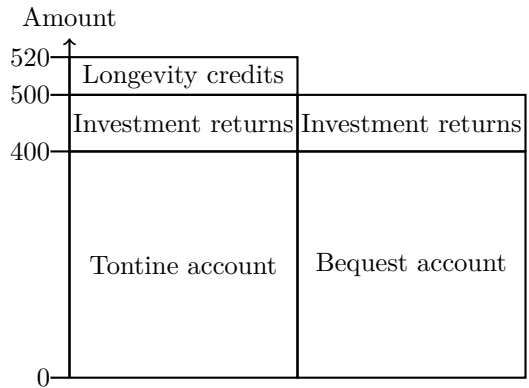
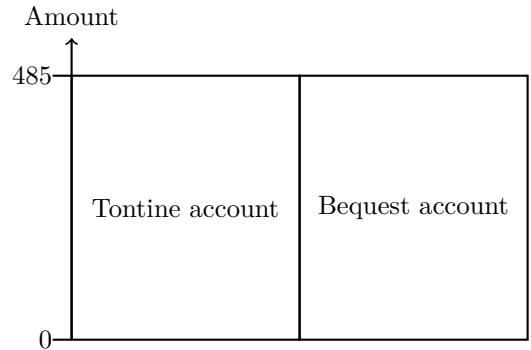
\begin{figure}[ht!]
    \centering
    \begin{subfigure}[b]{0.5\textwidth}
        \centering
    \begin{tikzpicture}
		        \draw [thick, black] (0,0) rectangle (3,3) node[pos=.5] {Tontine account}; 
        \draw [thick, black] (3,0) rectangle (6,3) node[pos=.5] {Bequest account};
				\draw [thick, black] (0,3) rectangle (3,3.75) node[pos=.5] {Investment returns}; 
				\draw [thick, black] (3,3) rectangle (6,3.75) node[pos=.5] {Investment returns};
        \draw [thick, black] (0,3.75) rectangle (3,4.25) node[pos=.5] {Longevity credits}; 
				\draw[->] [thick, black] (0,0) -- (0,4.5) node[above=1] {Amount}; 
			  \draw [thick, black] (-0.25,0) -- (0,0) node[left=3.5] {$0$}; 
				\draw [thick, black] (-0.25,3) -- (0,3) node[left=3.5] {$400$}; 
				\draw [thick, black] (-0.25,3.75) -- (0,3.75) node[left=3.5] {$500$}; 
				\draw [thick, black] (-0.25,4.25) -- (0,4.25) node[left=3.5] {$520$}; 
    \end{tikzpicture}
        \caption{Before re-balancing.}
				\label{FIGdiagbeforerebal}
    \end{subfigure}%
    \begin{subfigure}[b]{0.5\textwidth}
        \centering
    \begin{tikzpicture}
				\draw [thick, black] (0,0) rectangle (3,3.5) node[pos=.5] {Tontine account}; 
        \draw [thick, black] (3,0) rectangle (6,3.5) node[pos=.5] {Bequest account};
				\draw[->] [thick, black] (0,0) -- (0,4) node[above=1] {Amount}; 
				\draw [thick, black] (-0.25,0) -- (0,0) node[left=3.5] {$0$}; 
				\draw [thick, black] (-0.25,3.5) -- (0,3.5) node[left=3.5] {$485$}; 
    \end{tikzpicture}
        \caption{After consuming 50 units and re-balancing.}
				\label{FIGdiagafterrebal}
    \end{subfigure}
    \caption{Illustration of the tontine account and the bequest account before re-balancing (Figure \ref{FIGdiagbeforerebal}) and after re-balancing (Figure \ref{FIGdiagafterrebal}) of the account values to the chosen fixed proportion, here assumed to be $50$\%.   Before re-balancing and before consumption is deducted, the value of the tontine account is $520$ units and the value of the bequest account is $500$ units, giving a total pension savings value of $1\,020$.  After consuming $50$ units ($25$ units from each account), the total value of pension savings is $970$ units.  Re-balancing the tontine account to the chosen proportion of $50$\% of total pension savings, each account has value $485$ units.}
				\label{FIGaccounts}
\end{figure}

The retiree's savings are allocated to two accounts: the tontine account and the bequest account.  The tontine account is ear-marked as part of a tontine, in which there are infinitely many participants and longevity risk is fully diversified.  There is no uncertainty about the distribution of future lifetimes.  The bequest account is not part of any pooling arrangement.  The accounts are illustrated in Figure \ref{FIGaccounts}.  Both accounts earn investment returns at the same rate and have money withdrawn at the same rate to meet the retiree's spending needs.  Thus both the bequest and the tontine account are used to fund the retiree's spending requirements.  The bequest account is so-named since it is paid to the retiree's estate upon their death.
 
Only the tontine account attracts longevity credits.  The longevity credits are distributions of other participants' accounts held within the tontine, as members of the tontine die.  In turn, when the retiree themselves dies, the retiree's tontine account will be re-distributed among the tontine members.  Only the value of the retiree's bequest account is left to their estate.  Supposing that the force of mortality of retiree is $\lambda (t)$ and their tontine account value is $Y(t)$ at time $t$, then longevity credits are paid into their tontine account at the rate $\lambda (t) Y(t)$.  This is the perfect pool version of \citep{donnellyetal2014, stamos2008}.

We assume that the force of mortality is a deterministic function and that the true force of mortality is the same as the one used to calculate the longevity credits.  It is outside the scope of this paper to analyze the uncertainty about the force of mortality function and is, instead, left to future work.

The key idea for the proposed structure is that the proportion of the retiree's savings in the tontine account is kept constant relative to the total value of the two accounts.  The retiree withdraws money from their accounts to live on, with the amount withdrawn at the same rate from each account.  Similarly, as the two accounts are invested the same way, the account values grow at the same rate of investment return.  This means that, as the two account values change at the same rate due to withdrawals and investment returns, some of the longevity credits earned on the tontine account are transferred across to the bequest account in order to maintain the relative ratio of the two accounts.  

The longevity credits that are transferred out of the tontine account and into the bequest account will not be shared among the tontine participants upon the retiree's death.  For our mathematical set-up, the perfect pool assumption is very important.  It means that this transfer, which can occur among all the tontine participants' accounts, does not affect the rate at which longevity credits are paid to the retiree.  It is left for further work, in a finite pool setting, to determine the magnitude of the impact on the longevity credits of the transfer.  Here, our focus is on understanding the broad implications of the proposed tontine structure.  By assuming a perfect pool, we can obtain analytical solutions which give us high-level insights into the demand for the new product from retirees.

We assume that the tontine mechanism is actuarially fair at all moments in time.  In other words, each retiree's expected value of longevity credits (which are paid to the retiree as long as that retiree survives) equals the expected loss of that retiree's tontine account if they do die, over each very short time period.  This is the justification for allowing the re-balancing the tontine and bequest accounts of a retiree - the longevity credits transferred from the tontine account to the bequest account are owned by the retiree and are not the collective property of the tontine members.  Tontine mechanisms which are actuarially fair at all moments in time are proposed in \citet{donnellyetal2014} and \citet{sabin2010}.

The transfer of some of the longevity credits to the bequest account has interesting potential consequences, which may or may not occur.  The longer-lived retiree is more likely to end up with a higher income and more likely to end up with an income which is paid for life, compared to being in income drawdown alone.  If they live a long time, they may even end up with a higher bequest than in income drawdown situation, assuming that the same investment strategy and the same amounts are withdrawn in both situations.

Now we show how the values of the two accounts develop.  Let
\begin{itemize}
    \item $X(t) =Y(t)+Z(t) =$ Total value of tontine and bequest accounts at time $t$, namely the total value of the retiree's pension savings, where
    \item $Y(t) =$ Value of tontine account at time $t$, and
    \item $Z(t) =$ Value of bequest account at time $t$.
\end{itemize}

To illustrate the functioning of the accounts, suppose the retiree earns a constant annual return $r$ on their accounts and withdraws money at the rate $c(t)$ of their total pension savings at time $t$.  For example, if $c(t)=0.12$ and $X(t)=100\,000$ then the retiree consumes at the rate $c(t)X(t)=12\,000$ per annum at time $t$.  The retiree is assigned a force of mortality function $\lambda(t)$ at each time $t$.  This is where adverse selection can come in.  Adverse selection effects cannot be avoided in a tontine, just as in a life annuity.  However, the financial impact on the tontine member is not the same as in a life annuity, since the tontine members bear their own longevity risk and not an insurer.

The retiree's true force of mortality function can be different to their assigned one.  If their true force of mortality is lower than the assigned one (i.e. they are expected to live longer than implied by their assigned force of mortality) then the retiree will get larger longevity credits than they should get if we used their true force of mortality.

However there is a downside risk for the retiree in being assigned a lower force of mortality than their true one.  The sustainable lifetime income, namely the retiree's total account value divided by a life annuity factor, would be calculated based on their assigned force of mortality.  The retiree will draw down their account value faster than they should, based on their true force of mortality.   The retiree's accounts may be exhausted before their death (assuming a constant amount is withdrawn) or their income could fall to a very low level (assuming a constant percentage of the total account value is withdrawn), if the retiree lives long enough.  

Then, with the force of mortality of the retiree represented by $\lambda(t)$ at time $t$, the dynamics of the total value of the retiree's pension savings are
\[
\mathrm{d}X(t) = \left( r X(t) + \lambda(t) Y(t) - c(t)X(t) \right) \mathrm{d}t.
\]
Fix a constant $\alpha \in [0,1]$ and allocate the amount $\alpha X(t)$ to the tontine account at all times, i.e. $Y(t) = \alpha X(t)$.  This is the heart of the innovative feature of our proposed structure.  By substituting for $Y(t)$ in the last equation we get
\[
\mathrm{d}X(t) = \left( r + \alpha \lambda(t) - c(t) \right) X(t) \mathrm{d}t.
\]
Now consider the dynamics of the tontine account,
\[
\mathrm{d}Y(t) = \alpha d X (t) = \left( r + \alpha \lambda(t) - c(t) \right) Y(t) \mathrm{d}t.
\]
The reason why the term $\alpha \lambda(t) Y(t)$ on the right-hand side of the last equation is not $\lambda(t) Y(t)$ is due to the re-balancing of value from the tontine account to the bequest account.  The longevity credit rate earned on the tontine account, $\lambda(t) Y(t)$, is partially diverted to the bequest account.

The structure interpolates between a pure tontine ($\alpha=1$) to an income drawdown situation with no longevity pooling ($\alpha=0$).  However, the structure is not simply a pure tontine account and a separate income drawdown account, due to the re-balancing feature. 

Since the bequest account value is $Z(t)=(1-\alpha)X(t)$, its dynamics are
\[
\mathrm{d}Z(t) = \left( r + \alpha \lambda(t) - c(t) \right) Z(t) \mathrm{d}t.
\]
Thus, in a sense, both the tontine and the bequest accounts gain longevity credits, one directly and one indirectly through the re-balancing.  The tontine account gains longevity credits at the monetary rate $\lambda(t) Y(t)$, but must then divert part of it, namely the monetary rate $(1-\alpha) \lambda(t) Y(t)$, to the bequest account in order to maintain the desired account ratio.

Having detailed the new tontine structure, a natural question is how much money should be allocated to the tontine account versus the bequest account.  To answer this, we specify an optimization problem in which we allow for different strengths of a bequest motive.  Our aim in this paper is for explicit solutions in order to understand the new tontine structure better.   We give the optimal solution for different possible utility functions, in addition to increasing strengths of bequest motive.  We do not use the $Y$ and $Z$ notation hereafter.

\section{Model and problem set-up}\label{section:model}
Here we being by describing the financial market model and the mortality model within which we solve the utility maximization problem specified in Section \ref{subsection:problem} to determine the optimal strategies.  The parameterization of these models used in the numerical illustrations are detailed in Section \ref{SEC:numerical model parameters}.

\subsection{Notation, financial market model and the retiree's characteristics}\label{subsection:notation}
Let $(\Omega,\mathcal{F},(\mathcal{F}_t),\mathbb{P})$ be a filtered probability space over time $\,t\geq0\,$ measured in years that fulfills the usual conditions and supports a Brownian motion $W\!$, which drives the risky asset price.  The probability space also supports a random variable $\tau>0\,$ which is independent of $(\mathcal{F}_t)$ and represents the random time of death of the representative retiree.

There are two traded assets in the market: a risk-free bond with price process $P$ and a risky stock with price process $E$.  The dynamics of the prices are
\[
\mathrm{d}P(t)=r P(t)\mathrm{d}t \qquad \textrm{ and } \qquad \mathrm{d}E(t) =\mu E(t) \mathrm{d}t+\sigma E(t) \mathrm{d}W(t),
\]
where $\,\mu>r\,$ and $\,\sigma>0\,$ are constants.

The total value of the tontine account and bequest account is represented by the process $X$.  Let $(\mathcal{G}_t)$ be the smallest filtration that makes $\tau$ a stopping time and contains $(\mathcal{F}_t)$.  Denote by $\hat{\mathcal{G}}_n$ the set of all locally $n$-integrable, $(\mathcal{G}_t)$-predictable processes, for $n \in \{1,2\}$.  Let
\begin{itemize}
    \item$\omega\in\hat{\mathcal{G}}_2$ be the proportion of the retiree's total pensions savings that are invested in the risky stock.
    \item$c\in\hat{\mathcal{G}}_1$ be the rate at which total pensions savings are withdrawn by the retiree to meet their consumption needs.  We refer to $c$ as the consumption rate.  It is important to note that $c$ must be multiplied by the value of the total pension savings to find the monetary consumption rate.  For example, if $c(t)=0.1$ and $X(t)=200\,000$ then the monetary rate of consumption is $20\,000$ per annum at time $t$.
    \item the constant $\alpha \in [0,1]$ be the proportion of total pensions savings invested in the tontine account.  The tontine account and bequest account are continuously re-balanced so that this proportion is maintained at all times.
    \item$\lambda$ be the locally integrable force of mortality assigned to the retiree.  The probability of the retiree surviving for at least $t$ years is given by the survival function
    \[
		S(t) := \mathbb{P}[\tau>t]=\exp\Big(-\!\int_0^t\!\lambda(s)\,\mathrm{d}s\Big),\quad\mbox{for $\,t\geq0$}.
		\]
    The retiree receives longevity credits at the rate $\,\alpha\lambda(t)X(t)$ since they have the amount $\alpha X(t)$ invested in the tontine account at each time $\,t\geq 0$.
    \item constant $x_{0}>0$ be the initial total pension savings amount of the retiree.  The amount $\alpha x_{0}$ is initially allocated to the tontine account and the remaining amount $(1-\alpha)x_{0}$ is initially allocated to the bequest account.
\end{itemize}
The total pension savings $X$, i.e. the sum of the tontine account and the bequest account, is a self-financing process with dynamics
\begin{equation}\label{eq:$mathrmdX$}
\frac{\mathrm{d}X(t)}{X(t)} = \Big(r + ( \mu - r ) \omega(t) + \alpha \lambda(t) - c(t)\Big) dt + \sigma \omega(t) \mathrm{d}W(t),
\end{equation}
subject to $X(0)=x_{0}$ a.s.

\subsection{Specification of the retiree's problem}\label{subsection:problem}

To determine how much of their pension savings the retiree should allocate to the tontine account, we take account of the utility that the retiree gains from consumption and their bequest motive.  The general form of the function to be maximized is
\begin{equation}\label{EQNgeneralfn}
\mathbb{E}\Big[\int_{0}^{\tau}\!e^{-\rho s}U\big(c(s)X(s)\big)\,\mathrm{d}s + b e^{-\rho \tau} B\big((1-\alpha)X(\tau)\big)\Big],
\end{equation}
in which $U$ and $B$ are utility functions and the constant $\,b\geq0\,$ measures the strength of the bequest motive relative to the desire for consumption.  The higher the value of $b$, the stronger the bequest motive.  The constant $\,\rho>0\,$ represents the fixed time preference rate of the retiree.  To explain the expression (\ref{EQNgeneralfn}), at the random time of death of the retiree, $\tau$, the value of the bequest account is $(1-\alpha)X(\tau)$, i.e. it is a fraction $1-\alpha$ of the total pension savings account value $X(\tau)$.  The utility derived from the bequest is $B\big((1-\alpha)X(\tau)\big)$, calculated at the time of death.  Before their death, the retiree consumes their pension savings at the rate $c(t)$, $t \in [0,\tau)$.  The utility derived from consumption is, effectively, the sum of the utility of consumption at each time point allowing for the time preference of money.   In our paper, the utility functions $U$ and $B$ are chosen to be either both of the power utility form or both of the logarithmic form.  

\subsection{Model parameterization for the numerical illustrations} \label{SEC:numerical model parameters}
\begin{figure}[H]
    \centering
    \begin{subfigure}[b]{0.5\textwidth}
        \centering
    \includegraphics[trim=0 0.5cm 0 1.5cm,clip,width=\textwidth]{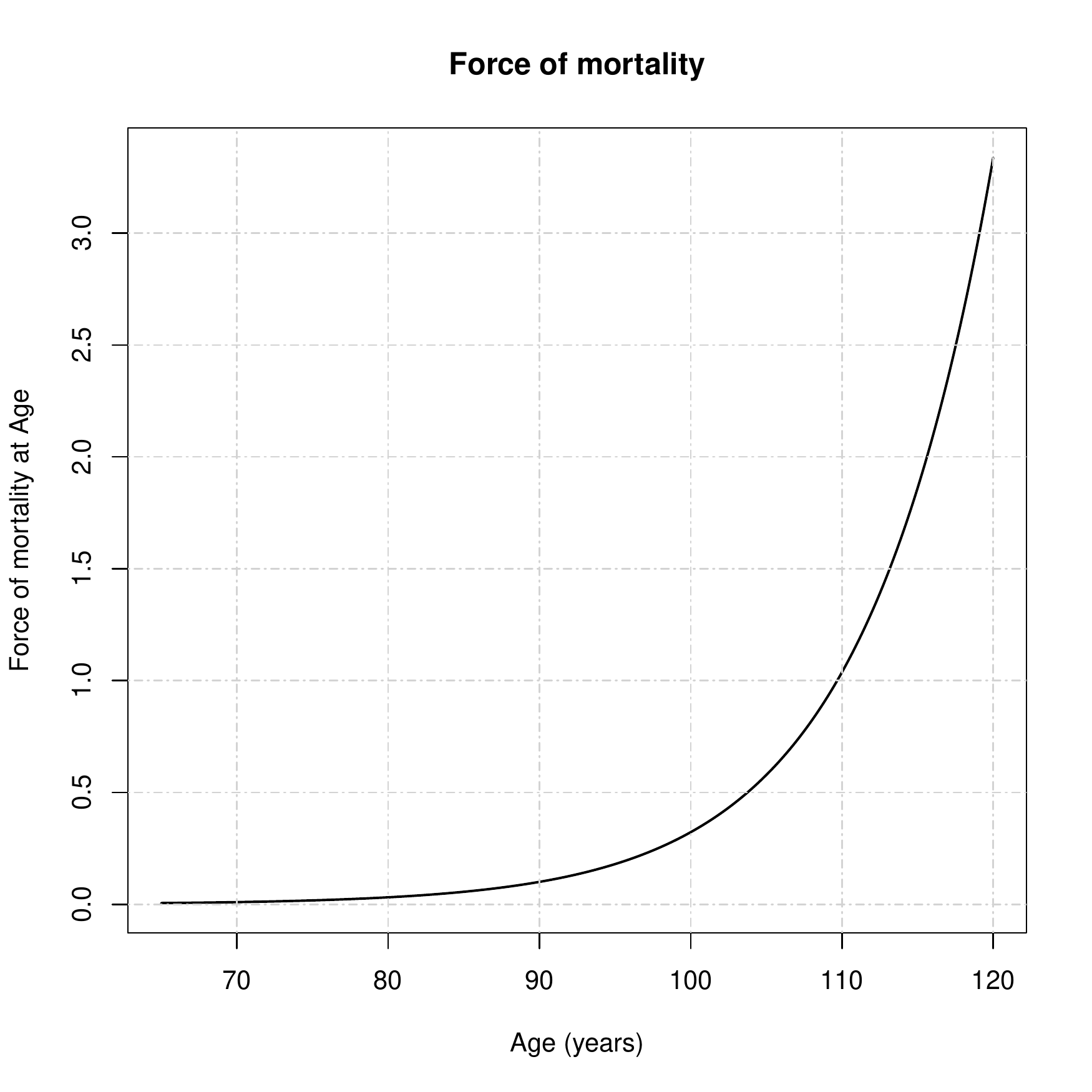}
		\caption{Force of mortality by age.}
		\label{fig:forcemort}
    \end{subfigure}%
    \begin{subfigure}[b]{0.5\textwidth}
        \centering
\includegraphics[trim=0 0.5cm 0 1.5cm,clip,width=\textwidth]{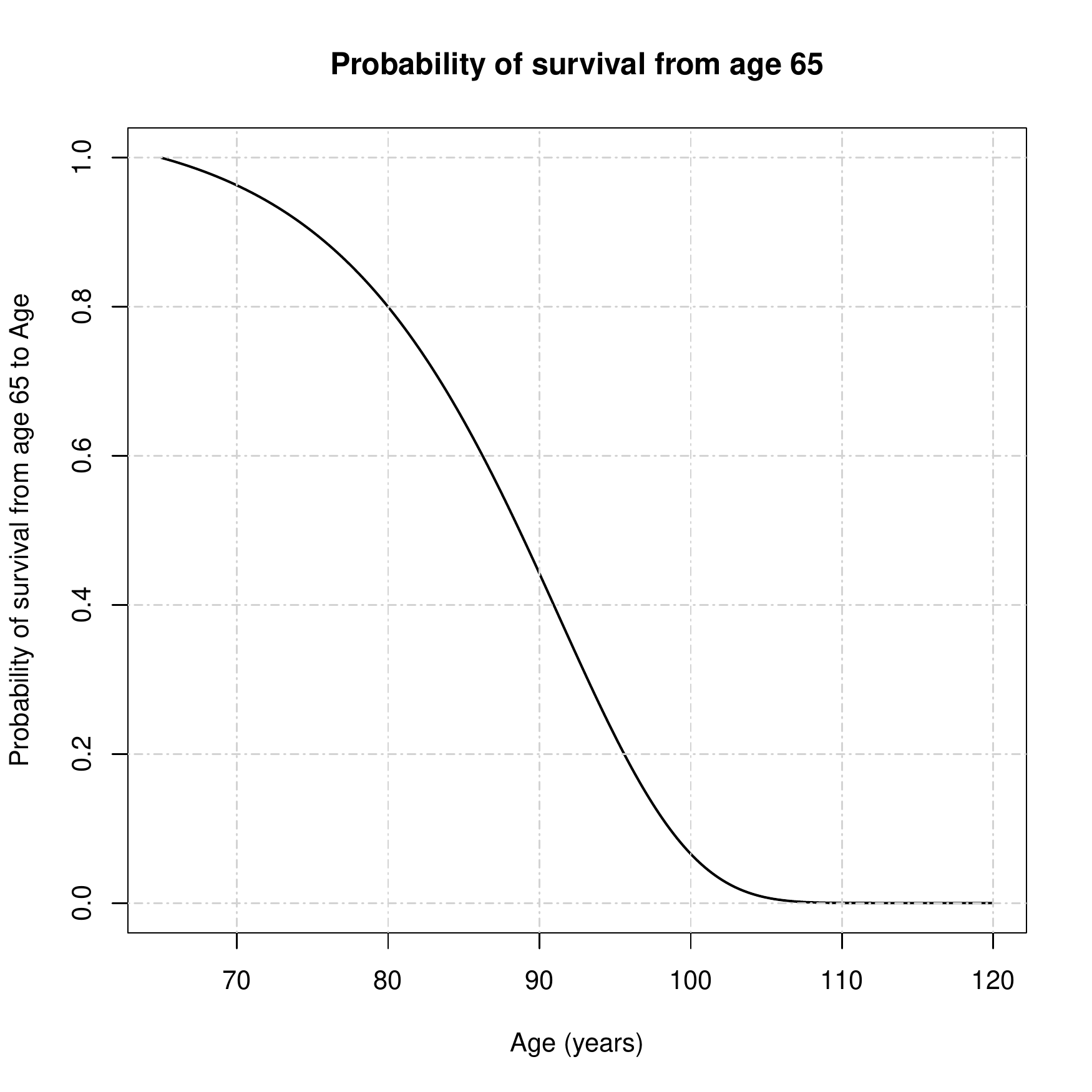}
\caption{Probability of survival from age 65.}
\label{fig:probsurv}
    \end{subfigure}
    \caption{Force of mortality and probability of survival from age $65$ for Makeham's Law of mortality detailed in Section \ref{SEC:numerical model parameters}.}
		\label{FIGMakeham}
\end{figure}

In the numerical illustrations, we assume that the retiree is age $65$ at time $0$.  Their force of mortality follows Makeham's Law so that $\,\lambda(t)=A+BC^{65+t}\,$ with $\,A=2.2\times 10^{-4}$, $\,B=2.7\times 10^{-6}$ and $\,C=1.124\,$ at age $65+t$, for $t\geq0$ \citep[Example 3.13, p.\,$65$]{Dicksonetal2013}.  This mortality law implies that a 65-year-old has a $80\%$ chance of surviving to age 80, a $22\%$ chance of surviving to age 95 and a less than $1\%$ chance of being alive at age 110.  Plots of the force of mortality and the probability of survival from age $65$ are shown in Figure \ref{FIGMakeham}.

The financial market model is a Black-Scholes model, which is detailed in Section \ref{subsection:notation}.  For the parameters, we choose $r=0.05$, $\mu=0.085$ and $\sigma=0.2$.  These are reasonable choices in the sense that the values lie in the range of estimations from empirical studies like \citet{JPMorgan2017}.

For the time preference rate of money, we set $\,\rho=r$.

\section{How to invest in the new tontine structure under power utility}\label{section:Power utility}

To study the maximization of expression (\ref{EQNgeneralfn}) under a power utility function, fix a constant $\,\gamma\in(-\infty,1)\setminus\{0\}$ and choose
\[
U(x):=x^\gamma/\gamma \qquad \qquad \textrm{and} \qquad \qquad B(x):=x^\gamma/\gamma, \qquad \mbox{for $\,x\geq 0$}.
\]
The constant $\gamma$ represents the risk aversion of the retiree and $1-\gamma$ is called the constant relative risk aversion (CRRA) in the literature.   The retiree becomes more and more risk averse as $1-\gamma \rightarrow + \infty$.

The objective function in formula (\ref{EQNgeneralfn}) is well-defined with these choices of utility function. We consider two possible interpretations of the retiree's problem.  These correspond to two sets over which to optimize expression (\ref{EQNgeneralfn}).

In the first interpretation, the retiree has already decided their consumption rate as well as how much to invest in the tontine account.  Thus both the proportion $\alpha$ of savings invested in the tontine account and the consumption rate $c$ are fixed constants.  We determine the optimal investment strategy for each value of $\alpha \in [0,1]$ and $c\geq0$.   While this is a restricted optimization set, since the values of $\alpha$ are constrained to be constant at all times, it may be a more realistic reflection of how the modern tontine with bequest is likely to operate in real life. 
It turns out that the optimal investment strategy is a constant proportion in the risky stock -- the same proportion as in the classical power utility optimization problem \citep[Section 6]{Merton1971} -- for each $\alpha \in [0,1]$.  In this case, we are able to write down an expression for the maximum value of (\ref{EQNgeneralfn}) for fixed $\alpha\in [0,1]$ and $c\geq0$.   We can then find the constant $\alpha\in [0,1]$ and $c\geq0$ that maximizes expression (\ref{EQNgeneralfn}) through a numerical procedure.  The analytical results are shown in Section \ref{subsection:Power utility fixed prop alpha and consumption c} and the numerical results are shown in Section \ref{subsec:Power utility numerical}, with further sensitivity testing in Section \ref{SUBSECsensitivity}. 

In the second interpretation, the retiree has already decided how much to invest in the tontine account.  In other words, the proportion $\alpha$ of savings invested in the tontine account is fixed in advance.  The problem is then to determine the optimal investment and consumption strategy corresponding to each value of $\alpha \in [0,1]$.  This is a less restrictive optimization set than the first interpretation, since the consumption rate is not constrained to be a constant.

Once again, the optimal investment strategy is a constant proportion in the risky stock, the same proportion as in the first interpretation.  The optimal consumption rate is a deterministic function of time.  The optimal strategies are given in Section \ref{subsection:Power utility fixed prop alpha}.  However, we do not run the numerical results as we wish to keep the focus on the more realistic first interpretation; retirees are more likely to choose the simple consumption rule in the first interpretation than a time-varying one as in this second interpretation.  We include this example more for mathematical completeness - this is the closest we can get to an analytical solution through the relaxation on the consumption rate.

We look next at the results of these two optimizations.  The technically complicated proofs are relegated to the Appendix, with some preliminary technical results shown in Appendix \ref{APPSUBSECpreliminaries}.

\subsection{Power utility with a fixed proportion $\alpha$ and a fixed consumption rate $c$}\label{subsection:Power utility fixed prop alpha and consumption c}

As we seek an explicit answer to understand the impact of the bequest motive, we avoid the difficulty of finding the optimal consumption strategy and assume that $c$ is an unknown constant.  We optimize over a restricted set of constant consumption rate strategies.  This means that the expression to maximize (\ref{EQNgeneralfn}) stays the same, but the set of controls over which it is maximized changes.  Our new objective is to find an investment strategy $\,\omega\in\hat{\mathcal{G}}_2$, a positive constant for consumption $\,c\geq0\,$ and a constant proportion in the tontine $\alpha\in[0,1]$ for the random lifespan $\tau$ so that
\begin{equation}\label{eq:$sup_omega,constants$}\sup_{\substack{c,\alpha\\\mbox{constant}}}\;\sup_{\omega\in\hat{\mathcal{G}}_2}\mathbb{E}\Big[\int_{0}^{\tau}\!e^{-\rho s}\big(cX(s)\big)^\gamma/\gamma\;\mathrm{d}s+be^{-\rho \tau}\big((1-\alpha)X(\tau)\big)^\gamma/\gamma\Big]\end{equation}
is attained.  The proofs for this section are shown in Appendix \ref{APPSUBSECfixedconsumandprop}.

With the power utility function, the optimal proportion of total pension savings to invest in the risky stock is
\begin{equation}\label{eq:$omega^*_gamma=$}
\omega^*(t):=w^*:=\frac{1}{1-\gamma}\frac{\mu-r}{\sigma^2}.
\end{equation}
The proof of the optimality of the investment strategy in equation (\ref{eq:$omega^*_gamma=$}) is shown in Section \ref{APPSUBSECfixedconsumandprop}.

To examine the finiteness of the expression (\ref{eq:$sup_omega,constants$}), define
\begin{itemize}
    \item A positive random variable $A$ with the tail distribution
    \[
		\mathbb{P}[A>t]=S(t)^{1-\gamma\alpha},
		\]
    in which the survival function $\,S(t)=\mathbb{P}[\tau>t]=\exp(-\int_0^t\!\lambda(s)\,\mathrm{d}s)\,$ for $\,t\geq0$.
    \item The exponential moment of $A$ for any $\,p\in\mathbb{R}$, i.e.
    \[
		M_A(p)=\mathbb{E}[\mathrm{e}^{pA}].
		\]
    \item A constant $\,k\in\mathbb{R}\,$ defined as
    \begin{equation}\label{eq:$k=$}
		k:= \frac{1}{2} \frac{\gamma}{\gamma-1} \left( \frac{\mu-r}{\sigma} \right)^{2} + \gamma(c-r) + \rho.
		\end{equation}
\end{itemize}
For fixed $c$ and $\alpha$, expression (\ref{eq:$sup_omega,constants$}) is finite if and only if one of two possible cases hold.  Either we are in the very special case that $\,\alpha=1,\,c=0,\,\gamma>0\,$ which results in (\ref{eq:$sup_omega,constants$}) equalling zero.  However, this is not of interest since it means that the consumption rate is zero.  The more interesting case is that the expression is finite when
\begin{equation}\label{eq:$M_A(-k)<infty$}
\mbox{either, for $\,k\neq0$,}\quad M_A(-k)<\infty \qquad \mbox{or,} \qquad \mbox{for $\,k=0$,} \quad\mathbb{E}[A]<\infty.
\end{equation}
 The condition (\ref{eq:$M_A(-k)<infty$}) is a technical assumption.  In practice, the force of mortality increases exponentially, so that $\,M_A(-k)<\infty\,$ no matter the value of $k$.  For example, the Makeham's Law of mortality fulfills this condition.

Furthermore, the technical condition is only a true restriction on $k$ in highly simplified models. For example, under a constant force of mortality $\lambda>0$, the condition (\ref{eq:$M_A(-k)<infty$}) holds true if and only if $\,k>-(1-\gamma\alpha)\lambda$.

When the condition (\ref{eq:$M_A(-k)<infty$}) holds and the optimal strategy (\ref{eq:$omega^*_gamma=$}) is followed, we find
\begin{align}\nonumber\mathbb{E}\Big[\int_{0}^{\tau}\!e^{-\rho s} & \big(cX^{\star}(s)\big)^{\gamma} \,\mathrm{d}s+be^{-\rho \tau} \big((1-\alpha)X^{\star}(\tau)\big)^{\gamma} \Big]
\\\label{eq:$sup=$SumOfMoments}=&\begin{cases}\displaystyle X^{\star}(0)^\gamma \Bigg( b \frac{(1-\alpha)^\gamma}{1-\gamma\alpha} M_{A_\gamma}(-k) + \gamma c^\gamma \frac{1-M_{A_\gamma}(-k)}{k} \Bigg)&\mbox{if $\,k\neq0$},
\\[1em]\displaystyle X^{\star}(0)^\gamma \Bigg(b \frac{(1-\alpha)^\gamma}{1-\gamma\alpha} + \gamma  c^\gamma\mathbb{E}[A_\gamma]\Bigg)&\mbox{if $\,k=0$},\end{cases}\end{align}
in which the total pension savings $X^{\star}(t)$ at $\,t\geq0\,$ is given by
\[
X^*(t) = x_{0} \exp \left( \Big( r + \Big(\frac{1}{2}-\Big(\frac{\gamma}{1-\gamma}\Big)^2 \Big)\Big( \frac{\mu-r}{\sigma} \Big)^2 \Big) t + \int_0^{t} \alpha\lambda(u)-c\;\mathrm{d}u + \frac{\mu-r}{\sigma} W(t) \right).
\]

The consumption rate has been replaced by a constant rate $c$ in the dynamics of the pension savings $X^{\star}$ as it is assumed to be a constant in the optimization problem.

Even though equation (\ref{eq:$sup=$SumOfMoments}) is explicit, there is no analytic way to solve for the point of global maximum with respect to $\alpha$ and $c$, with the chosen Makeham's Law mortality model (detailed in Section \ref{SEC:numerical model parameters}).   These global maximums correspond to the optimal constant proportion of savings invested in the tontine and the optimal constant rate of consumption of the problem (\ref{eq:$sup_omega,constants$}).  We can only find them numerically, which we do next in Section \ref{subsec:Power utility numerical}.

\subsubsection{Numerical results under power utility}\label{subsec:Power utility numerical}
In this section, we investigate numerically how the optimal constant proportion in the tontine account and the optimal constant consumption rate, calculated by an evaluation of equation (\ref{eq:$sup=$SumOfMoments}), change as the retiree's risk aversion and strength of bequest motive changes.  Our aim is to understand the behaviour of the optimal constant proportion of pensions savings in the tontine account -- its volatility and magnitude -- as $\gamma$ and $b$ are varied.  We also study the optimal constant consumption rate.

\begin{figure}[H]\centering
	\includegraphics[trim=0 0.5cm 0 1.5cm,clip,scale=0.6]{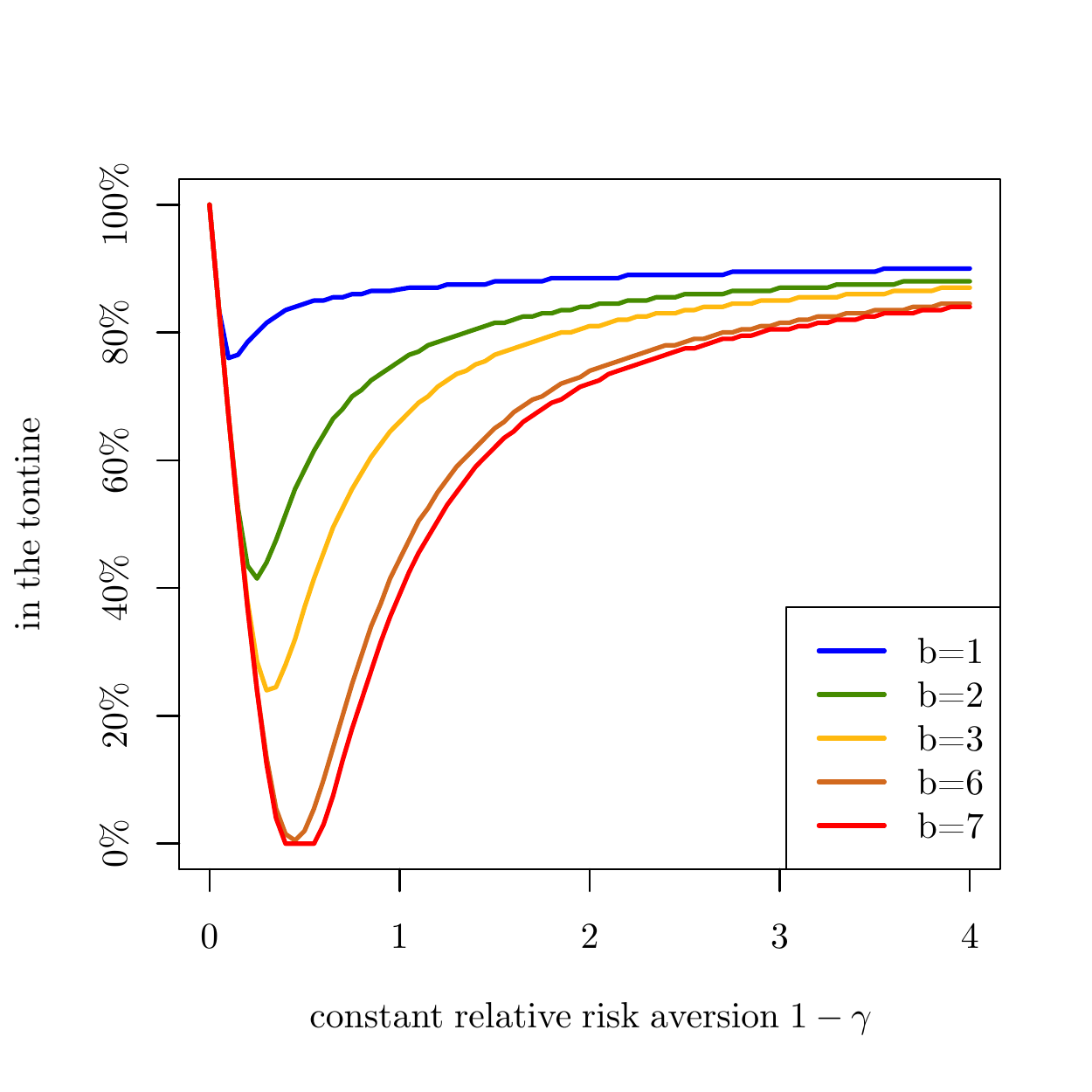}
	\caption{The optimal constant percentage $100\alpha$ of total pension savings in the tontine account as risk aversion $1-\gamma$ increases, for various strengths of bequest motive $b$, under power utility.}
		\label{fig:PowerTontineCRRA}
\end{figure}

We calculate how the optimal constant percentage $100\alpha$ of pension savings to invest in the tontine account varies with the risk aversion coefficient in the power utility function, allowing for different strengths of the bequest motive (Figure \ref{fig:PowerTontineCRRA}).  The optimal percentage is constrained to lie in $[0,1]$ but it turns out that only the lower bound of zero is binding in our numerical study.  With the objective of maximizing the sum of the expected discounted utility of consumption and the bequest account, there are some unexpected effects.  These effects are strongest with the highest considered strength of bequest motive ($b=7$), so we focus the reader's attention on those results first.

First note that, if there was no bequest motive (corresponding to $b=0$), then it is optimal for the retiree to put all of their pension savings in the tontine account.  This result is clear without doing any mathematical optimization.  With no bequest motive, the retiree can maximize their consumption as they do not want to leave any money to their estate.  However, once we allow additionally for a bequest motive, the optimal percentage in the tontine account shows unusual behaviour.

As the retiree becomes less risk averse (i.e.\ $1-\gamma$ falls from value $4$ to around value $0.5$), the percentage in the tontine account falls.  This makes sense; the retiree wishes to leave money to their estate rather than to the tontine members.  They are willing to sacrifice higher consumption in order to leave a bigger bequest account.  However, as the retiree becomes even less risk averse (i.e.\ $1-\gamma$ falls from value $0.5$ to $0$), this effect is sharply reversed.  The retiree allocates more money to the tontine account.

\begin{figure}[H]
    \centering
    \begin{subfigure}[b]{0.5\textwidth}
        \centering
\includegraphics[trim=0 0.5cm 0 1.5cm,clip,width=\textwidth]{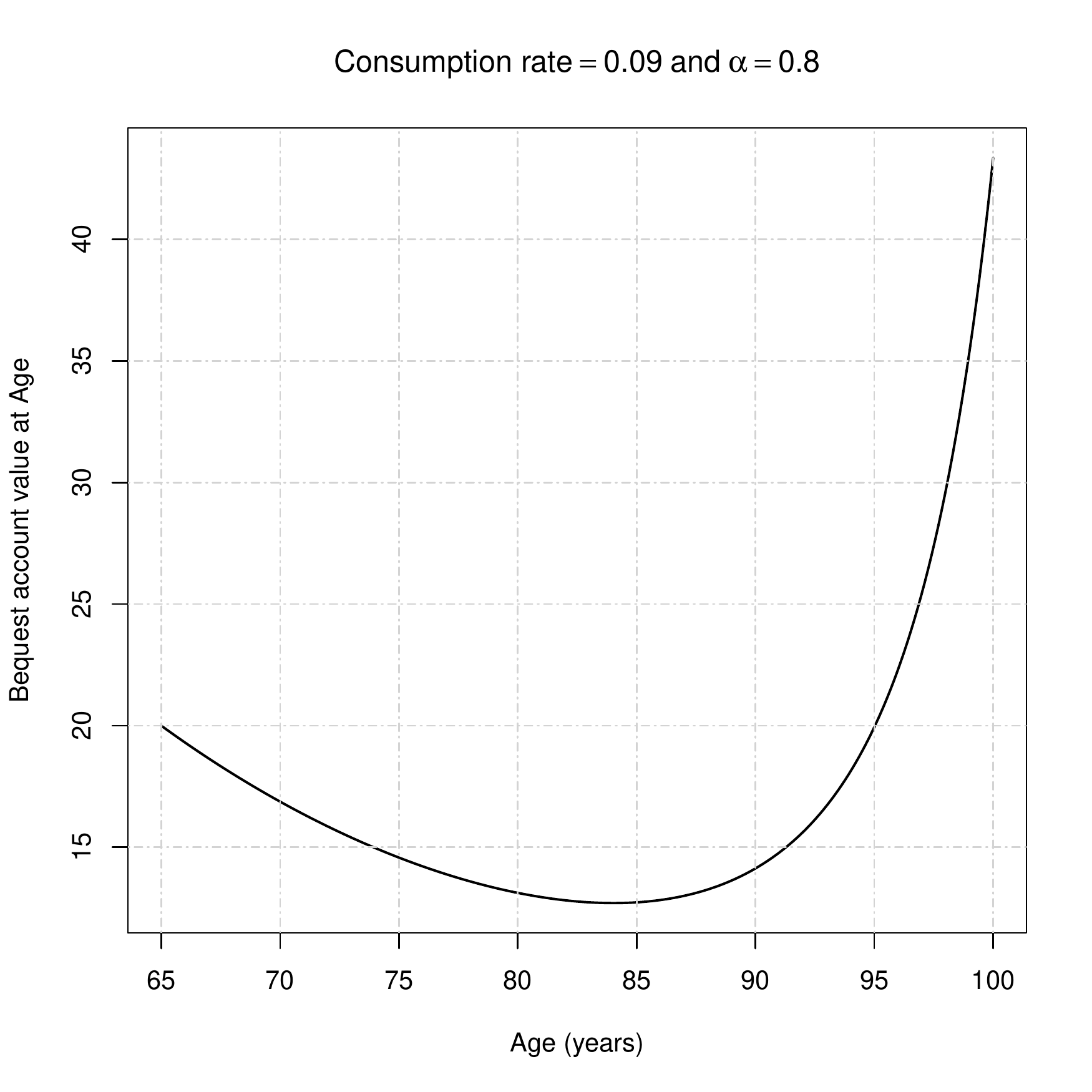}
\caption{Bequest account value up to age 100.}
\label{fig:beqaccrestrict100}
    \end{subfigure}%
    \begin{subfigure}[b]{0.5\textwidth}
        \centering
\includegraphics[trim=0 0.5cm 0 1.5cm,clip,width=\textwidth]{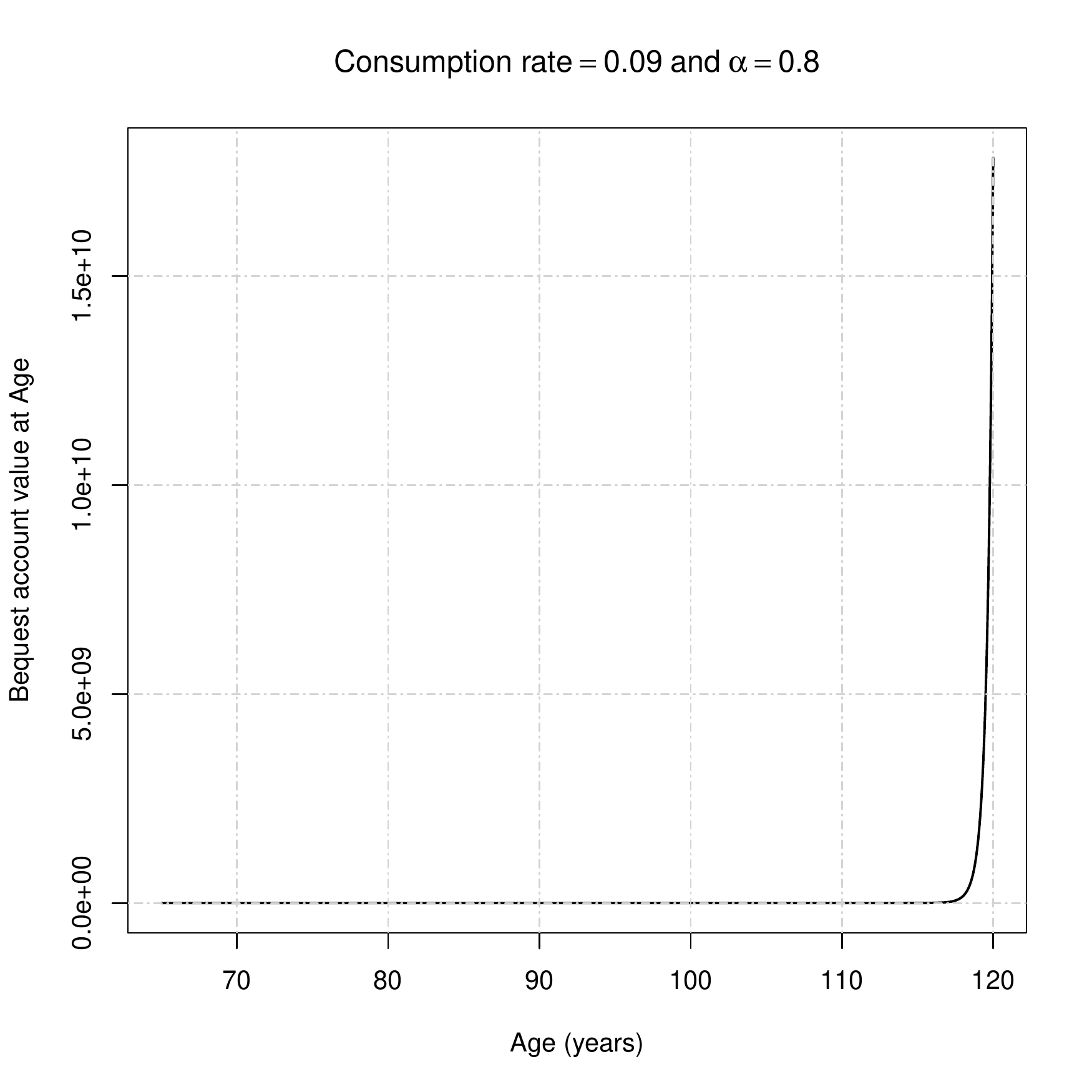}
\caption{Bequest account value up to age 120.}
\label{fig:beqacc}
    \end{subfigure}
    \caption{Bequest account values up to different ages, for the numerical analysis in Section \ref{subsec:Power utility numerical}..}
		\label{FIGBequestAcc}
\end{figure}

The explanation for this surprising answer is that the bequest account value gets very large, if the retiree lives a very long time.  A sample bequest account value from age $65$ to age $100$ is shown in Figure \ref{fig:beqaccrestrict100}.  This illustrative figure assumes that the accounts are invested $100$\% in the risk-free bond which accumulates interest at the annual rate of $5$\% (unlike our numerical results which allow for investment in the risky stock), consumption is a constant rate of $9$\% of the total pension savings value and $80$\% of the pension savings are allocated to the tontine account.  The bequest account value falls from value $20$ at age $65$ to just below value $13$ at age $85$.  It then climbs to about $43$ at age $100$.  There is a $6.6\%$ chance of a retiree surviving from age $65$ to age $100$ in our model.  Past age $100$, the bequest account value starts increasing rapidly (Figure \ref{fig:beqacc}), due to the longevity credits increasing as the retiree's force of mortality grows (Figure \ref{fig:forcemort}).  At age $110$, the bequest account value is close to $4\,000$ while at age $120$ the value is $17.84$ billion.  

\begin{figure}[H]\centering
	\includegraphics[trim=0 0.5cm 0 1.5cm,clip,scale=0.6]{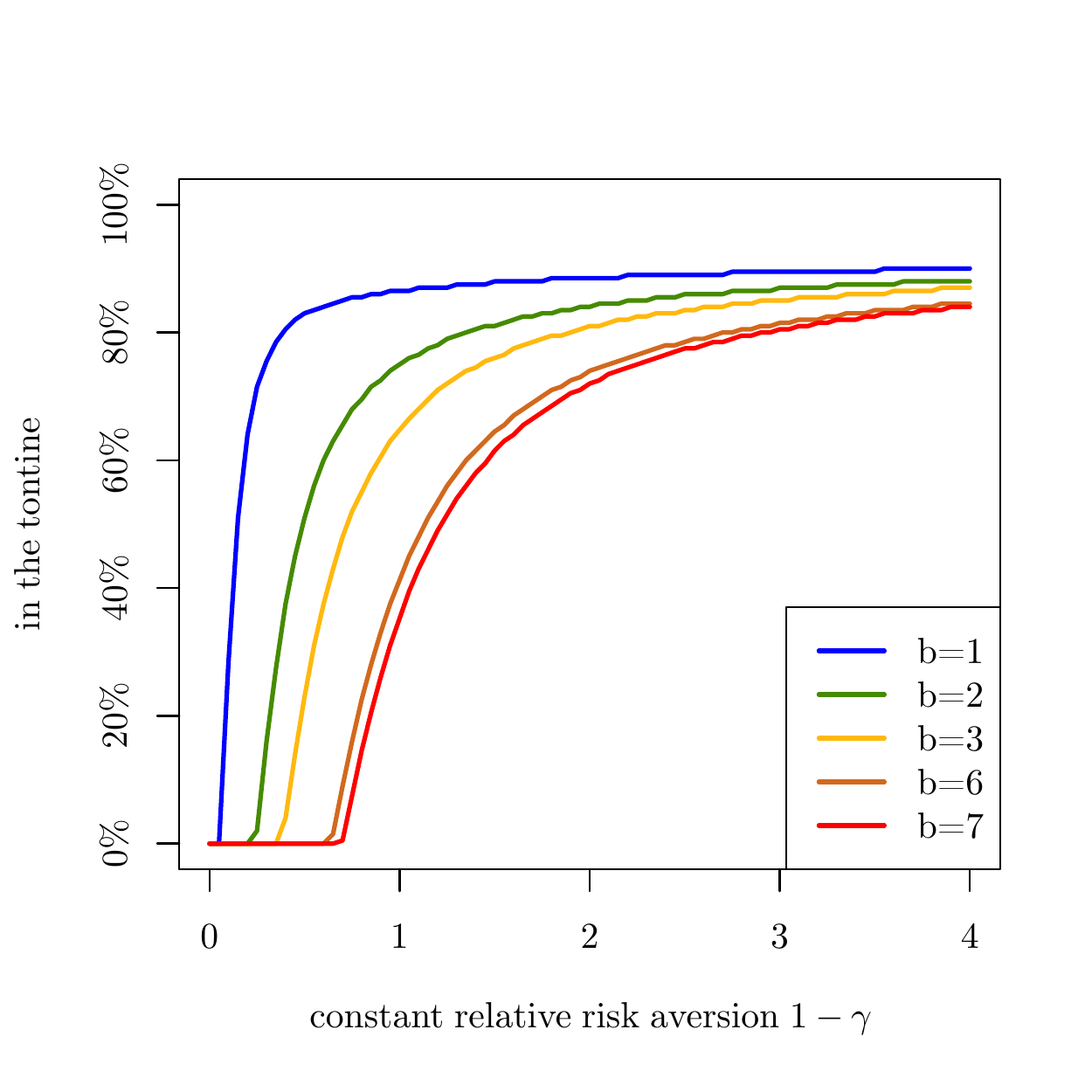}
	\caption{The optimal constant percentage $100\alpha$ of total pension savings in the tontine account as risk aversion $1-\gamma$ increases, for various strengths of bequest motive $b$, under power utility.  The optimization takes only the first $35$ years from age $65$ until age $100$ into account.  Thus the participants do not gain utility from what may happen after age $100$.}
	    \label{fig:PowerTontineCRRA35years}
\end{figure}

To further support our explanation of the impact of extremely high bequest account values obtained at very old ages, we re-did our calculations but removed any ``value'' from the value function past age $100$.  This eliminates the utility gain from anything that may happen after age $100$ and thus eliminates the utility gain from the extremely high bequest account values seen at very old ages.  The result of this adjusted calculation show in Figure \ref{fig:PowerTontineCRRA35years} that, indeed, we do not see the percentage allocated to the tontine increasing towards $100\%$ for the least risk averse retirees ($1-\gamma$ roughly between $0$ and $0.5$).  Instead, the tontine account percentage falls to zero.

It is also important to note that very few retirees survive long past age $100$: the survival probability from age $65$ to age $110$ is $0.015 \times 10^{-2}$ and to age $120$ is $4.15 \times 10^{-13}$.  However, the impact of retiree longevity on all the tontine participants should be studied in a more realistic set-up than the one assumed in this paper, which assumes that the mortality is always perfectly pooled.

With low levels of risk aversion, the retiree is gambling on living a very long time and thus leaving a very large bequest to their estate.  This effect is compounded by the desire to maximize the expected discounted utility of consumption, which also leads the retiree to put even more money into the tontine account.  These two effects working in the same direction results in the optimal percentage to allocate to the tontine account increasing sharply as $1-\gamma$ falls from value $0.5$ to $0$.  As the strength of the bequest motive diminishes, the same overall pattern is observed although it is less pronounced.

Furthermore, the results are very sensitive for retirees with low levels of risk aversion and a desire to leave a bequest, as can be seen from the rapid change in the allocation to the tontine account for these retirees (\ref{fig:PowerTontineCRRA}).  This suggests that, for these retirees, a set up that is less sensitive to the precise choice of parameter values should be found.

The other interesting effect seen in Figure \ref{fig:PowerTontineCRRA} is that as the retiree becomes more risk averse, it is optimal to invest a little more than $80\%$ of pension savings in the tontine account.  Even as the strength of bequest motive increases, this percentage is relatively stable as risk aversion increases.  The convergence to around $80$\% of savings in the tontine account as the risk aversion increases  suggests that the bequest motive plays a much smaller role in the expected utility gains for more risk averse retirees.

\begin{figure}[H]\centering
\includegraphics[trim=0 0.5cm 0 1.5cm,clip,scale=0.6]{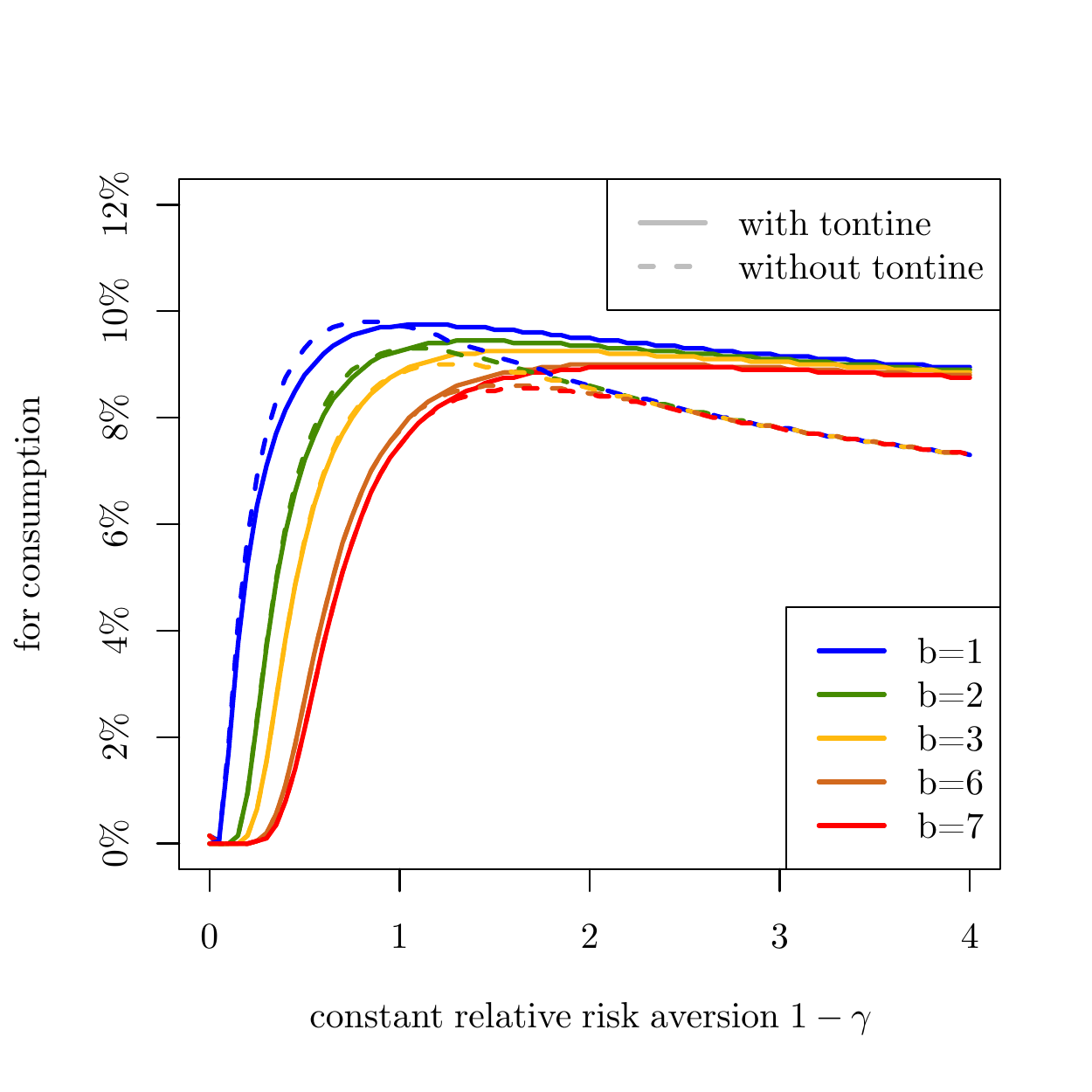}
\caption{The optimal consumption rate as a percentage of the total pension savings as risk aversion increases, for various strengths of bequest motive, under power utility.  The results when the retiree can allocate a constant proportion of total pension savings to a tontine account are shown by the solid lines.  The results when the retiree can allocate money to a bequest account only (the classical income drawdown problem) are shown by the dashed lines.}
\label{fig:PowerConsumptionCRRA}
\end{figure}

We also calculate how the optimal constant rate $c$ to withdraw from the pensions savings varied with the risk aversion coefficient (the solid lines in Figure \ref{fig:PowerConsumptionCRRA}, labelled ``with tontine'') for different strengths of the bequest motive.   As a benchmark, we calculated the optimal consumption rate when $\alpha=0$, i.e. a version of the classical consumption-investment power utility maximization problem, which is an income drawdown situation (the dashed lines in Figure \ref{fig:PowerConsumptionCRRA}, labelled ``without tontine'').

The change in the optimal consumption rate as the risk aversion increases is similar in both structures - the modern tontine with bequest (``with tontine'') and income drawdown (``without tontine'').  However, the optimal consumption rate is higher for the modern tontine with bequest contract as is expected.  The increase in the optimal consumption rate is higher, the more risk averse is the retiree.

\subsubsection{Sensitivity analysis for the percentage in the tontine} \label{SUBSECsensitivity}
The numerical results for the power utility function in Section \ref{subsec:Power utility numerical} showed that around $80$\% of pensions savings are allocated to the tontine account, for a risk averse retiree.  This percentage is surprisingly stable across the different strengths of bequest motive.  How sensitive is this result to variations in the parameters?

\begin{figure}[H]
    \centering
    \begin{subfigure}[b]{0.49\textwidth}
        \centering
	\includegraphics[trim=0 0.5cm 0 1.5cm,clip,width=\textwidth]{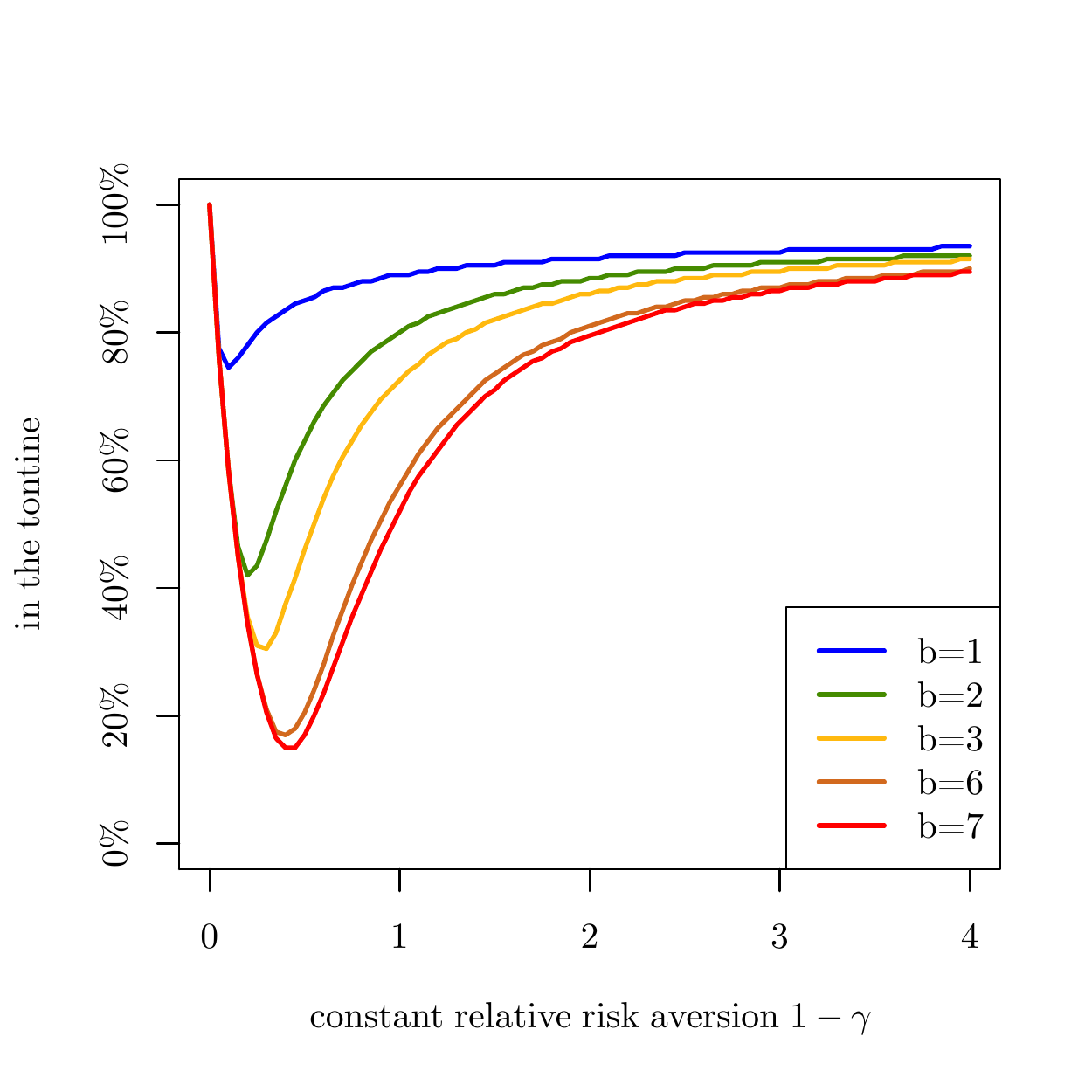}
	\caption{$(r,\mu,\sigma)$$=$$(1\%, 3\%, 15\%)$}
	\label{fig:C1.116r0.01m0.03s0.15}
    \end{subfigure}%
    \begin{subfigure}[b]{0.49\textwidth}
        \centering
	\includegraphics[trim=0 0.5cm 0 1.5cm,clip,width=\textwidth]{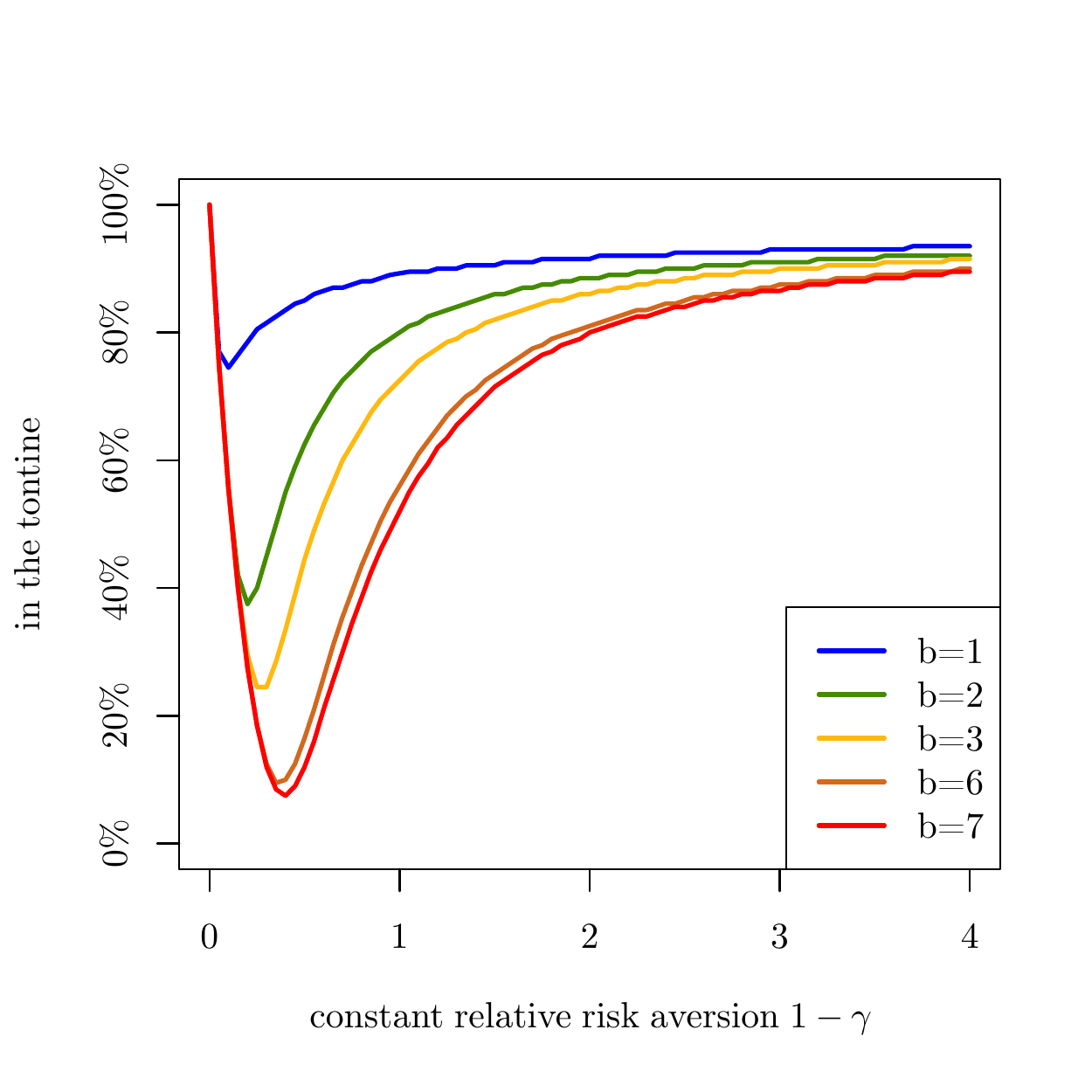}
	\caption{$(r,\mu,\sigma)$$=$$(1\%, 3\%, 25\%)$}
	\label{fig:C1.116r0.01m0.03s0.25}
    \end{subfigure}
		\vfill
    \begin{subfigure}[b]{0.49\textwidth}
        \centering
	\includegraphics[trim=0 0.5cm 0 1.5cm,clip,width=\textwidth]{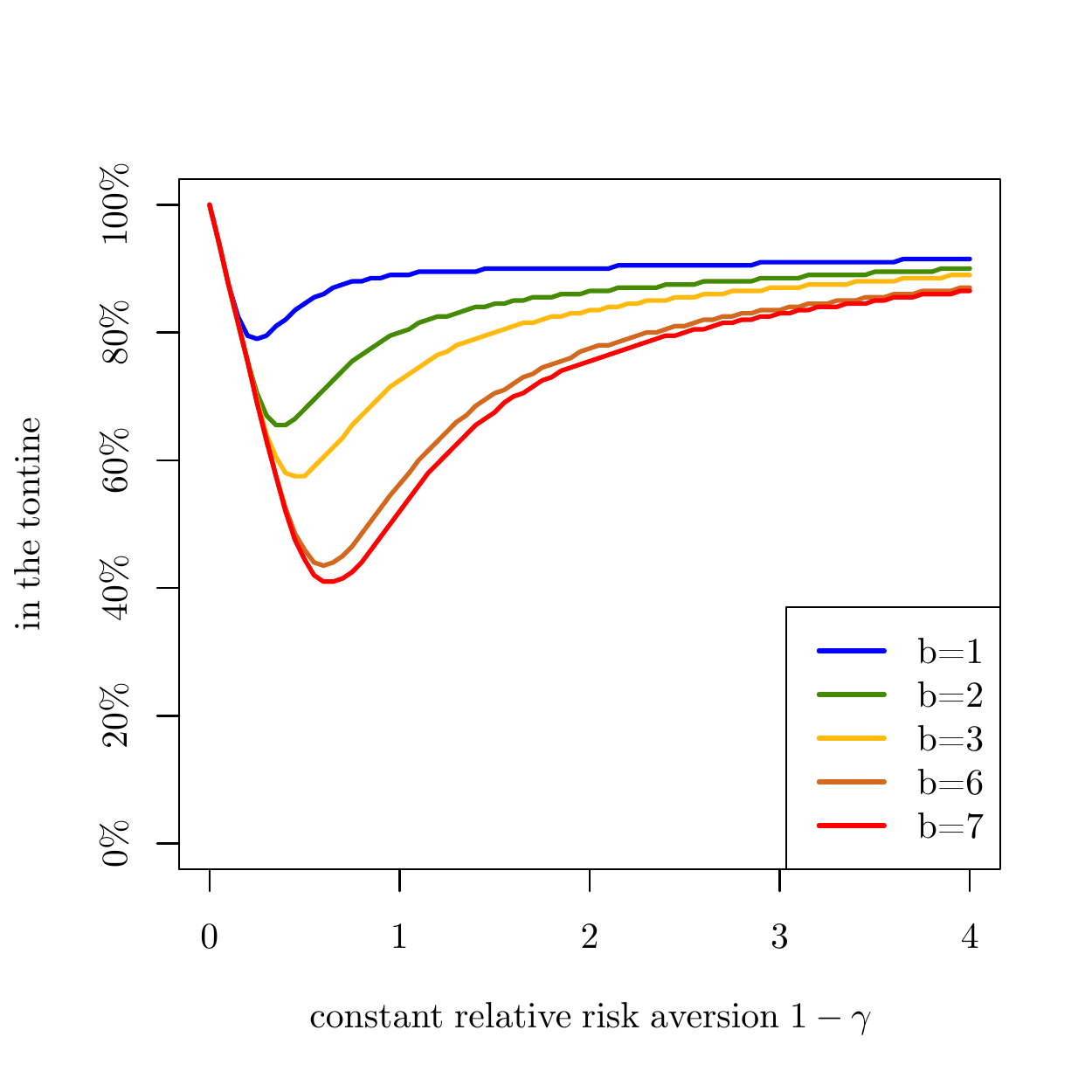}
	\caption{$(r,\mu,\sigma)$$=$$(1\%, 10\%, 15\%)$}
	\label{fig:C1.116r0.01m0.10s0.15}
    \end{subfigure}
    \begin{subfigure}[b]{0.49\textwidth}
        \centering
	\includegraphics[trim=0 0.5cm 0 1.5cm,clip,width=\textwidth]{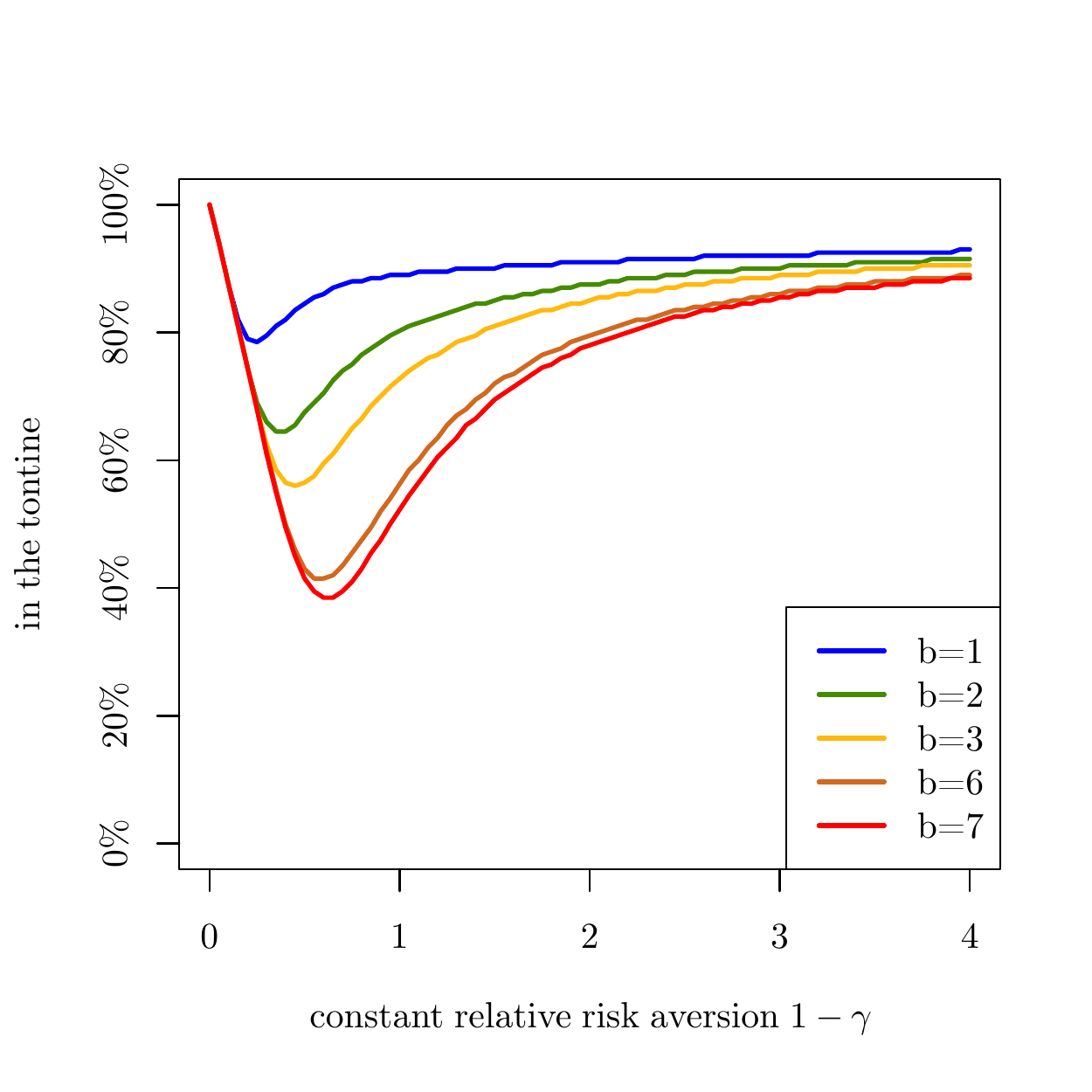}
	\caption{$(r,\mu,\sigma)$$=$$(1\%, 10\%, 25\%)$}
	\label{fig:C1.116r0.01m0.10s0.25}
    \end{subfigure}
		\vfill
	    \begin{subfigure}[b]{0.49\textwidth}
        \centering
	\includegraphics[trim=0 0.5cm 0 1.5cm,clip,width=\textwidth]{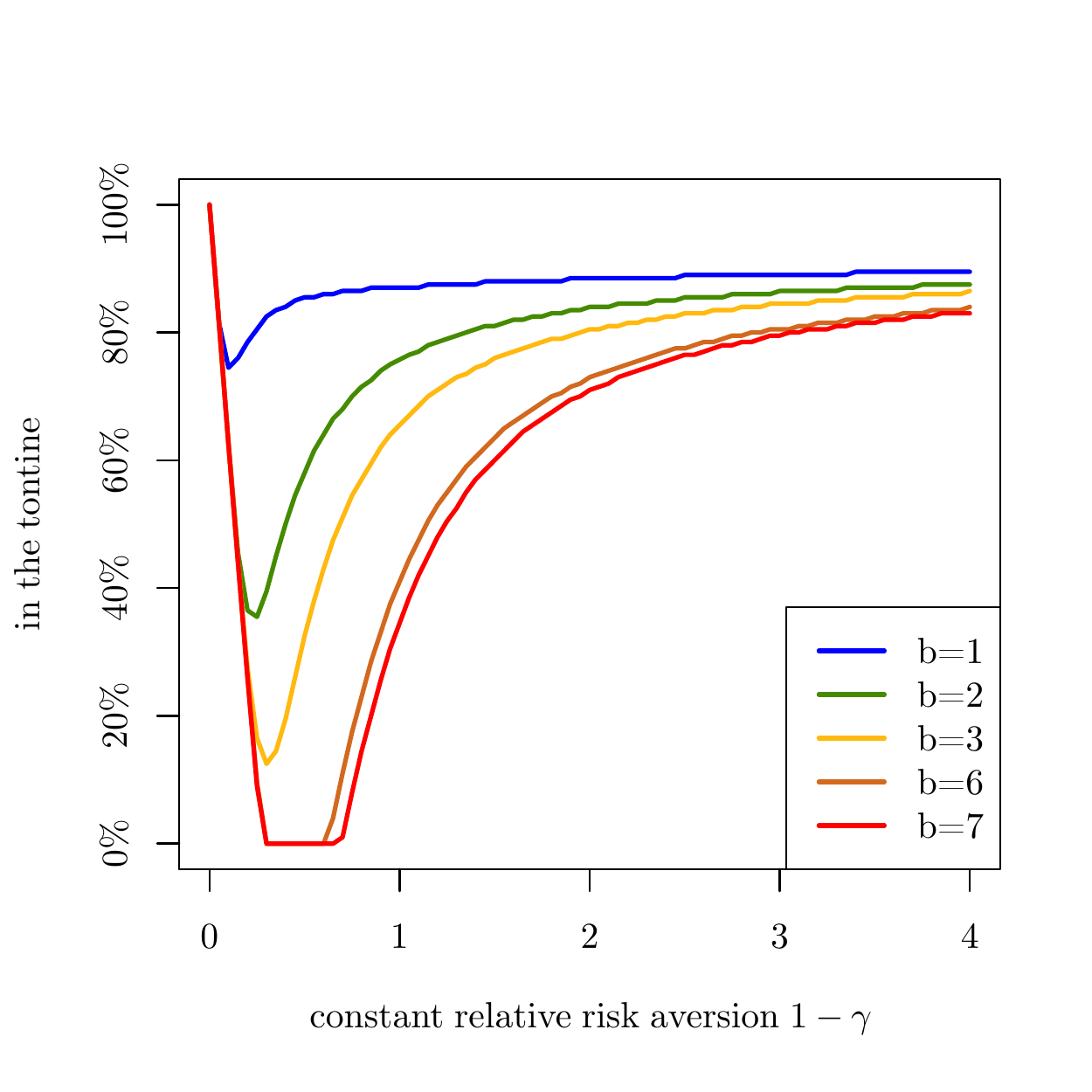}
	\caption{$(r,\mu,\sigma)$$=$$(7\%, 10\%, 15\%)$}
	\label{fig:C1.116r0.07m0.10s0.15}
    \end{subfigure}%
    \begin{subfigure}[b]{0.49\textwidth}
        \centering
	\includegraphics[trim=0 0.5cm 0 1.5cm,clip,width=\textwidth]{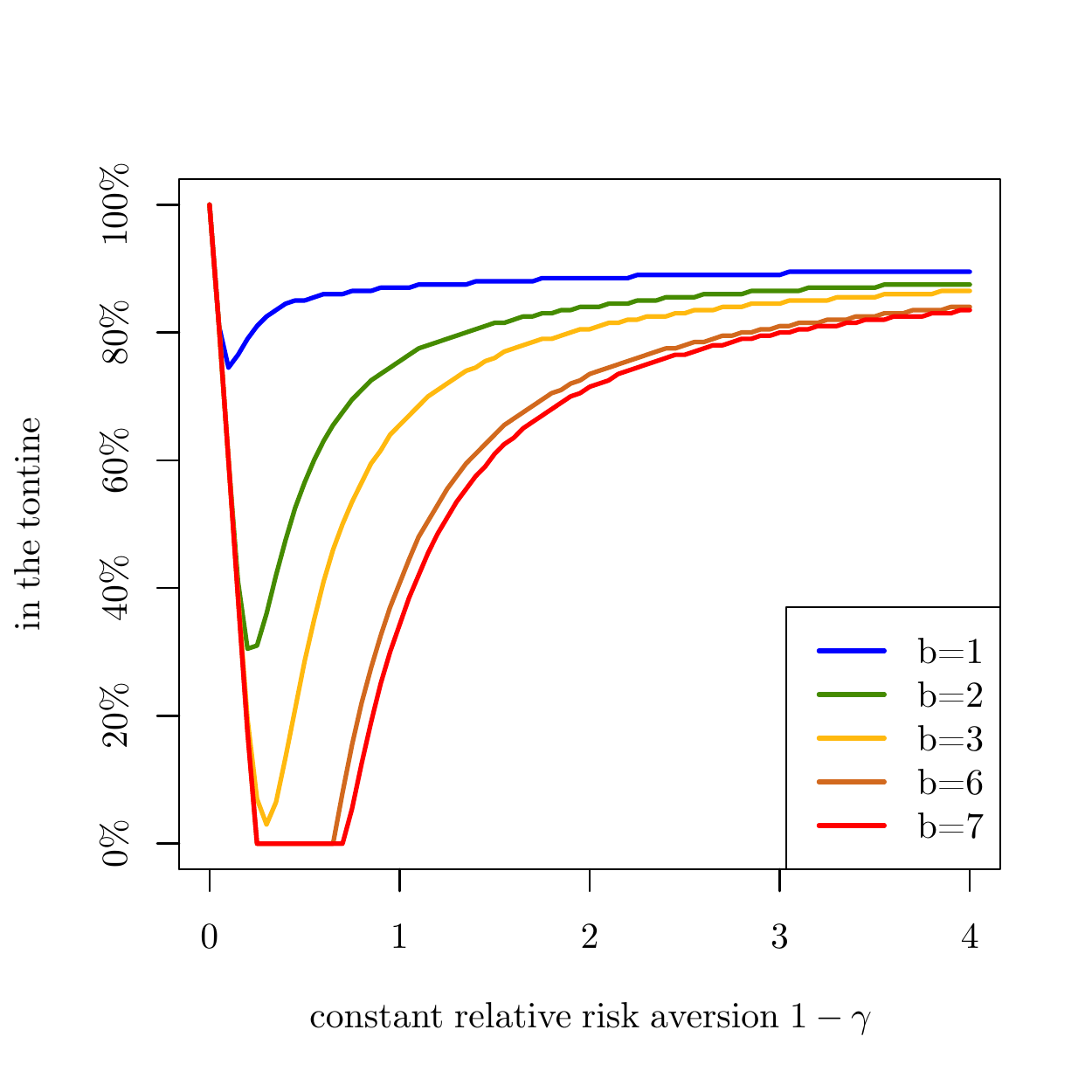}
	\caption{$(r,\mu,\sigma)$$=$$(7\%, 10\%, 25\%)$}
	\label{fig:C1.116r0.07m0.10s0.25}
    \end{subfigure}
    \caption{Sensitivity analysis of the optimal percentage allocated to the tontine account when the life expectancy from age $65$ is increased by $5$ years.  This is the first analysis discussed in Section \ref{SUBSECsensitivity}, and the relevant financial market model parameter values are shown below each figure.  The mortality model parameter $C=1.134$ in the table, which corresponds to an increased life expectancy from age $65$ of $5$ years, compared to the original life expectancy when $C=1.124$.}
		\label{FIGSensHigherLifeExp}
\end{figure}

To answer our question, we do a sensitivity analysis and find that, in general, it is not very sensitive.  Again we constrain the proportion of total pension savings allocated to the tontine account to the range $[0,1]$.  We begin by increasing the life expectancy from age $65$ by $5$ years, through changing one of the mortality model parameters from $C=1.124$ to $C=1.116$.  Then the optimal proportion in the tontine account is calculated under various financial market model parameters.  The full numerical results are displayed in Figure \ref{FIGSensHigherLifeExp}.  Note that the corresponding original parameters given in Section \ref{SEC:numerical model parameters} are $r=5$\%, $\mu=8.5$\%, $\sigma=20$\% and $C=1.124$.   

For different sets of the parameters, the optimal percentage of total pension savings $100\alpha$ to put in the tontine account is quite stable, for the more risk averse retirees (i.e. $1-\gamma \geq 3$).  It falls roughly in the range $80\%-95$\%.  However, the variability in the optimal percentage increases as investors become less risk averse and it also increases for those with a strong bequest motive.

A lower risk-free interest rate (the parameter $r$) results in a higher optimal percentage in the tontine account.  The longevity credits compensate for the lower interest rate, incentivizing the retiree to invest more in the tontine account.

Less volatility in the stock price (the parameter $\sigma$) results in a higher optimal percentage in the tontine account.  As the stock price volatility decreases, uncertainty about the financial market outcomes decreases.  Consequently, uncertainty about the monetary consumption rate and the bequest amount decreases.  This means that as $\sigma$ decreases, more money can be allocated to the tontine account to increase consumption while maintaining a similar bequest amount.

\begin{figure}[H]
    \centering
    \begin{subfigure}[b]{0.49\textwidth}
        \centering
	\includegraphics[trim=0 0.5cm 0 1.5cm,clip,width=\textwidth]{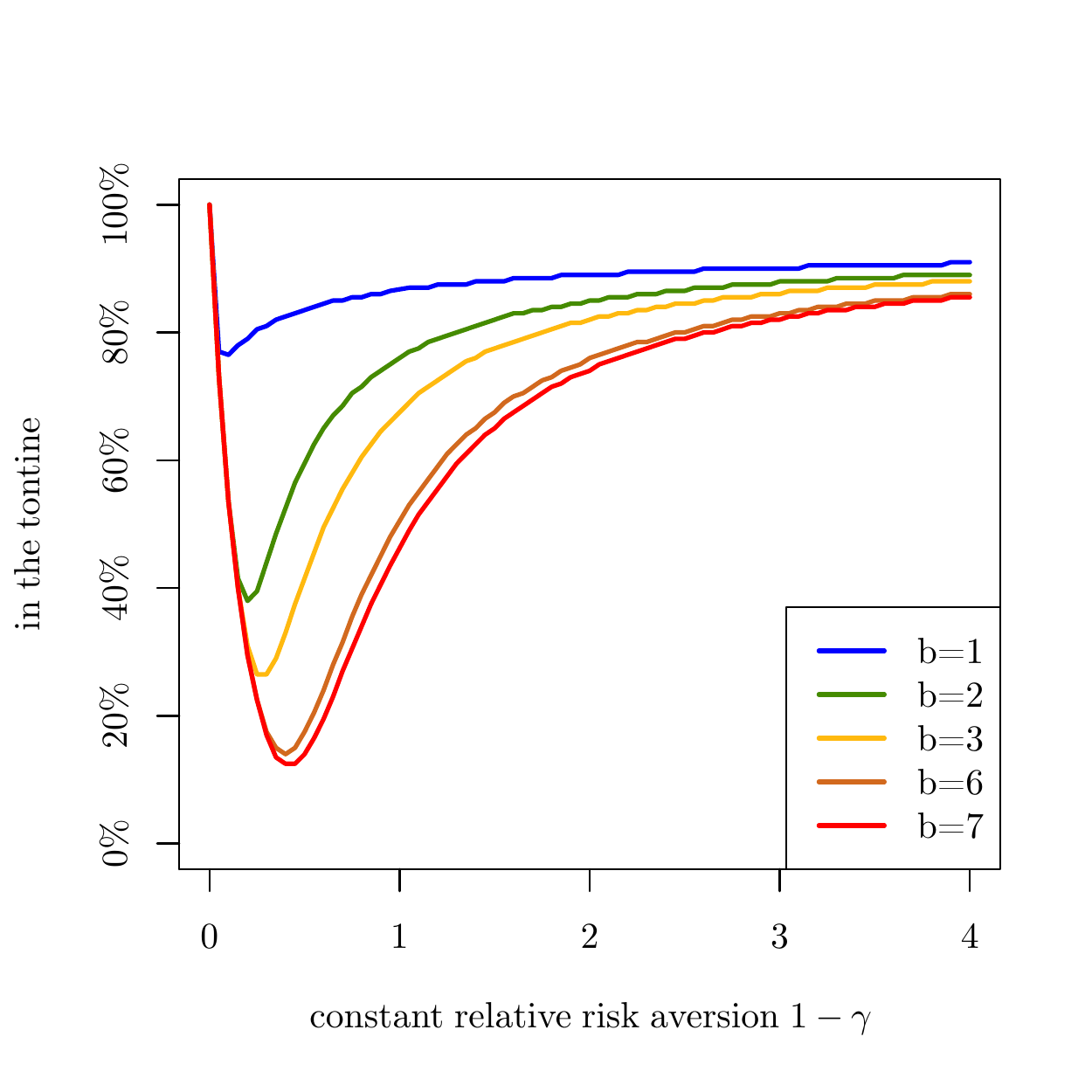}
	\caption{$(r,\mu,\sigma)$$=$$(1\%, 3\%, 15\%)$}
	\label{fig:C1.134r0.01m0.03s0.15}
    \end{subfigure}%
    \begin{subfigure}[b]{0.49\textwidth}
        \centering
	\includegraphics[trim=0 0.5cm 0 1.5cm,clip,width=\textwidth]{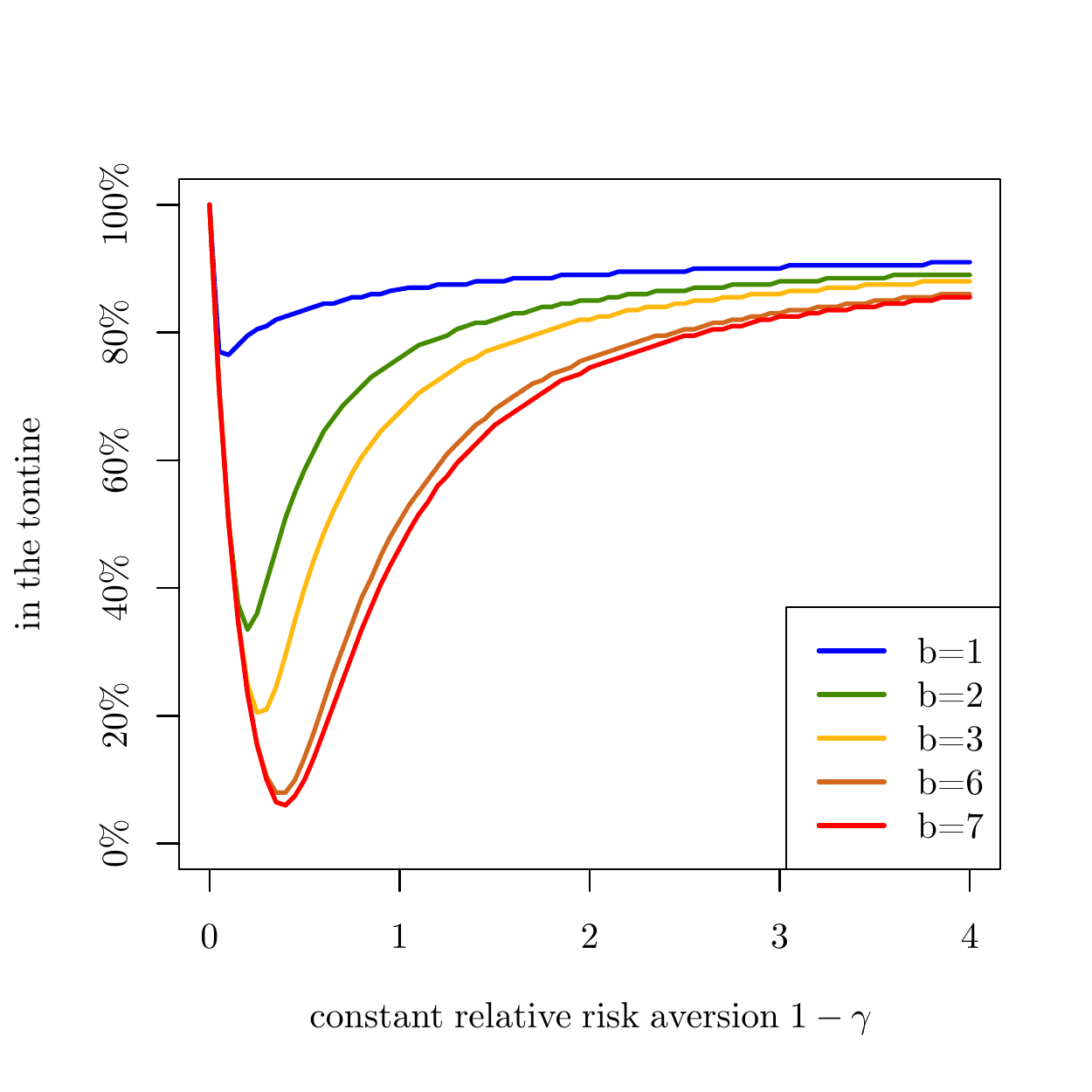}
	\caption{$(r,\mu,\sigma)$$=$$(1\%, 3\%, 25\%)$}
	\label{fig:C1.134r0.01m0.03s0.25}
    \end{subfigure}
		\vfill
    \begin{subfigure}[b]{0.49\textwidth}
        \centering
	\includegraphics[trim=0 0.5cm 0 1.5cm,clip,width=\textwidth]{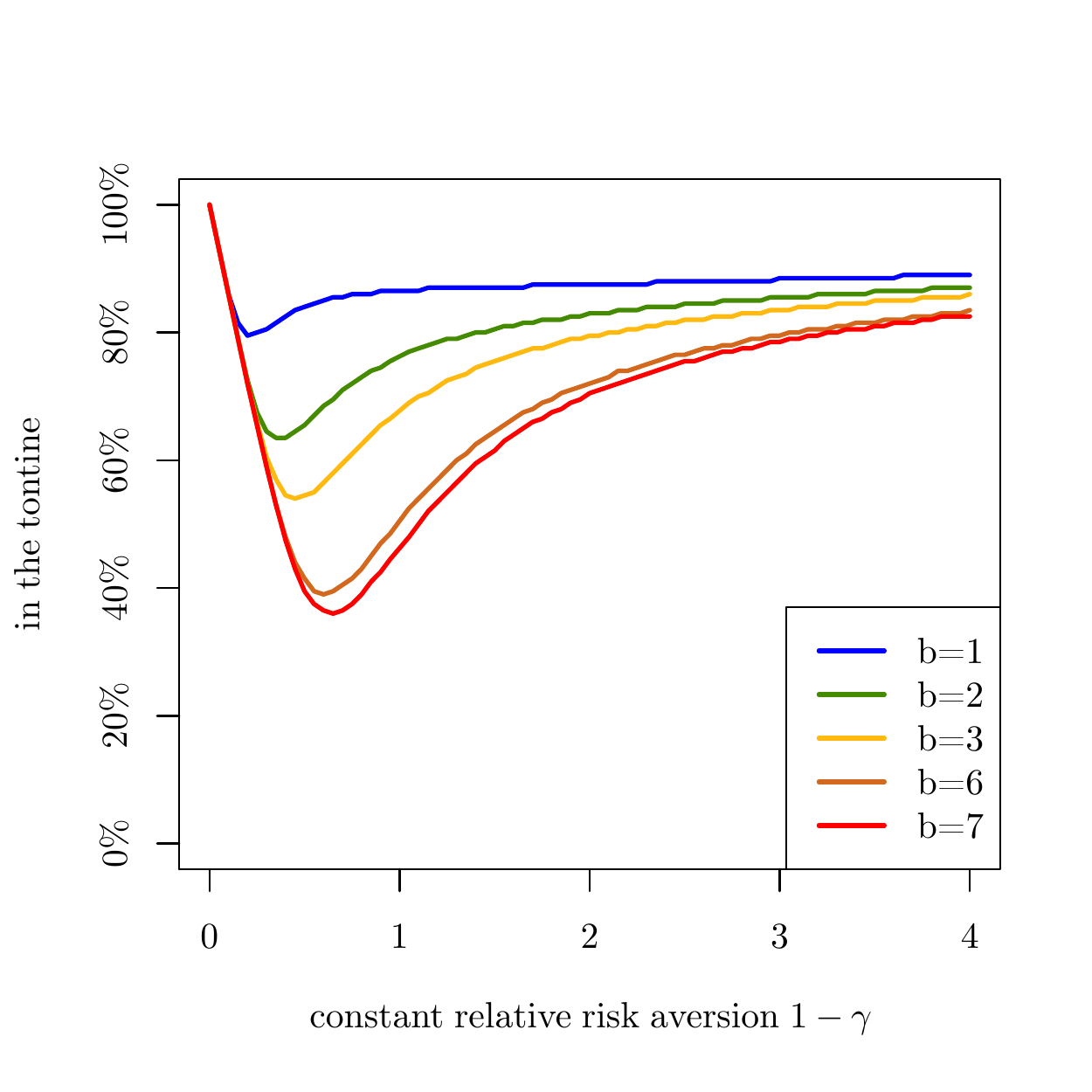}
	\caption{$(r,\mu,\sigma)$$=$$(1\%, 10\%, 15\%)$}
	\label{fig:C1.134r0.01m0.10s0.15}
    \end{subfigure}
    \begin{subfigure}[b]{0.49\textwidth}
        \centering
	\includegraphics[trim=0 0.5cm 0 1.5cm,clip,width=\textwidth]{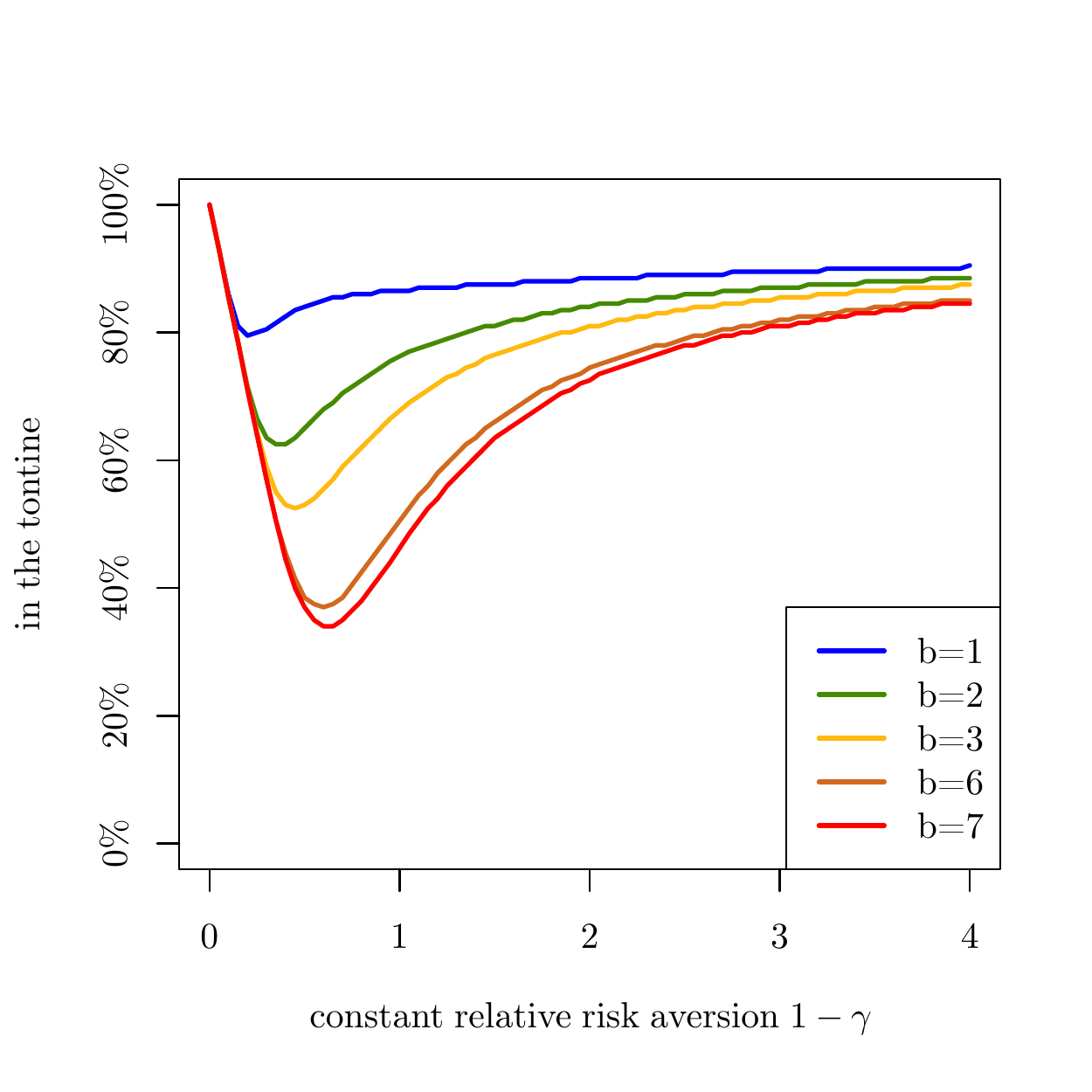}
	\caption{$(r,\mu,\sigma)$$=$$(1\%, 10\%, 25\%)$}
	\label{fig:C1.134r0.01m0.10s0.25}
    \end{subfigure}
		\vfill
	    \begin{subfigure}[b]{0.49\textwidth}
        \centering
	\includegraphics[trim=0 0.5cm 0 1.5cm,clip,width=\textwidth]{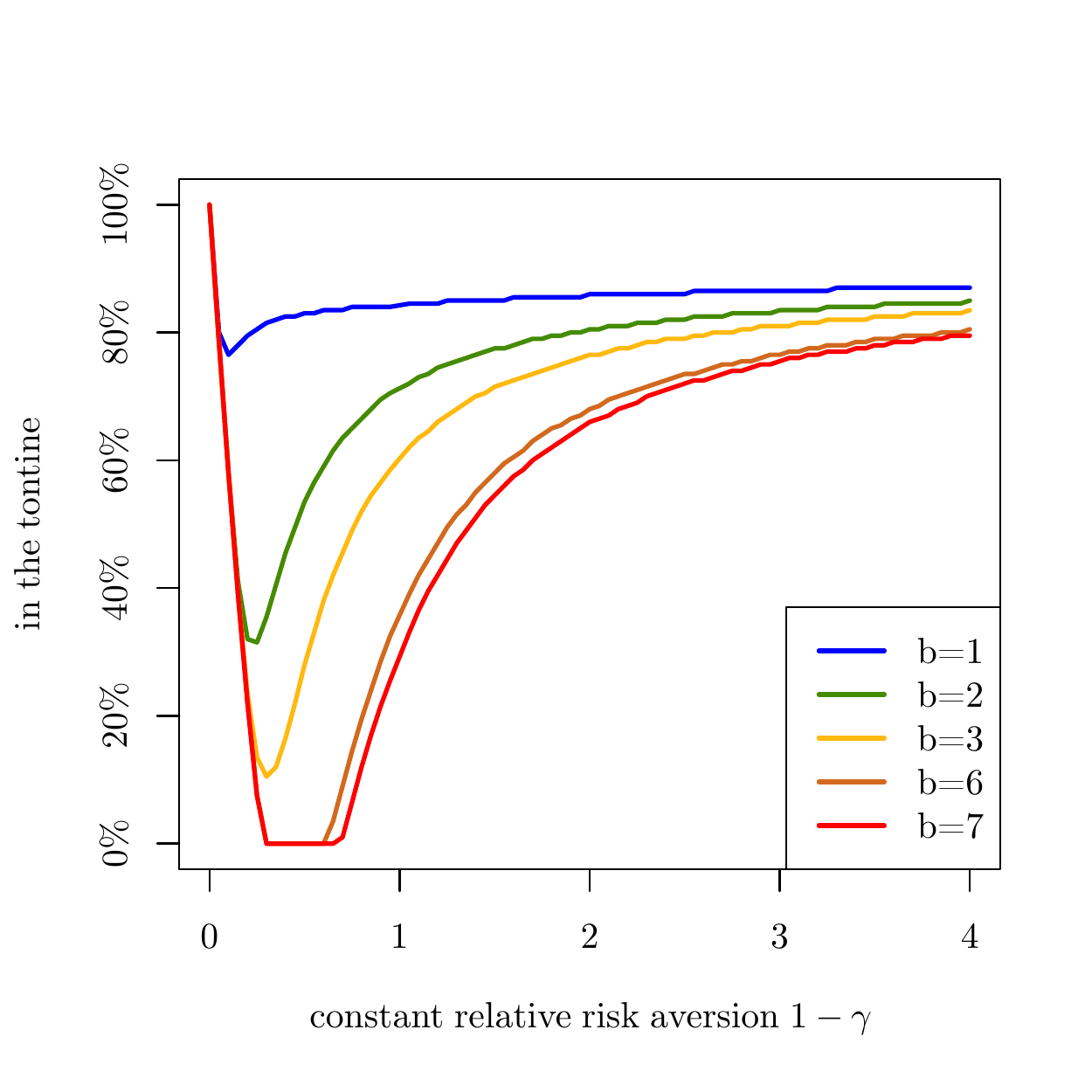}
	\caption{$(r,\mu,\sigma)$$=$$(7\%, 10\%, 15\%)$}
	\label{fig:C1.134r0.07m0.10s0.15}
    \end{subfigure}%
    \begin{subfigure}[b]{0.49\textwidth}
        \centering
	\includegraphics[trim=0 0.5cm 0 1.5cm,clip,width=\textwidth]{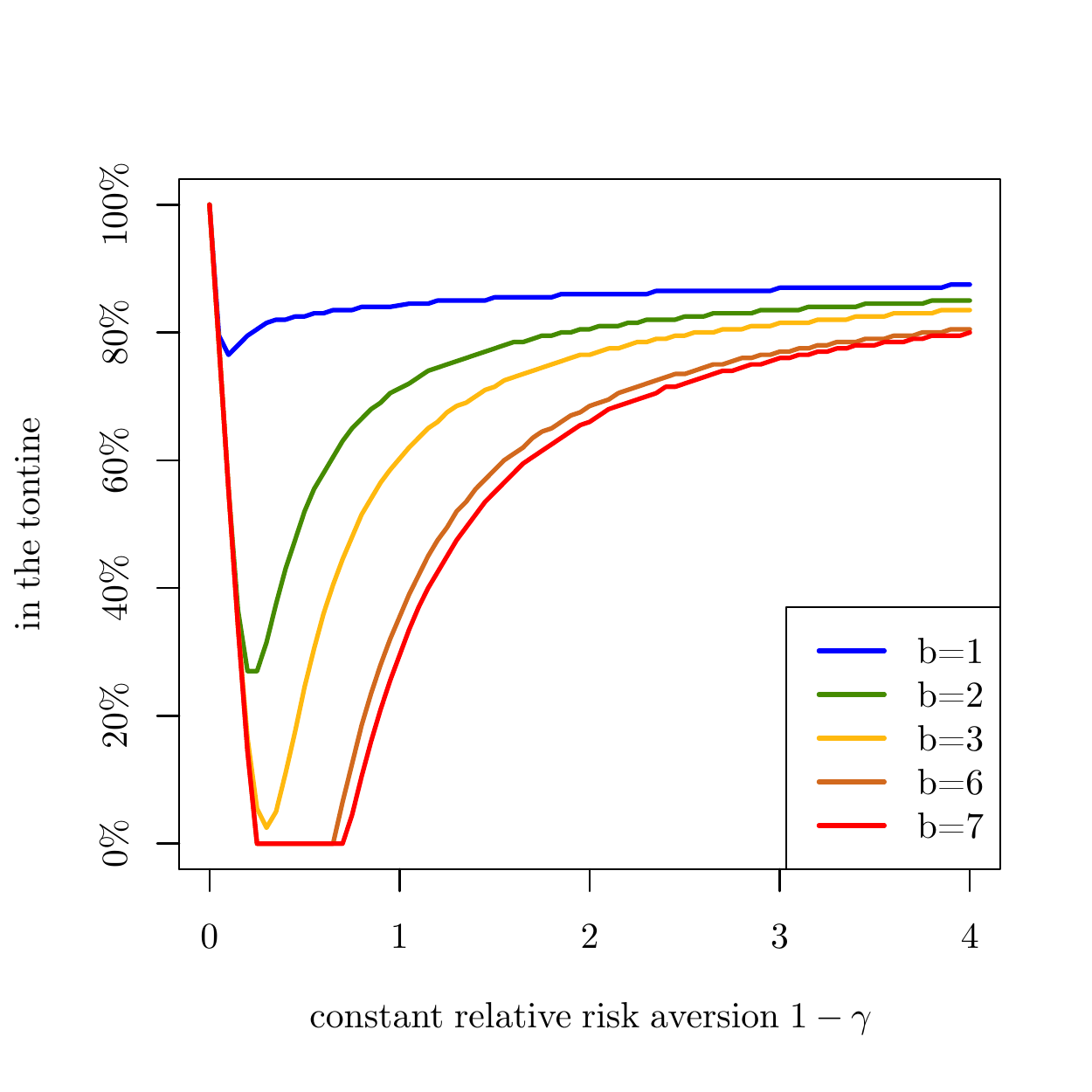}
	\caption{$(r,\mu,\sigma)$$=$$(7\%, 10\%, 25\%)$}
	\label{fig:C1.134r0.07m0.10s0.25}
    \end{subfigure}
    \caption{Sensitivity analysis of the percentage allocated to the tontine account when the life expectancy from age $65$ is decreased by $5$ years.  This is the second analysis discussed in Section \ref{SUBSECsensitivity}, and the relevant financial market model parameter values are shown below each figure.  The mortality model parameter $C=1.116$ in the table, which corresponds to a decreased life expectancy from age $65$ of $5$ years, compared to the original life expectancy when $C=1.124$.}
		\label{FIGSensLowerLifeExp}
\end{figure}

Next we by decrease the life expectancy from age $65$ by $5$ years, through setting the mortality model parameter $C=1.134$.  Then the optimal proportion in the tontine account is calculated under the same sets of financial market model parameters as before.  The full numerical results are displayed in Figure \ref{FIGSensLowerLifeExp}.  Our interest is in how these results differ from the results with the higher life expectancy (Figure \ref{FIGSensHigherLifeExp}). 

A higher life expectancy (derived from the parameter $C$) results in a higher optimal percentage in the tontine account as the retiree expects to gain longevity credits for longer.  However, this is not a significant effect as can be seen by comparing the figures with the same values of parameters $(r,\mu, \sigma)$.  For example, consider Figures \ref{fig:C1.116r0.01m0.03s0.15} and \ref{fig:C1.134r0.01m0.03s0.15}, between which there is a 10 year difference in life expectancy from age $65$, with Figure \ref{fig:C1.116r0.01m0.03s0.15} having the higher life expectancy.  There is a relatively small drop in the optimal percentage allocated to the tontine account as the life expectancy drops by 10 years.

\subsection{Power utility with a fixed proportion $\alpha$ and a variable consumption rate $c$}\label{subsection:Power utility fixed prop alpha}
Now we examine the second interpretation of the problem, which was outlined in Section \ref{section:Power utility}.  The retiree has already decided how much to invest in the tontine account, which means that the proportion $\alpha$ of savings invested in the tontine account is known and fixed.  The problem is then to determine the optimal investment and consumption strategy corresponding to each value of $\alpha \in [0,1]$.  Thus we allow for a variable consumption rate, unlike Section \ref{subsection:Power utility fixed prop alpha and consumption c} in which the consumption rate is a constant.  We show only the analytical solutions here.

For a given proportion $\alpha$ of savings invested in the tontine account, we can find a candidate for the optimal controls to maximize formula (\ref{EQNgeneralfn}) using the Hamilton-Jacobi-Bellman equation.  That is, fix $\alpha \in [0,1]$ and find the controls $(c,\omega)\in (\hat{\mathcal{G}}_1,\hat{\mathcal{G}}_2)$ that attains the expression
\begin{equation}\label{eq:$sup_c,omega$}\sup_{c,\omega}\mathbb{E}\Big[\int_{0}^{\tau}\!e^{-\rho s}\big(cX(s)\big)^\gamma/\gamma\;\mathrm{d}s+be^{-\rho \tau}\big((1-\alpha)X(\tau)\big)^\gamma/\gamma\Big].\end{equation}
A sketch of the Hamilton-Jacobi-Bellman approach is outlined in Section \ref{APPSUBSECfixedprop}.

In this example, we are able to determine explicitly the investment strategy and the consumption rate as the solution of a particular equation, called the Chini equation.  We do not numerically solve for the consumption rate.

With the power utility function, the proportion of total pension savings to invest in the risky stock is
\[
\omega^*(t):=w^*:=\frac{1}{1-\gamma}\frac{\mu-r}{\sigma^2}
\]
and the rate at which to consume the total pension savings amount is
\[
c^*(t):=\left( \gamma h(t) \right)^{\frac{1}{1-\gamma}},
\]
at each $t\geq0$, where $h$ is the solution to the Chini equation
\begin{gather}\label{eq:$0=h'+$}
0=\partial_{t} h(t) +(1-\gamma) \gamma^{\frac{1}{\gamma-1}} \, h(t)^{\gamma/(\gamma-1)}+h(t)\psi(t)+\varphi(t)
\end{gather}
with
\[
\varphi(t):=\frac{b}{\gamma} \lambda(t) \left( 1-\alpha\right)^{\gamma}\qquad\textrm{and} \qquad\psi(t):=\gamma \left( r+\alpha \lambda(t) \right) + \frac{1}{2} \frac{\gamma}{1-\gamma} \left( \frac{\mu-r}{\sigma} \right)^{2} - \lambda(t) - \rho.
\]

If $\,b=0$, then equation (\ref{eq:$0=h'+$}) becomes a Bernoulli differential equation, which has an explicit solution \citep[Exercise 19.2, page 308]{Bjork2009}.  However if $\,b\neq0$, which are the cases of interest in this paper, then the solution to the equation must be determined numerically.  The starting value $h(0)$ needs to be determined either by a suitable transversality condition or by the initial condition
\[
\sup_{\omega,c}\mathbb{E}\Big[\int_{0}^{\tau}\!e^{-\rho s}\big(c(s)X(s)\big)^\gamma\,\mathrm{d}s + b e^{-\rho \tau} \big((1-\alpha)X(\tau)\big)^\gamma\Big]=h(0)\,X(0)^\gamma.
\]

\section{How to invest in the new tontine structure under logarithmic utility}\label{section:Logarithmic utility}
Here we choose logarithmic utility functional forms in the general expression to be maximized.  To study the logarithmic utility version means that we set
\[
U(x)=\log(x)\ \qquad \qquad \textrm{and} \qquad \qquad B(x)=\log(x)\qquad\mbox{for $\,x \geq 0$}
\]
in the objective function, shown by expression (\ref{EQNgeneralfn}).  As before, our objective is to find an investment strategy $\omega\in\hat{\mathcal{G}}_2$, a positive percentage for consumption $c\in\hat{\mathcal{G}}_1$ and a constant proportion of total pension savings in the tontine $\alpha\in[0,1]$ for the remaining lifespan $\tau$ so that
\begin{equation}\label{eq:$sup_omega,c,alpha log$}
\sup_{\omega,c,\alpha}\mathbb{E}\Big[\int_{0}^{\tau} e^{-\rho s}\log(c(s)X(s)) \mathrm{d}s + b e^{-\rho \tau} \log\big((1-\alpha)X(\tau)\big)\Big]
\end{equation}
is attained.   The proofs for this section are shown in Appendix \ref{APPSUBSUBSECfixedconsumandproplog}.

Define
\begin{itemize}
    \item A positive random variable $A$ with the tail distribution
    \begin{equation}\label{eq:$mathbbP[A>t]=$}\mathbb{P}[A>t]=S(t)(1-\log S(t)),\end{equation} 
    in which the survival function $\,S(t)=\mathbb{P}[\tau>t]=\exp(-\int_0^t\!\lambda(s)\,\mathrm{d}s)\,$ for $\,t \geq 0$.
    \item The exponential moment of $\,\tau-t\,$ conditioned on $\,\tau>t$, i.e.
    \[
		M_\tau(p,t)=\frac{\mathbb{E}[\mathrm{e}^{p(\tau-t)}\mathbbm{1}_{\tau>t}]}{S(t)},
		\]
		in which $\mathbbm{1}_{D}$ denotes the zero-one indicator function of the set $D\subseteq\Omega$.
    \item The function $\,\kappa_{A,\tau}:\mathbb{R} \rightarrow \mathbb{R}$
    \begin{equation}\label{eq:$kappa=$} \kappa_{A,\tau}(p)=\frac{M_A(p)}{M_\tau(p,0)-M_A(p)},\end{equation}
    where $M_A(p):=\mathbb{E}[\mathrm{e}^{pA}]$ is the exponential moment-generating function of $A$.
\end{itemize}

Next we give the optimal strategies to follow under logarithmic utility. The proof is shown in the Appendix.  

The optimal strategy in the tontine structure is for the retiree to invest the constant proportion 
\begin{equation}\label{eq:$omega^*=$}
\omega^*(t)=w^*=\frac{\mu-r}{\sigma^2}
\end{equation}
of pension savings in the risky stock, a proportion which depends only on the market parameters.   Short-selling of the risky stock is avoided, i.e. the proportion is never negative due to the assumption that $\mu > r$.  However, reasonable values for $r,\mu,\sigma$ can lead to portfolios with more than $100\%$ in the risky stock.  If borrowing of the risk-free bond is prohibited then, in these cases, the optimal investment strategy is $100\%$ in the risky stock and the optimal consumption and tontine allocation are unchanged.

Under logarithmic utility, the optimal rate at which to consume the pension savings is at the deterministic rate
\begin{equation}\label{eq:$c^*=$}
c^*(t)=\frac{\rho}{1-(1-b \rho)\,M_\tau(-\rho,t)},
\end{equation}
at time $t\geq0$, which depends only on the characteristics of the retiree, namely the strength $b$ of their bequest motive relative to their own consumption, their time preference rate $\rho$ and their force of mortality function (through the function $M_\tau(-\rho,t)$).  The following bounds and limiting features of the optimal consumption rate are noted.  The optimal proportion for consumption $c^*$ satisfies the inequalities
\[
\min(\rho,1/b)\leq c^*\!\leq\max(\rho,1/b),
\]
where the convention $\,1/0=\infty\,$ is used for $b=0$. In particular, if $\,\rho=1/b$, then $\,\rho=c^*=1/b\,$ is a constant.  Moreover, if $\,\lim_{t\uparrow\infty}\lambda(t)=\lambda(\infty)<\infty$, then $\,c^*(t)\rightarrow(\rho+\lambda(\infty))/(1+b\lambda(\infty))$.  Similarly, if $\,\lim_{t\uparrow\infty}\lambda(t)=\lambda(\infty)>0$, then $\,c^*(t)\rightarrow(\rho/\lambda(\infty)+1)/(1/\lambda(\infty)+b)$.

It is optimal to allocate the constant proportion
\begin{equation}
\label{eq:$alpha^*=$}\alpha^*=\frac{1-b \rho}{1+b \rho\,\kappa_{A,\tau}(-\rho)}
\end{equation}
of the retiree's pension savings to the tontine account.  This optimal proportion depends only on the characteristics of the retiree.  In general, $\,\alpha^*\leq1\,$ holds true, which follows from the fact that $A$ has a heavier tail than $\tau$.  However, there are values of $b$ and $\rho$ such that $\alpha^*$ is negative.  To ensure the non-negativity of the optimal $\alpha^*$, its expression above implies that we must choose $\rho<1/b$.  Otherwise, the optimal proportion of pension savings allocated to the tontine is $\,\alpha^*=0$.  If there is no bequest motive, i.e. $b=0$, then everything is allocated to the tontine account, $\alpha^*=1$.

The total pension savings process resulting from following the above optimal strategies is
\[
X^*(t) = x_{0} \exp \left( \Big( r + \frac{1}{2} \Big( \frac{\mu-r}{\sigma} \Big)^2 \Big) t + \int_0^{t} \left( \alpha^*\lambda(u)-c^*(u) \right) \mathrm{d}u + \frac{\mu-r}{\sigma} W(t) \right),
\]
which is a geometric Brownian motion.

\subsection{Explicit results under logarithmic utility}\label{subsection:Log utility numerical}
In this section, we investigate how the optimal constant proportion in the tontine account varies with the strength of the bequest motive.  We calculate how the optimal consumption rate changes as the retiree ages, for different strengths of the bequest motive.  We find that less is invested in the tontine account as the strength of the bequest motive increases.  Further, the optimal consumption rate as a fraction of the total pension savings account value increases as the retiree ages. 

\begin{figure}[H]\centering
	\includegraphics[scale=0.6]{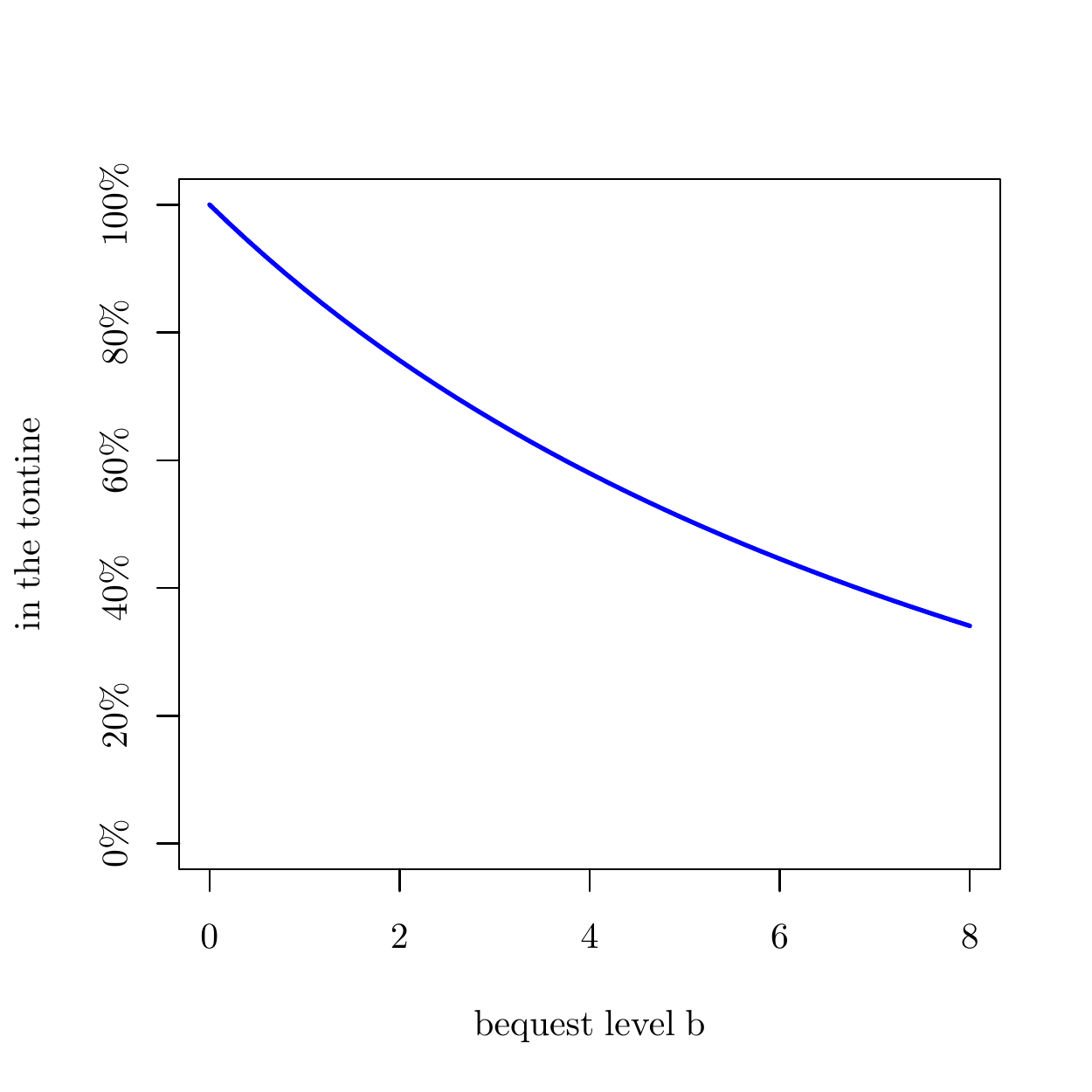}
	\caption{The optimal percentage $100\alpha$ in the tontine account as the strength of bequest motive $b$ is varied, under logarithmic utility.}
		\label{fig:LogTontine}
\end{figure}

For the market parameters $r=0.05$, $\mu=0.085$ and $\sigma=0.2$, the optimal investment strategy under logarithmic utility (given by equation \ref{eq:$omega^*=$}) is to invest $87.5\%$ of pension savings in the risky stock and the remainder in the risk-free bond at all times.  We calculate how much should a retiree allocate to the tontine account as the strength of the bequest motive increases (Figure \ref{fig:LogTontine}).  With no bequest motive ($b=0$), the retiree allocates all of their wealth to the tontine, i.e. the retiree is a member of a pure tontine.  As soon as a bequest motive exists ($b>0$), the percentage in the tontine account falls, reaching $50\%$ at $b=5$.  However, the rate of fall decreases with the strength of the bequest motive, as can be seen from the curvature of the line in Figure \ref{fig:LogTontine}.

\begin{figure}[H]\centering
\includegraphics[scale=0.6]{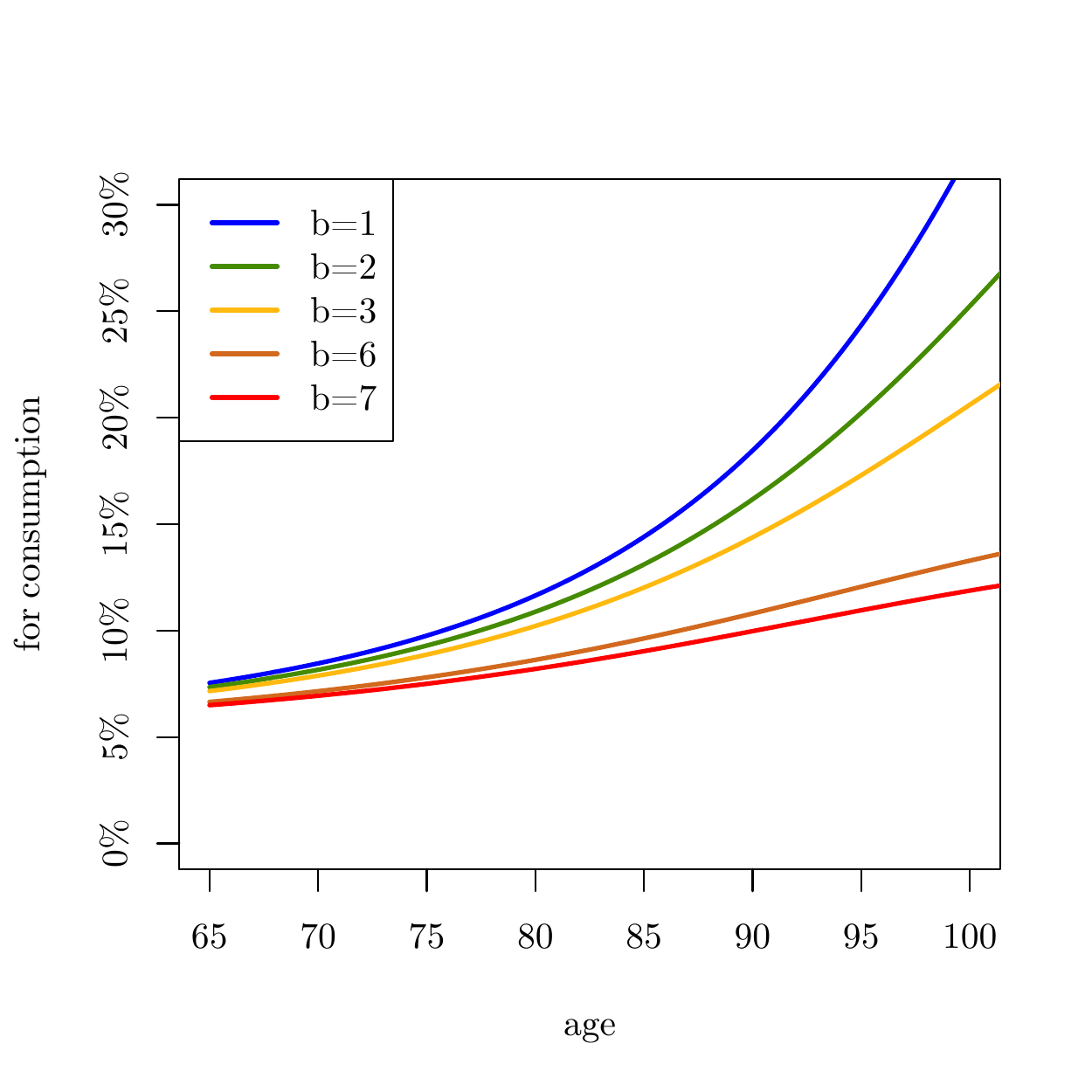}
	\caption{The optimal consumption rate $c(y-65)$ as a percentage of the total pension savings $X(y-65)$ as retiree's age $y$ increases from $65$ to $100$, for various strengths of bequest motive $b$, under logarithmic utility.}
	\label{fig:LogConsumption}
\end{figure}

With the logarithmic utility function, we are able to calculate the optimal consumption rate at each age, i.e. the rate at which the remaining pension savings are consumed.  It is a deterministic rate given by equation (\ref{eq:$c^*=$}).  We evaluate it at different ages, in which the retiree is age $65$ at time $0$, for different strengths of the bequest motive (Figure \ref{fig:LogConsumption}).  The highest consumption rate occurs with a low bequest motive ($b=1$).  The retiree has more incentive to consume their wealth than leave it to their estate.  As the strength of the bequest motive increases, the retiree consumes less in order to leave money to their estate.  However, the rates of consumption diverge with the bequest motive strength as the retiree ages.  At age 65, the consumption rate is close to $7\%$ for all strengths of bequest motive ($b \in \{ 1,2,\ldots,7\}$).  At age 90, it ranges from $8\%$ for a relatively very strong bequest motive ($b=7$) to about $18\%$ for a relatively weak one ($b=1$).  It is important to note that the pension savings are likely to fall in value over time, as the retiree depletes their fund through consumption.

\section{Conclusion}
We introduce a new decumulation product, called a modern tontine with bequest, that gives retirees a chance of having both a lifelong income and leaving a bequest to their estate.  The pension savings of each retiree is split between two accounts, one called the tontine account and another called the bequest account.  The tontine account is part of a wider tontine pool which means two things for the retiree.  One is that the tontine account attracts longevity credits, which are the share of the funds released by deaths in the tontine pool.  As long as the retiree is alive, these longevity credits are, in the worst case, zero -- corresponding to no-one dying in the tontine pool -- and otherwise they are strictly positive.  The second implication of the tontine account is that the retiree's tontine account value is lost to that retiree's estate upon their death.  This is the price paid for receiving longevity credits while alive.

Both the tontine account and the bequest account grow at the same rate due to investment returns and shrink at the same rate due to consumption.  However, only the tontine account gains longevity credits.  Thus it will grow faster than the bequest account.  A key feature of the product is that the account values are continuously re-balanced in order to maintain their relative values in a fixed ratio.  The consequence of re-balancing is that some of the longevity credits are effectively transferred from the tontine account to the bequest account.  

The transfer of longevity credits to the bequest account value means that, although its value will probably fall for several years (depending on the investment returns achieved and the consumption rate), eventually its value starts increasing.  This is due to the increase in the size of longevity credits as the retiree gets older, which are themselves proportional to the product of the retiree's force of mortality and their total account value.

The modern tontine with bequest is based on a tontine mechanism that is actuarially fair over short time periods.  This is critical to the justification for re-balancing the two accounts, and thus indirectly transferring longevity credits from the tontine account to the bequest account.  The longevity credits are the ``positive'' reward given to the retiree for risking their tontine account value over a short time period and not dying.  The loss of the tontine account upon the death of the retiree over the short time period is the ``negative'' reward.  In an actuarially-fair mechanism, allowing for the events of the retiree surviving or dying over each short time period and the value of the reward on each event, the expected value of the reward is zero.  The longevity credits belong to the retiree and the wider tontine pool has no claim on them unless they stay within the tontine account.

The product as presented has risks and uncertainties.  It does not guarantee that a reasonable level of retirement income will be paid for life, nor does it guarantee a certain level of bequest to their estate.  However, income drawdown, when a retiree decides how much to withdraw from their pension savings to live on and is in no pooling arrangement, guarantees neither a reasonable retirement income for life nor does it guarantee a certain level of bequest.  A conventional life annuity does guarantee a lifelong retirement income but it pays no bequest.

However, the product may be very attractive to retirees.  In a poll of around $180$ actuaries, $50$\% of them chose the proposed product after it had been described to them.  A further $25$\% chose a pure tontine product, with the remainder split evenly between a conventional life annuity contract and income drawdown.

Having presented our new, innovative product, our main research question is to ask how much should the retiree allocate to the tontine account.  We frame this as a utility maximization problem, where the retiree wishes to maximize the sum of the discounted expected utility of consumption and the bequest amount.  A weight applied to the discounted expected utility of the bequest amount allows us to vary the strength of the bequest motive.

Using a power utility function, we find two interesting results.  First, as the level of risk aversion increases, the retiree puts a proportion of their pension savings in the tontine account that is fairly constant regardless of the strength of the bequest motive.  This proportion falls initially as the retiree becomes less risk averse.  The rate of fall is greatest for the strongest bequest motive.

However, our second interesting result is that for retirees with very weak levels of risk aversion, the proportion allocated to the tontine account increases.  The reason is that the bequest account value can be very large at old ages.  The least risk averse retirees gamble on living a long time and thus gaining not only utility from consumption but also utility from a large bequest account value.

Our results rest on strong assumptions.  We assume that longevity is perfectly pooled and there is no uncertainty about the future mortality distribution of a retiree.  Neither of these assumptions are likely to hold in practice.  It remains to show how well the new proposed structure, the modern tontine with bequest, could operate in practice with a finite group of tontine members and uncertainty about the future mortality distribution.

\section*{Acknowledgements}
This research was funded by the Institute and Faculty of Actuaries' Actuarial Research Centre grant ``Minimizing longevity and investment risks while optimising future pension plans''.  The authors are grateful for this funding.

The authors thank an anonymous referee for very helpful comments on the submitted manuscript.

\bibliographystyle{apa} 


\appendix
\section*{Appendix}
\section{Proofs and verification}
Here we prove that the claimed optimal strategies -- how much to invest in the risky stock, how much to allocate to the tontine account and how to consume the pension savings -- are indeed the optimal strategies.

\subsection{Technical preliminaries} \label{APPSUBSECpreliminaries}

Lemmas \ref{lem:$omega=nu$} and \ref{lem:$sup=sup$} seem to be used across the relevant literature for optimal investment strategies for pension funds. In a nutshell, there is a technical difference between expression (\ref{EQNgeneralfn}) when $c$ and $\omega$ are adapted to $(\mathcal{G}_t)$ and the same problem when $c$ and $\omega$ are adapted to $(\mathcal{F}_t)$. The independence between $\tau$ and $(\mathcal{F}_t)$ can be directly applied to compute (\ref{EQNgeneralfn}) when they are adapted to $(\mathcal{F}_t)$. However, justification is needed when they are adapted to $(\mathcal{G}_t)$. To the best of our knowledge, it is only stated partially in \citet[p.\,$370$]{Protter2005} and \citet[Remark 2.2]{DeAngelisStabile2017}.  In lack of a reference, we give the necessary results for completeness.

First, we introduce a technical assumption on $\tau$. We assume that 
\begin{equation}\label{eq:$P>0$}\mathbb{P}[\tau>t]>0\quad\mbox{for all $\,t\geq0$}.\end{equation}

Furthermore, we denote with $\hat{\mathcal{F}}_n$ the set of all locally $n$-integrable, $(\mathcal{F}_t)$-predictable processes.

\begin{lem}\label{lem:$omega=nu$}
    Whether $(\mathcal{F}_t)$ and $\tau$ are independent or not, for a $(\mathcal{G}_t)$-predictable process $\,\omega$, there exists an $(\mathcal{F}_t)$-predictable process $\,\nu\,$ such that
    \begin{equation}\label{eq:$omega=nu$}\omega(t)\mathbbm{1}_{\tau\geq t}=\nu(t)\mathbbm{1}_{\tau\geq t}\quad\mbox{for all $\,t\geq0$}.\end{equation}
    In case of $(\mathcal{F}_t)$ and $\tau$ independent such that (\ref{eq:$P>0$}) holds true, for $\,\omega\in\hat{\mathcal{G}}_n$, there exists $\,\nu\in\hat{\mathcal{F}}_n\,$ such that (\ref{eq:$omega=nu$}) holds true.
\begin{proof}

    As mentioned in \citet[p.\,$370$]{Protter2005}, the first statement follows from an application of the Monotone Class Theorem.  For the second statement, consider the set
    \begin{equation}\label{eq:$D_t=$}
		D_t=\Big\{\exists\,u\leq t\,\colon\!\int_0^u\!|\nu(x)|^n\,\mathrm{d}x=\infty\Big\}\quad\mbox{for $\,t\geq0$}.
		\end{equation}
	In view of (\ref{eq:$omega=nu$}), the local integrability of $\omega$ ensures the local integrability of $\nu$ until time $\tau$, hence
    \begin{equation}\label{eq:$D_tsubseteq$}D_t\subseteq\{\tau    <t\}\quad\mbox{for all $\,t\geq0$}.\end{equation}
    From \citet[p.\,$43$, Theorem\,$13$ and p.58, Theorem\,$33$]{DellacherieMeyer1978}, the measurable projection theorem ensures that $D_t$ is a measurable set, which belongs to $\mathcal{F}_\infty$.  In particular, assumption (\ref{eq:$P>0$}) implies
    \begin{equation}\label{eq:$1>P[D>t]$}
		1>\mathbb{P}[D_t]\quad\mbox{for all $\,t\geq0$}.
		\end{equation}
    
    Now, we argue via contradiction that $v$ is a locally integrable process.  Therefore, we assume that $\,v\notin\hat{\mathcal{G}}_n$ and in turn assume that there exists an $\,s\geq0\,$ such that $\,\mathbb{P}[D_s]>0$.  In view of (\ref{eq:$1>P[D>t]$}), 
    \begin{equation}\label{eq:$1>P[D]>0$}
		1>\mathbb{P}[D_{s}]>0.
		\end{equation}
    As of (\ref{eq:$D_tsubseteq$}) and (\ref{eq:$1>P[D]>0$}), $\,\mathbb{P}[\tau<s]>0$.  Combining this with (\ref{eq:$P>0$}) shows that
    \[
		1>\mathbb{P}[\tau<s]>0.
		\]
    Using these two last results together with (\ref{eq:$D_tsubseteq$}), we find that
    \[
		\mathbb{P}[D_s\cap\{\tau\leq s\}]=\mathbb{P}[\tau\leq s]>\mathbb{P}[D_s]\,\mathbb{P}[\tau\leq s],
		\]
    contradicting the independence between $(\mathcal{F}_t)$ and $\tau$. 
\end{proof}    
\end{lem}

\begin{rmk}
    In Lemma \ref{lem:$omega=nu$}, an assumption like (\ref{eq:$P>0$}) is necessary to ensure that local integrability of $\omega$ implies local integrability of $\nu$ as the following example shows.
    Let $\,\mathcal{F}_t\,=\{\Omega,\emptyset\}$, let $\tau\sim\mathcal{U}(0,1)$ be an independent random variable, and let $\,\omega(t)=(1-\tau\wedge t)^{-1}$ for all $\,t\geq0$.
    Then $\,\nu(t)=(1-t)^{-1}$ for all $\,1>t\geq0$ and $\nu$ is not locally integrable, however $\omega$ is locally integrable.
\end{rmk}

The next lemma shows that the problem formulation for the logarithmic case in Section \ref{section:Logarithmic utility} is well-defined.

\begin{lem}\label{lem:quasi-integrable}
    Let $\,\omega\in\hat{\mathcal{G}}_2$ and $\,0\leq c\in\hat{\mathcal{G}}_1$, $\,\alpha\leq1$. 
    Then the integrand inside of the expectation of the objective function in the logarithmic case, Problem (\ref{eq:$sup_omega,c,alpha log$}), is quasi-integrable and
    \begin{equation}\label{eq:$E<infty$}
		\mathbb{E}\Big[\int_{0}^{\tau}\!e^{-\rho s}\log\big(c(s)X(s)\big)\,\mathrm{d}s + b e^{-\rho \tau} \log\big((1-\alpha)X(\tau)\big)\Big]<\infty.
		\end{equation}
    More precisely, both summands in the above integral are quasi-integrable and have an expectation that is smaller than $\infty$.
\begin{proof}
    In view of \citet[p.\,$84$, Theorem\,$37$]{Protter2005}, equation (\ref{eq:$mathrmdX$}) has a unique solution that can be expressed in the following way,
    \[
		X(t)=x_0\exp\Big(\int_0^t\!(1-\omega(s))r+\omega(s)\mu+\alpha\lambda(s)-c(s)-\frac{1}{2}\sigma^2\omega(s)^2\,\mathrm{d}s+\int_0^t\!\sigma\omega(s)\,\mathrm{d}W(s)\Big).
		\]
    Taking the logarithm yields
    \begin{align}\nonumber&\log X(t)=\log x_0+
    \\\label{eq:$logX=$}&\quad\;\;\int_0^t\!(1-\omega(s))r+\omega(s)\mu+\alpha\lambda(s)-c(s)\,\mathrm{d}s\,+\!\int_0^t\!\sigma\omega(s)\,\mathrm{d}W(s)\,-\frac{1}{2}\int_0^t\!\sigma^2\omega(s)^2\,\mathrm{d}s.\end{align}
    
    Now, we estimate the terms occurring in (\ref{eq:$E<infty$}).  As $x_0$ is a constant, the following estimates hold true.
    \begin{align}\label{eq:estimate1a}&\mathbb{E}\Big[\int_0^\infty\!\mathrm{e}^{-\rho t}\Big(\log x_0\Big)^+\mathbbm{1}_{t\leq \tau}\,\mathrm{d}t\Big]\leq\Big(\log x_0\Big)^+\!\int_0^\infty\!\mathrm{e}^{-\rho t}\,\mathrm{d}t=\frac{(\log x_0)^+}{p}<\infty,
    \\[1em]
    \label{eq:estimate1b}&\mathbb{E}\Big[\mathrm{e}^{-\rho \tau}\Big(\log\big((1-\alpha)x_0\big)\Big)^+\Big]\leq\Big(\log\big((1-\alpha)x_0\big)\Big)^+<\infty.
    \end{align}
    The following function is bounded above, as it is quadratic and concave,
    \[
		f(u)=(1-u)r+u\mu-\frac{1}{4}u^2\sigma^2\quad\mbox{for $\,u\in\mathbb{R}$}.
		\]
    Hence, the following estimates hold true.
    \begin{align}\nonumber&\mathbb{E}\Big[\int_0^\infty\!\mathrm{e}^{-\rho t}\Big(\int_0^t\!(1-\omega(s))r+\omega(s)\mu-\frac{1}{4}\omega(s)^2\sigma^2\,\mathrm{d}s\Big)^+\mathbbm{1}_{t\leq \tau}\,\mathrm{d}t\Big]\leq||f^+||_\infty\!\int_0^\infty\!\mathrm{e}^{-\rho t}t\,S(t)\,\mathrm{d}t
    \\\label{eq:estimate2a}&\quad\;\;\leq\frac{||f^+||_\infty}{p^2}<\infty,
    \\[1em]
    \nonumber&\mathbb{E}\Big[\mathrm{e}^{-\rho \tau}\Big(\int_0^\tau\!(1-\omega(s))r+\omega(s)\mu-\frac{1}{4}\omega(s)^2\sigma^2\,\mathrm{d}s\Big)^+\Big]\leq||f^+||_\infty\,\mathbb{E}[\mathrm{e}^{-\rho \tau}\tau]
    \\\label{eq:estimate2b}&\quad\;\;\leq||f^+||_\infty\frac{2}{p}M_\tau(-p/2)<\infty.\end{align}
    Let $Z$ be a continuous local martingale with $\,Z(0)=0$. Using the supermartingale property of a stochastic exponential $\mathcal{E}$, Fatou's Lemma and $\,\exp(u)\geq u\,$ for all $\,u\geq0$, we observe that
    \begin{equation*}1\geq\liminf_{t\rightarrow\infty}\mathbb{E}\Big[\mathcal{E}\Big(\frac{Z(t\wedge \tau)}{2}\Big)\Big]\geq\mathbb{E}\Big[\mathcal{E}\Big(\frac{Z(\tau)}{2}\Big)\Big]\geq\mathbb{E}\Big[\mathcal{E}\Big(\frac{Z(\tau)}{2}\Big)\mathbbm{1}_{\mathcal{E}(\frac{Z(\tau)}{2})>1}\Big]
    \geq\mathbb{E}\Big[\Big(\frac{Z(\tau)}{2}-\frac{[Z](\tau)}{8}\Big)^+\Big]\end{equation*}
    with $[Z]$ being the quadratic variation of $Z$. 
    In particular, we observe that
    \begin{equation}\label{eq:$2leqZ$}2\geq\mathbb{E}\Big[\Big(Z(\tau)-\frac{1}{4}[Z](\tau)\Big)^+\Big].\end{equation}
    Now, because of Lemma \ref{lem:$omega=nu$}, we can assume without loss of generality that $\omega$ is independent from $\tau$.
    In particular, a product of positive independent functionals of $\omega$ and $\tau$ commutes with the expectation.
    Applying (\ref{eq:$2leqZ$}) to the resulting product of expectations yields the following estimates,
    \begin{align}\nonumber&\mathbb{E}\Big[\int_0^\infty\!\mathrm{e}^{-\rho t}\Big(\int_0^t\!\sigma\omega(s)\,\mathrm{d}W(s)\,-\frac{1}{4}\int_0^t\!\sigma^2\omega(s)^2\,\mathrm{d}s\Big)^+\mathbbm{1}_{t\leq \tau}\,\mathrm{d}t\Big]
    \\\label{eq:estimate4a}&\quad\;\;=\int_0^\infty\!\mathrm{e}^{-\rho t}\mathbb{E}\Big[\Big(\int_0^t\!\sigma\omega(s)\,\mathrm{d}W(s)\,-\frac{1}{4}\int_0^t\!\sigma^2\omega(s)^2\,\mathrm{d}s\Big)^+\Big]S(t)\,\mathrm{d}t\leq2\int_0^\infty\!\mathrm{e}^{-\rho t}\,\mathrm{d}t=\frac{2}{p}<\infty,
    \\[1em]
    \nonumber&\mathbb{E}\Big[\mathrm{e}^{-\rho \tau}\Big(\int_0^\tau\!\sigma\omega(s)\,\mathrm{d}W(s)\,-\frac{1}{4}\int_0^\tau\!\sigma^2\omega(s)^2\,\mathrm{d}s\Big)^+\Big]\leq\mathbb{E}\Big[\Big(\int_0^\tau\!\sigma\omega(s)\,\mathrm{d}W(s)\,-\frac{1}{4}\int_0^\tau\!\sigma^2\omega(s)^2\,\mathrm{d}s\Big)^+\Big]
    \\\label{eq:estimate4b}&\quad\;\;\leq2.\end{align}
    In view of (\ref{eq:$mathbbP[A>t]=$}), we observe that
    \begin{align}\nonumber&\mathbb{E}\Big[\int_0^\infty\!\mathrm{e}^{-\rho t}\Big(\int_0^t\!\alpha\lambda(s)\,\mathrm{d}s\Big)^+\mathbbm  {1}_{t\leq \tau}\,\mathrm{d}t\Big]\,=\alpha^+\!\int_0^\infty\!\mathrm{e}^{-\rho t}(-1)S(t)\log S(t)\,\mathrm{d}t\leq\alpha^+\!\int_0^\infty\!\mathrm{e}^{-\rho t}\,\mathrm{d}t
    \\\label{eq:estimate3a}&\quad\;\;=\frac{\alpha^+}{p}<\infty,
    \\[1em]
    \label{eq:estimate3b}&\mathbb{E}\Big[\mathrm{e}^{-\rho \tau}\Big(\int_0^\tau\!\alpha\lambda(s)\,\mathrm{d}s\Big)^+\Big]=\alpha^+\!\int_0^\infty\!\mathrm{e}^{-\rho t}S'(t)\log S(t)\,\mathrm{d}t=\alpha^+M_A(-p)<\infty.\end{align}
    Using $\,\exp(u)\geq u^+\,$ for all $\,u\geq0$ and $\,c\geq0$, we find that
    \begin{align}\nonumber&\int_0^\tau\!\mathrm{e}^{-\rho t}\Big(\log c(t)-\int_0^t\!c(s)\,\mathrm{d}s\Big)^+\mathrm{d}t\leq\int_0^\infty\! c(t)\exp\Big(-\int_0^t\!c(s)\,\mathrm{d}s\Big)\,\mathrm{d}t
    \\\label{eq:estimate5}&\quad\;\;=\int_0^\infty\! \partial_t\Big(-\exp\Big(-\int_0^t\!c(s)\,\mathrm{d}s\Big)\Big)\,\mathrm{d}t=1-\exp\Big(-\int_0^t\!c(s)\,\mathrm{d}s\Big)\leq1.\end{align}
    In view of $\,b\geq0$, adding the above estimates (\ref{eq:estimate1a})--(\ref{eq:estimate1b}), (\ref{eq:estimate2a})--(\ref{eq:estimate2b}) and (\ref{eq:estimate4a})--(\ref{eq:estimate5}) yields the statement.
\end{proof}
\end{lem}

As mentioned before, Lemma \ref{lem:$sup=sup$} below seems to be used across the relevant literature for optimal investment strategies for pension savers.  Since we could find no reference proving this statement, we show it below for completeness.

\begin{lem}\label{lem:$sup=sup$}
    \begin{align}
		\sup_{\substack{\omega\in\hat{\mathcal{G}}_2\\\,c\in\bar{\mathcal{G}}_1}}\mathbb{E}\Big[&\int_{0}^{\tau}\!e^{-\rho s}U\big(c(s)X(s)\big)\,\mathrm{d}s+be^{-\rho \tau} B\big((1-\alpha)X(\tau)\big)\Big] \nonumber
    \\
		&=\sup_{\substack{\omega\in\hat{\mathcal{F}}_2\\\,c\in\bar{\mathcal{F}}_1}}\mathbb{E}\Big[\int_{0}^{\tau}\!e^{-\rho s}U\big(c(s)X(s)\big)\,\mathrm{d}s+be^{-\rho \tau} B\big((1-\alpha)X(\tau)\big)\Big]  \label{eq:$sup_omegainmathcalFT$}
    \\ 
		&=\sup_{\substack{\omega\in\hat{\mathcal{F}}_2\\\,c\in\bar{\mathcal{F}}_1}}\mathbb{E}\Big[\int_{0}^\infty\!e^{-\rho s}S(s)\Big(U\big(c(s)X(s)\big)+b\lambda(s)B\big((1-\alpha)X(s)\big)\Big)\,\mathrm{d}s\Big]. \label{eq:$sup_omegainmathcalF$}
		\end{align}
    More precisely, for any given $\omega,c$, from $\hat{\mathcal{F}}_1,\hat{\mathcal{F}}_2$ respectively, the equality between lines (\ref{eq:$sup_omegainmathcalFT$}) to (\ref{eq:$sup_omegainmathcalF$}) holds even without the supremum for both summands in the above integral.
\begin{proof}
    In view of Lemma \ref{lem:$omega=nu$}, for given $\,\omega\in\hat{\mathcal{G}}_2\,$ and $\,0\leq c\in\hat{\mathcal{G}}_1$, let $\,\nu\in\hat{\mathcal{F}}_2\,$ and $\,0\leq\varepsilon\in\hat{\mathcal{F}}_1\,$ be such that (\ref{eq:$omega=nu$}) holds true, respectively.  Let $Y$ be a stochastic process with dynamics given by (\ref{eq:$mathrmdX$}) in which $X:=Y$, $\omega:=\nu$ and $c:=\epsilon$.
    
		As $\nu$ and $\varepsilon$ are independent of $\tau$, the process $Y$ is independent of $\tau$.  As the involved processes are quasi-integrable (see Lemma \ref{lem:quasi-integrable} for the logarithmic case) we are able to apply Fubini's theorem, which reveals that
    \[
		\begin{split}
		\mathbb{E}\Big[\int_{0}^{\tau}\!&e^{-\rho s}U\big(c(s)X(s)\big)\,\mathrm{d}s+be^{-\rho \tau} B\big((1-\alpha)X(\tau)\big)\Big]
    \\
		&=\mathbb{E}\Big[\int_{0}^\infty\!e^{-\rho s}U\big(c(s)Y(s)\big)\mathbbm{1}_{\tau\geq t}\,\mathrm{d}s+be^{-\rho \tau} B\big((1-\alpha)Y(\tau)\big)\Big]
    \\
		&=\int_{0}^\infty\!e^{-\rho s}\mathbb{E}\big[U\big(c(s)Y(s)\big)\big]\mathbb{E}\big[\mathbbm{1}_{\tau\geq t}\big]\,\mathrm{d}s+\mathbb{E}\Big[\int_0^\infty\!be^{-\rho s} B\big((1-\alpha)Y(\tau)\big)\lambda(s)S(s)\,\mathrm{d}s\Big]
    \\
		&=\mathbb{E}\Big[\int_{0}^\infty\!e^{-\rho s}U\big(c(s)X(s)\big)S(s)+be^{-\rho s}B\big((1-\alpha)X(s)\big)\lambda(s)S(s)\Big)\,\mathrm{d}s\Big].
		\end{split}
		\]
    This together with $\,\hat{\mathcal{F}}_n\subseteq\hat{\mathcal{G}}_n\,$ and a simple inequality argument shows the statement.
\end{proof}
\end{lem}

\subsection{Power utility: Optimal strategies for a fixed consumption rate $c$} \label{APPSUBSECfixedconsumandprop}
Proposition \ref{pro:$c$deterministic$w^*$optimal} proves the optimality of the investment strategies stated for the power utility and the log utility in Sections \ref{subsection:Power utility fixed prop alpha and consumption c} and \ref{section:Logarithmic utility}, respectively.  Proposition \ref{pro:$X(0)^gamma/gammafrac$} completes the proofs required for Section \ref{subsection:Power utility fixed prop alpha and consumption c}.

\begin{pro}\label{pro:$c$deterministic$w^*$optimal}
    If $c$ is a deterministic control, then the optimal investment strategy $\,\omega^*\in\hat{\mathcal{F}}_2$ for the Problem (\ref{EQNgeneralfn}) with $U$ and $B$ both of the power utility is given by (\ref{eq:$omega^*_gamma=$}).  With $U$ and $B$ both of the log utility form, the optimal investment strategy is given by (\ref{eq:$omega^*=$}).
\begin{proof}
    We give the proof in case of logarithmic utility, the proof for the power utility follows in a similar way. Let $\,\omega\in\hat{\mathcal{F}}_2$ and $\,t\geq0$. In view of \citet[p.$84$, Theorem $37$]{Protter2005}, equation (\ref{eq:$mathrmdX$}) has a unique solution that can be expressed with the stochastic exponential $\mathcal{E}$ in the following way,
    \[
		X(t)=x_0\exp\Big(\int_0^t\!(1-\omega(s))r+\omega(s)\mu+\alpha\lambda(s)-c(s)\,\mathrm{d}s\Big)\mathcal{E}\Big(\int_0^t\!\sigma\omega(s)\,\mathrm{d}W(s)\Big).
		\]
    Taking the power to $\gamma \in (0,1)$ yields
    \[
		X(t)^\gamma=x_0^\gamma\exp\Big(\gamma\int_0^t\!f_\gamma(\omega(s))+\alpha\lambda(s)-c(s)\,\mathrm{d}s\Big)\mathcal{E}\Big(\int_0^t\!\gamma\sigma\omega(s)\,\mathrm{d}W(s)\Big)
		\]
		with
		\[
		f_\gamma(u)=(1-u)r+\mu u-(1-\gamma)u^2\sigma^2/2\quad\mbox{for $\,u\in\mathbb{R}$}.
		\]
    The first and second order conditions for $f_\gamma$ with respect to $u$ are 
    \[
		\begin{split}
		\partial_{u} f_{\gamma}(u)= & (\mu-r)-(1-\gamma)u\sigma^2=0, \\
		\partial_{uu} f_{\gamma}(u) = &-(1-\gamma)\sigma^2<0\quad\mbox{as $\,\gamma<1$}.
		\end{split}
		\]
    In particular, $\,u^*(\gamma)=(\mu-r)/(\sigma^2(1-\gamma))\,$ is the unique point of global maximum of the strictly concave function $f_\gamma$.
    As $\,\gamma>0$, this implies that
    \[
		X(t)^\gamma\leq x_0^\gamma\exp\Big(\gamma\int_0^t\!f_\gamma(u^*(\gamma))+\alpha\lambda(s)-c(s)\,\mathrm{d}s\Big)\mathcal{E}\Big(\int_0^t\!\gamma\sigma\omega(s)\,\mathrm{d}W(s)\Big).
		\]
    Taking expectations on both sides and using $c$ deterministic to apply the supermartingale property of $\mathcal{E}$ yields that
    \begin{equation}\label{eq:$X^gammaleq$}\mathbb{E}\big[X(t)^\gamma\big]\leq x_0^\gamma\exp\Big(\gamma\int_0^t\!f_\gamma(u^*(\gamma))+\alpha\lambda(s)-c(s)\,\mathrm{d}s\Big).\end{equation}
    Taking the logarithm on both sides, applying Jensen's inequality, dividing by $\gamma$ and, in a second step, taking the limit as $\,\gamma\downarrow0\,$ followed by a comparison with (\ref{eq:$omega^*=$}) yields
    \[
		\mathbb{E}\big[\log X(t)\big]\leq\log\mathbb{E}\big[X(t)^\gamma\!/\gamma\big]\leq\log x_0+\!\int_0^t\!f_\gamma(u^*(\gamma))+\alpha\lambda(s)-c(s)\,\mathrm{d}s,
		\]
    \begin{equation} \label{eq:$mathbblogEX(t)leg$}
		\mathbb{E}\big[\log X(t)\big]\leq \log x_0+\!\int_0^t\!f_0(w^*)+\alpha\lambda(s)-c(s)\,\mathrm{d}s.
		\end{equation}
    Inequality (\ref{eq:$mathbblogEX(t)leg$}) becomes an equality when $\omega,X$ are replaced with $\omega^*,X^*$.  In this case, $\mathcal{E}$ is a true martingale as can be seen from Novikov's condition and $\omega^*$ being deterministic.
    
		Since $\,B(x)=U(x)=\log(x)$, the integrand in (\ref{eq:$sup_omegainmathcalF$}) is linear in $\log X$.  Applying (\ref{eq:$mathbblogEX(t)leg$}) to (\ref{eq:$sup_omegainmathcalF$}) yields
		\[
    \begin{split}
		& \sup_{\omega\in\hat{\mathcal{F}}}\mathbb{E}\Big[\int_0^\infty\!\!\!\mathrm{e}^{-\rho t}S(t)\Big(U\big(c(t)X(t)\big)+bB\big((1-\alpha)X(t)\big)\lambda(t)\Big)\,\mathrm{d}t\Big] \\
		& \leq\mathbb{E}\Big[\int_0^\infty\!\!\!\mathrm{e}^{-\rho t}S(t)\Big(U\big(c(t)X^*(t)\big)+bB\big((1-\alpha)X^*(t)\big)\lambda(t)\Big)\,\mathrm{d}t\Big].
		\end{split}
		\]
    In fact, as $\,\omega^*\in\hat{\mathcal{F}}_2$, the above inequality yields equality.  This implies the statement as the supremum of (\ref{EQNgeneralfn}) coincides with (\ref{eq:$sup_omegainmathcalF$}) according to Lemma \ref{lem:$sup=sup$}.
\end{proof}
\end{pro}

\begin{pro}\label{pro:$X(0)^gamma/gammafrac$}
    In the power utility Problem (\ref{eq:$sup_omega,constants$}), the statements (\ref{eq:$M_A(-k)<infty$}) and (\ref{eq:$sup=$SumOfMoments}) hold true.
\begin{proof}
    We follow up on the proof of Proposition \ref{pro:$c$deterministic$w^*$optimal}.  The inequality (\ref{eq:$X^gammaleq$}) becomes an equality when $\omega,X$ is replaced with $\omega^*,X^*$, respectively; see equation (\ref{eq:$omega^*_gamma=$}). This is true for all $\,\gamma\in(-\infty,1)\setminus\{0\}$. Thus, using (\ref{eq:$k=$}), (\ref{eq:$mathbbP[A>t]=$}) and (\ref{eq:$X^gammaleq$}), we can compute (\ref{eq:$sup_omegainmathcalF$}) in the following way.
		    \begin{align}
		\nonumber &\mathbb{E}\Big[\int_{0}^\infty\!S(t)e^{-\rho t}\Big(c^\gamma X(t)^\gamma/\gamma+b(1-\alpha)^\gamma X(t)^\gamma/\gamma\,\lambda(t)\Big)\,\mathrm{d}t\Big] \\
		\nonumber &=\frac{X(0)^\gamma}{\gamma}\int_0^\infty\!S(t)\mathrm{e}^{-\rho t}\Big(c^\gamma+b(1-\alpha)^\gamma \lambda(t)\Big)\exp\Big(\gamma\int_0^t\!f_\gamma(u^*(\gamma))\,\mathrm{d}s\Big)\mathrm{d}t \\
		\nonumber &=\frac{X(0)^\gamma}{\gamma}\int_0^\infty\!\mathrm{e}^{-kt}\mathrm{e}^{-(1-\gamma\alpha)\int_0^t\!\lambda(s)\,\mathrm{d}s}c^\gamma+\mathrm{e}^{-kt}\mathrm{e}^{-(1-\gamma\alpha)\int_0^t\!\lambda(s)\,\mathrm{d}s}b(1-\alpha)^\gamma\lambda(t)\,\mathrm{d}t \\
		\label{eq:$c^gammaS+b(1-alpha)^gammag$}&=\frac{X(0)^\gamma}{\gamma}\int_0^\infty\!\mathrm{e}^{-kt}\mathbb{P}[A>t]c^\gamma+\mathrm{e}^{-kt}\frac{\partial_t\mathbb{P}[A\leq t]}{1-\gamma\alpha}b(1-\alpha)^\gamma\,\mathrm{d}t.\end{align}
    Notice that $\,\gamma<1\,$ and $\,\alpha\leq1$, hence $\,1-\gamma\alpha>0$, and therefore $1/(1-\gamma\alpha)$ is well-defined.
    In case of $\,k\neq0$, using integration-by-parts, we can see that
    \begin{equation}\label{eq:$int_0^u=frac-frac$}\int_0^u\!\mathrm{e}^{-kt}\mathbb{P}[A>t]\,\mathrm{d}t=\frac{1-\int_{[0,u]}\!\mathrm{e}^{-kt}\,\mathbb{P}_A(\mathrm{d}t)}{k}-\frac{\mathrm{e}^{-ku}\mathbb{P}[A>u]}{k},\quad\mbox{for $\,u\geq0$}.\end{equation}
    
		If $\,M_A(-k)=\infty$, as $A$ is positive, it holds true that $\,k<0$.  In particular, $\,(1-M_A(-k))/k=\infty\,$ and $\,-\mathrm{e}^{-ku}\mathbb{P}[A>u]/k>0\,$ for all $\,u\geq0$. Hence
    \begin{equation}\label{eq:if$M_A(-k)=infty$}\int_0^\infty\!\mathrm{e}^{-kt}\mathbb{P}[A>t]\,\mathrm{d}t=\frac{1-M_A(-k)}{k},\quad\mbox{if $\,M_A(-k)=\infty$}.\end{equation}
    Next we consider the case of $\,M_A(-k)<\infty$. It always holds for $\,k>0\,$ that $\,\mathrm{e}^{-ku}\mathbb{P}[A>u]\rightarrow0\,$ for $\,u\uparrow\infty$. Furthermore, in case of $\,k<0$, the following estimate holds true,
    \[M_A(-k)\geq\int_{[0,u]}\!\mathrm{e}^{-kt}\,\mathbb{P}_A(\mathrm{d}t)+\mathrm{e}^{-ku}\mathbb{P}[A>u].\]
    In particular, rearranging the terms and taking the limit shows that
    \begin{equation}\label{eq:$lime-kuP[A>u]=0$}\lim_{u\rightarrow\infty}\mathrm{e}^{-ku}\mathbb{P}[A>u]=0\quad\mbox{if $\,M_A(-k)<\infty\,$ and $\,k\neq0$}.\end{equation}
    Applying (\ref{eq:$lime-kuP[A>u]=0$}) to (\ref{eq:$int_0^u=frac-frac$}) and combining it with (\ref{eq:if$M_A(-k)=infty$}) shows then that
    \begin{equation}\label{eq:$kneq0$}\int_0^\infty\!\mathrm{e}^{-kt}\mathbb{P}[A>t]\,\mathrm{d}t=\frac{1-M_A(-k)}{k}\quad\mbox{for $\,k\neq0$}.\end{equation}
    On the other hand, using Fubini's theorem, we can show
    \begin{equation}\label{eq:$k=0$}\int_0^\infty\!\mathrm{e}^{-kt}\mathbb{P}[A>t]\,\mathrm{d}t=\mathbb{E}[A]\quad\mbox{for $\,k=0$}.\end{equation}
    Moreover, it always holds true that
    \begin{equation}\label{eq:$=fracM_A(-k)1-gammaalpha$}\int_0^\infty\!\mathrm{e}^{-kt}\frac{\partial_t\mathbb{P}[A\leq t]}{1-\gamma\alpha}\,\mathrm{d}t=\frac{M_A(-k)}{1-\gamma\alpha}\quad\mbox{for $\,k\in\mathbb{R}$}.\end{equation}
    The statement follows from combining (\ref{eq:$c^gammaS+b(1-alpha)^gammag$}) with (\ref{eq:$kneq0$})--(\ref{eq:$=fracM_A(-k)1-gammaalpha$}).
\end{proof}
\end{pro}

\subsection{Power utility: the candidate for a variable consumption rate $c$} \label{APPSUBSECfixedprop}

In this section, we sketch a Hamilton-Jacobi-Bellman approach to find a candidate for the optimal investment and consumption strategy under a power utility function, assuming that a fixed proportion $\alpha$ of savings is allocated to the tontine account at all times.  We wish to maximize
\[
\mathbb{E} \left[ \int_{0}^{\tau} e^{-\rho s} U \left( c(s)X(s)\right) ds + b e^{-\rho \tau} B \left( (1-\alpha) X(\tau) \right)  \right]
\]
over the set of investment and consumption strategies, for the wealth dynamics given by
\begin{equation} \label{APPAwealthdynamics}
\frac{\mathrm{d}X(t)}{X(t)} = \Big(r + ( \mu - r ) \omega(t) + \alpha \lambda(t) - c(t)\Big) dt + \sigma \omega(t) \mathrm{d}W(t).
\end{equation}
We begin by defining the value function 
\[
\valuefn (t,x,\omega,c) = \mathbb{E} \left[ \int_{t}^{\tau} e^{-\rho (s-t)} U \left( c(s)X(s)\right) ds + b e^{-\rho (\tau-t)} B \left( (1-\alpha) X(\tau) \right) \, \bigg\vert X(t) = x, \tau > t \right].
\]
To integrate out $\tau$, the random time of death, note that its probability density function conditional on the individual surviving to time $t>0$ is $f_{\tau \, \vert \tau>t}(s) := \lambda(s)\exp(- \int_{t}^{s} \lambda(u) du)$ for $s \geq t$. Using Lemma \ref{lem:$sup=sup$} the value function can be re-written as 
\[
\begin{split}
\valuefn (t,x,\omega,c) = & \mathbb{E} \bigg[ \int_{t}^{\infty} f_{\tau \, \vert \tau>t}(T) \int_{t}^{T} e^{-\rho (s-t)} U \left( c(s)X(s)\right) ds \, dT \\
& \qquad \qquad \qquad \qquad + b \int_{t}^{\infty} f_{\tau \, \vert \tau>t}(T) e^{-\rho (T-t)} B \left( (1-\alpha) X(T) \right) \, \bigg\vert X(t) = x \bigg].
\end{split}
\]
Changing the order of integration and denoting by $\mathbb{E}^{t,x}$ the expectation conditional on $X(t) = x$,
\[
\begin{split}
\valuefn (t,x,\omega,c) = & \mathbb{E}^{t,x} \bigg[ \int_{t}^{\infty} e^{-\rho (s-t)} U \left( c(s)X(s)\right) \int_{s}^{\infty} f_{\tau \, \vert \tau>t}(T) dT \, ds \\
& \qquad \qquad \qquad \qquad + b \int_{t}^{\infty} f_{\tau \, \vert \tau>t}(T) e^{-\rho (T-t)} B \left( (1-\alpha) X(T) \right) dT \bigg] \\
= & \mathbb{E}^{t,x} \bigg[ \int_{t}^{\infty} e^{-\int_{t}^{s} \left( \lambda(u) + \rho \right) du } U \left( c(s)X(s)\right) ds \\
& \qquad \qquad \qquad \qquad + b \int_{t}^{\infty} \lambda(T) e^{- \int_{t}^{T} \left( \lambda(u) + \rho \right) du} B \left( (1-\alpha) X(T) \right) dT \bigg].
\end{split}
\]
Thus upon defining for notational ease the function
\begin{equation} \label{APPEQNutilityease}
\hat{U} (s) := U \left( c(s)X(s)\right) + b \lambda(s) B \left( (1-\alpha) X(s) \right),
\end{equation}
the value function can be re-expressed as
\[
\valuefn (t,x,\omega,c) = \mathbb{E}^{t,x} \left[ \int_{t}^{\infty} e^{-\int_{t}^{s} \left( \lambda(u) + \rho \right) du } \, \hat{U}(s) ds \right]
\]
We begin by deriving the relevant Hamilton-Jacobi-Bellman equation, following the technique in \citet[Chapter 19]{Bjork2009}.

Assume that from time $t$ to time $t+h$, the participant follows arbitrary investment and consumption strategies $(\omega(s), c(s))$, for $s \in (t,t+h]$.  After time $t+h$, the participant follows the optimal consumption and investment strategies.  Then 
\begin{equation} \label{APPAvalueinequality}
\valuefn (t,x) \geq \mathbb{E}^{t,x} \left[ \int_{t}^{t+h} e^{- \int_{t}^{s} (\lambda(u) + \rho) du} \, \hat{U}(s) ds \right] + \mathbb{E}^{t,x} \left[ e^{- \int_{t}^{t+h} (\lambda(u) + \rho) du} \, \valuefn(t+h, X(t+h) ) \right]. 
\end{equation}

Define the operator
\[
\mathcal{L}^{\omega, c}_{t} h (t,x) =\partial_t h(t,x) + x \partial_x h(t,x) \left[ r + \omega (\mu - r)  + \alpha \lambda(t)  - c(t) \right] + \frac{1}{2} x^{2} \partial_{xx} h(t,x) \sigma^{2} \omega^{2}.
\]

Assuming sufficient differentiability so that we can apply Ito's lemma to the product \\$e^{- \int_{t}^{t+h} (\lambda(u) + \rho) du} \valuefn(t+h, X(t+h))$, we use the wealth dynamics in (\ref{APPAwealthdynamics}) to find
\begin{equation} \label{APPAvalueito}
\begin{split}
e^{- \int_{t}^{t+h} (\lambda(u) + \rho) du} \, \valuefn(t+h, X(t+h))
= & \valuefn(t, X(t)) + \int_{t}^{t+h} \mathcal{L}^{\omega, c}_{s} e^{- \int_{t}^{s} (\lambda(u) + \rho) du} \valuefn(s,X(s)) ds \\
& \qquad + \int_{t}^{t+h} e^{- \int_{t}^{s} (\lambda(u) + \rho) du} X(s) \valuefn_{x} (s,X(s)) \sigma \omega(s) d W(s).
\end{split}
\end{equation}
Substituting (\ref{APPAvalueito}) into (\ref{APPAvalueinequality}), the expectation of the stochastic integrals vanishes, leaving
\[
\begin{split}
\valuefn(t,x) & \geq \mathbb{E}^{t,x} \left[ \int_{t}^{t+h} e^{- \int_{t}^{s} (\lambda(u) + \rho) du} \, \hat{U}(s) ds \right] \\
& \qquad \qquad + \mathbb{E}^{t,x} \left[ \valuefn(t,X(t)) + \int_{t}^{t+h} \mathcal{L}^{\omega, c}_{s} e^{- \int_{t}^{s} (\lambda(u) + \rho) du} \valuefn(s,X(s)) ds \right].
\end{split}
\]
Rearranging the last inequality and cancelling out the $\valuefn(t,x)$ terms, we obtain
\begin{equation} \label{APPAvalueinequalityii}
0 \geq \mathbb{E}^{t,x} \left[ \int_{t}^{t+h} e^{- \int_{t}^{s} (\lambda(u) + \rho) du} \, \hat{U}(s) ds \right] + \mathbb{E}^{t,x} \left[ \int_{t}^{t+h} \mathcal{L}^{\omega, c}_{s} e^{- \int_{t}^{s} (\lambda(u) + \rho) du} \valuefn(s,X(s)) ds \right]. 
\end{equation}
Divide by $h>0$, let $h$ go to zero and assume enough regularity so that we can take the limit within the expectation to obtain
\[
0 \geq \hat{U}(t) - ( \lambda(t) + \rho) \valuefn(t,x) + \partial_t \valuefn(t,x) + x \partial_x \valuefn(t,x) \left[ r + \omega (\mu - r)  + \alpha \lambda(t) - c(t) \right] + \frac{1}{2} x^{2} \partial_{xx} \valuefn(t,x) \sigma^{2} \omega^{2}.
\]
If we choose the optimal consumption and investment strategies, and substitute back in for $\hat{U}(t)$ from equation (\ref{APPEQNutilityease}), then we obtain equality:
\begin{equation} \label{APPAHJBinfty}
\begin{split}
& ( \lambda(t) + \rho) \valuefn(t,x) \\
 = & \sup_{(\omega, c)} \bigg\{ U \left( cx \right) + b \lambda(t) B \left( (1-\alpha) x \right) + \partial_t \valuefn (t,x) + x \partial_x \valuefn (t,x) \left[ r + \omega (\mu - r)  + \alpha \lambda(t)  - c \right] \\
& \qquad + \frac{1}{2} x^{2} \partial_{xx} \valuefn(t,x) \sigma^{2} \omega^{2} \bigg\}.
\end{split}
\end{equation}

Guessing the form of the value function
\begin{equation} \label{APPAvaluefninftyguess}
\valuefn (t,x) = h (t) x^{\gamma},
\end{equation}
the first order conditions suggest that a candidate optimal investment strategy is
\[
\hat{\omega} (t) := \hat{\omega} := \frac{1}{1-\gamma} \frac{\mu-r}{\sigma^{2}}
\]
and a candidate optimal consumption rate is
\[
\hat{c} (t) := \left( \gamma h(t) \right)^{\frac{1}{1-\gamma}}.
\]
Setting $\omega := \hat{\omega}$ and $c:=\hat{c} (t)$ in equation (\ref{APPAHJBinfty}), and using the candidate value function (\ref{APPAvaluefninftyguess}) results in
\[
\begin{split}
0= & \partial_{t} h(t) + h(t) \left( \gamma \left( r+\alpha \lambda(t) \right) + \frac{1}{2} \frac{\gamma}{1-\gamma} \left( \frac{\mu-r}{\sigma} \right)^{2} - \lambda(t) - \rho\right) + \frac{b}{\gamma} \lambda(t) \left( 1-\alpha\right)^{\gamma} \\
& \qquad + (1-\gamma) \gamma^{\frac{1}{\gamma-1}} h(t)^{\frac{\gamma}{\gamma-1}},
\end{split}
\]
which is precisely (\ref{eq:$0=h'+$}).

\subsection{Log utility: Optimal strategy for a variable consumption rate $c$} \label{APPSUBSUBSECfixedconsumandproplog}
In this section, we give most of the proofs that relate to Problem (\ref{eq:$sup_omega,c,alpha log$}).  First, we derive the solution using an heuristic argument, then we verify this solution.  The other proof required is to show the optimality of the investment strategy (\ref{eq:$omega^*=$}); this is proved in Proposition \ref{pro:$c$deterministic$w^*$optimal}.   

We begin by defining the corresponding value function for $\,x>0\,$ and $\,t\geq0\,$ as
\[
V(x,t)=\sup_{\omega,c,\alpha}\mathbb{E}\Big[\int_t^\tau\!e^{-\rho s}\log\big(c(s)X(s)\big)\,\mathrm{d}s+be^{-\rho \tau}\log\big((1-\alpha)X(\tau)\big)\mathbbm{1}_{t<\tau}\Big|X(t)=x\Big].
\]
We observe that $V(x,t)$ and $V(y,t)$ for $\,x,y>0\,$ only differ by a constant that is independent of $\,t\geq0$. In particular, no matter what amount $x$ at $t$ is given, the optimization problem leads to the same controls $\,\omega^*,c^*,\alpha^*$. In particular, this indicates that the optimal controls are independent of the optimal wealth process, hence this indicates that the optimal controls are deterministic. 

If $c$ is a deterministic control, then Proposition \ref{pro:$c$deterministic$w^*$optimal} implies that the optimal investment strategy $\omega^*$ is given by (\ref{eq:$omega^*=$}). Moreover, if $c,\omega$ are deterministic controls, then the optimal $\alpha^*$ can be written as a solution to an associated deterministic optimization problem.

\begin{pro}\label{pro:$alpha^*$optimal}
    If $c$ is a deterministic control, then the optimal percentage $\,\alpha^*\leq1$ for Problem (\ref{eq:$sup_omega,c,alpha log$}) is given by (\ref{eq:$alpha^*=$}).
\begin{proof}
    Under the optimal investment strategy, equation (\ref{eq:$mathbblogEX(t)leg$}) holds with equality. Thus,
    \begin{align}\nonumber\sup_{\substack{\omega\in\hat{\mathcal{F}}_2}}\mathbb{E}\Big[\int_0^\infty\!\mathrm{e}^{-\rho t}S(t)\Big(&\log\big(c(t)X(t)\big)+\log\big((1-\alpha)X(t)\big)b\lambda(t)\Big)\,\mathrm{d}t\Big]
    \\\label{eq:log-split}&\begin{aligned}
        =\int_0^\infty\!&S(t)\mathrm{e}^{-\rho t}\Big(b\lambda(t)\log(1-\alpha)+(1+b\lambda(t))\int_0^t\!\alpha\lambda(s)\,\mathrm{d}s\Big)
        \\[-0.4em]&+S(t)\mathrm{e}^{-\rho t}\Big(\log c(t)-(1+b\lambda(t))\int_0^t\!c(s)\,\mathrm{d}s\Big)
        \\&+S(t)\mathrm{e}^{-\rho t}(1+b\lambda(t))\Big(\log X(0)+f_0(w^*)\,t\Big)\,\mathrm{d}t.\end{aligned}
    \end{align}
    In particular, the optimal percentage $\,\alpha^*\leq1\,$ coincides with the point of global maximum of the following function,
    \[
		\begin{split}
		g(\alpha) & =\int_0^\infty\!S(t)\mathrm{e}^{-\rho t}\Big(b\lambda(t)\log(1-\alpha)+(1+b\lambda(t))\int_0^t\!\alpha\lambda(s)\,\mathrm{d}s\Big)\,\mathrm{d}t \\
		& = b\log(1-\alpha)M_\tau(-\rho)+\alpha\Big(\frac{M_\tau(-\rho)-M_A(-\rho)}{\rho}+bM_A(-\rho)\Big)\quad\mbox{for $\alpha\leq 1$},
		\end{split}
		\]
    with $\,M_\tau(-\rho)=\mathbb{E}[\mathrm{e}^{-\rho\tau}]\,$ and $\,M_A(-\rho)=\mathbb{E}[\mathrm{e}^{-\rho A}]$.  The first and second order condition for $g$ are 
     \[
		\begin{split}
		g'(\alpha) = & -\frac{bM_\tau(-\rho)}{1-\alpha}+\Big(\frac{M_\tau(-\rho)-M_A(-\rho)}{\rho}+bM_A(-\rho)\Big)=0, \\
    g''(\alpha)= & -\frac{bM_\tau(-\rho)}{(1-\alpha)^2}<0.
		\end{split}
		\]
    In particular, $\alpha^*$ from (\ref{eq:$alpha^*=$}) is the unique point of global maximum of the strictly concave function $g$ for $\,\alpha\in\mathbb{R}$.
    In view of (\ref{eq:$mathbbP[A>t]=$}), the inequality $\,\mathbb{P}[\tau\leq t]\geq \mathbb{P}[A\leq t]\,$ holds true for all $\,t\geq0$. 
    Thus, we see that
    \[
		M_\tau(-p)=\int_0^\infty\!\mathbb{P}[\tau\leq\log(t)/(-p)]\,\mathrm{d}t\geq\int_0^\infty\!\mathbb{P}[A\leq\log(t)/(-p)]\,\mathrm{d}t=M_A(-p).
		\]
    In view of $\,b,p\geq0$ and equations (\ref{eq:$kappa=$}) and (\ref{eq:$alpha^*=$}), the inequality $\,\alpha^*\leq1$ holds true.
    Hence, $\alpha^*$ from (\ref{eq:$alpha^*=$}) is the unique point of global maximum of $g$ for $\,\alpha\leq1$, and therefore it is the optimal percentage in the logarithmic case, Problem (\ref{eq:$sup_omega,c,alpha log$}).
\end{proof}
\end{pro}

In view of (\ref{eq:log-split}), if $c$ is deterministic, then the optimal $c^*$ for Problem (\ref{eq:$sup_omega,c,alpha log$}) coincides with the solution to a calculus of variations problem that is given by
\[\sup_{\substack{c\in\hat{\mathcal{F}}_1}}\int_0^\infty\!S(t)\mathrm{e}^{-\rho t}\Big(\log c(t)-(1+b\lambda(t))\int_0^t\!c(s)\,\mathrm{d}s\Big)\,\mathrm{d}t.
\]
Using the Euler-Lagrange equation and an appropriate transversality condition, it can be shown that a candidate for the solution is given by (\ref{eq:$c^*=$}).

In preparation for the verification of $w^*,c^*,\alpha^*$ defined by (\ref{eq:$omega^*=$})--(\ref{eq:$alpha^*=$}), respectively, to be the optimal solution for Problem (\ref{eq:$sup_omega,c,alpha log$}), we introduce a couple of new functions.  One of them we call $V$ and will prove that it is the value function of the underlying Problem. 

\begin{defi}
    \begin{equation}
        \label{eq:$V=$}
        V(t,x)=\varphi(t)\log(x)+\psi(t),\quad\mbox{for $\,t\geq0\,$ and $\,x>0$},
        \end{equation}    
        with
        \begin{align}
        \label{eq:$varphi=$}
        \varphi(t)=\int_t^\infty\!\!\!&\mathrm{e}^{-\rho s}S(s)\big(1+b\lambda(t)\big)\,\mathrm{d}s,
        \\
        \label{eq:$psi=$}
        \begin{split}\psi(t)=\int_t^\infty\!\!\!&\mathrm{e}^{-\rho s}S(s)\Big(\log c^*(s)+b\log(1-\alpha)\lambda(s)\Big)
            \\
            & + \varphi(s)\Big(r+(\mu-r)w^*+\alpha\lambda(s)-c^*(s)-(w^*)^2\sigma^2/2\Big)\,\mathrm{d}s.
            \end{split}
        \end{align}
\end{defi}

\begin{lem}[Solution to the HJB-equation]\label{lem:HJB}
    If $\,\alpha<1$, then $V(t,x)$ from (\ref{eq:$V=$}) is twice continuously differentiable in $x$, absolutely continuous in $t$ and fulfills the following partial differential equation for all $\,x>0\,$ at almost every $\,t\geq0$,
    \begin{equation}\label{HJB}\begin{aligned}0=\max_{c,w}\Big[&\mathrm{e}^{-\rho t}S(t)\Big(\log(cx)+b\log\big((1-\alpha)x\big)\lambda(t)\Big)+\partial_tV(t,x)
    \\&+x\,\partial_xV(t,x)\Big(r+(\mu-r)w+\alpha\lambda(t)-c\Big)+x^2\,\partial_{xx}V(t,x)\,w^2\sigma^2/2\Big].\end{aligned}\end{equation}
    Moreover, $\,c^*,\omega^*$ defined by (\ref{eq:$omega^*=$}),(\ref{eq:$c^*=$}), respectively, correspond to the solutions of the maximization problem inside of (\ref{HJB}).
\begin{proof}
    First, we establish that $\varphi,\psi$ are finite functions, which then implies that $V$ has the claimed regularity because of Lebesgue's Fundamental Theorem of Calculus. Using integration-by-parts and the fact that $S\lambda$ is the probability density function of the random variable $\tau$, we can see that
    \begin{equation}\label{eq:$int=M_T$}
		\int_t^\infty\!\frac{\mathrm{e}^{-\rho s}S(s)}{\mathrm{e}^{-\rho t}S(t)}\,\mathrm{d}s=\frac{1}{p}-\frac{1}{p}M_\tau(-p,t)\quad\mbox{and}\;\;\int_t^\infty\!\frac{\mathrm{e}^{-\rho s}S(s)}{\mathrm{e}^{-\rho t}S(t)}b\lambda(s)\,\mathrm{d}s=bM_\tau(-p,t).
		\end{equation}
    Thus, comparing the definitions (\ref{eq:$c^*=$}),(\ref{eq:$varphi=$}) of $c^*$,$\varphi$, respectively, with (\ref{eq:$int=M_T$}) shows that
    \begin{equation}\label{eq:$c^*=varphi$}c^*(t)=\frac{p}{1-(1-bp)M_\tau(-p,t)}=\frac{\mathrm{e}^{-\rho t}S(t)}{\varphi(t)},\end{equation}
    which implies that $\varphi$ is finite. 
    
    As $\,\rho>0\,$ and $\,1\geq M_\tau(-p,t)\geq0$, a bit of algebra applied to (\ref{eq:$c^*=varphi$}) shows that
    \begin{equation}\label{eq:$varphi<=$}
		\varphi(t)=\mathrm{e}^{-\rho t}S(t)\frac{1-(1-b\rho)M_\tau(-\rho,t)}{\rho}\leq\mathrm{e}^{-\rho t}S(t)\frac{1+b\rho}{\rho}.
		\end{equation}
    In particular,
    \begin{equation}
        \label{eq:$varphileq$}\varphi(t)\leq\mathrm{e}^{-\rho t}\frac{1+b\rho}{\rho},\qquad\varphi(t)\lambda(t)\leq S(t)\lambda(t)\frac{1+b\rho}{\rho} \qquad \textrm{and} \qquad \varphi(t)c^*(t)\leq\mathrm{e}^{-\rho t}.\end{equation}
    In view of (\ref{eq:$varphi=$}),(\ref{eq:$c^*=varphi$}) and $\varphi$ finite (as shown before) as well as $\,b\lambda\geq0$, we find for large enough $t$ that
	\begin{equation}\label{eq:$eSlogc^*leq$}
		\begin{gathered}\Big|\mathrm{e}^{-\rho t}S(t)\log c^*(t)\Big|\leq(1+\rho t)\mathrm{e}^{-\rho t}+\Big|\mathrm{e}^{-\rho t}S(t)\log\varphi(t)\Big|=(1+\rho t)\mathrm{e}^{-\rho t}-\mathrm{e}^{-\rho t}S(t)\log\varphi(t)
    \\
		\leq(1+\rho t)\mathrm{e}^{-\rho t}+\varphi'(t)\log\varphi(t)=(1+\rho t)\mathrm{e}^{-\rho t}+\partial_t\Big(\varphi(t)\big(\log\varphi(t)-1\big)\Big).
		\end{gathered}
		\end{equation}
    As $\,\alpha<1$, the term $\log(1-\alpha)$ is finite.  In particular,
    \begin{equation}
        \label{eq:$log(1-alpha)<infty$}\Big|\mathrm{e}^{-\rho t}S(t)b\log(1-\alpha)\lambda(t)\Big|\leq S(t)\lambda(t)\Big|b\log(1-\alpha)\Big|<\infty.
        \end{equation}
    The functions $S\lambda$ and $\,t\mapsto(1+pt)\mathrm{e}^{-\rho t}\,$ and $\,t\mapsto\partial_t(\varphi(t)(\log\varphi(t)-1))\,$ are integrable. In particular, (\ref{eq:$varphileq$})--(\ref{eq:$log(1-alpha)<infty$}) are finite and integrable, and $\varphi,\psi$ are in turn finite, hence $V$ has the claimed regularity.
    
    Next we define two functions that correspond to the two underlying maximization problems within (\ref{HJB}).
		\[
		h(c)=\mathrm{e}^{-\rho t}S(t)\log(c)-cx\,\partial_xV(t,x)=S(t)\log(c)-c\varphi(t)
		\]
		and
    \[
		i(w)=x\,\partial_xV(t,x)(\mu-r)w+x^2\,\partial_{xx}V(t,x)\,w^2\sigma^2/2=\varphi(t)(\mu-r)w-\varphi(t)w^2\sigma^2/2.
		\]
    The first and second order conditions for both of these functions are
    \[
		\partial_{c} h(c)=\mathrm{e}^{-\rho t}S(t)/c-\varphi(t)=0 \quad\mbox{and}\qquad \partial_{cc} h(c)=-\mathrm{e}^{-\rho t}S(t)/c^2<0,
		\]
    \[
		\partial_{w} i(w)=\varphi(t)(\mu-r)-\varphi(t)w\sigma^2=0 \qquad \mbox{and} \qquad \partial_{ww} i(w)=-\varphi(t)\sigma^2<0.
		\]
    In particular, $h$ and $i$ are strictly concave functions with global maximum at $c^*$ and $w^*$ as given as in (\ref{eq:$c^*=varphi$}) and (\ref{eq:$omega^*=$}) respectively.  
    
    It is straightforward to check that $V$ fulfills (\ref{HJB}).
\end{proof}
\end{lem}

\begin{lem}[Transversality condition]\label{lem:Transversality}
    Let $\,\alpha<1$. Let $c,\omega$ be elements from $\hat{\mathcal{F}}_1$, $\hat{\mathcal{F}}_2$, respectively. If the corresponding wealth process $X$, whose dynamics satisfy (\ref{eq:$mathrmdX$}), fulfills
    \begin{equation}\label{eq:$>-infty$}\mathbb{E}\Big[\int_{0}^\infty\!e^{-\rho t}S(t)\Big(\log\big(c(t)X(t)\big)+b\lambda(t)\log\big((1-\alpha)X(t)\big)\Big)\,\mathrm{d}t\Big]>-\infty,\end{equation}
    then there exists a sequence of real numbers $\,t_n\uparrow\infty$, depending on $c$ and $\omega$, such that
    \[ \lim_{n\rightarrow\infty}\mathbb{E}[V(t_n,X(t_n))]=0.\]
\begin{proof}
    In view of Lemma \ref{lem:quasi-integrable}, Lemma \ref{lem:$sup=sup$} and the assumption (\ref{eq:$>-infty$}), we can see that    
    \[
		\mathbb{E}\Big[\int_{0}^\infty\!e^{-\rho t}S(t)\Big(\log\big(c(t)X(t)\big)+b\lambda(t)\log\big((1-\alpha)X(t)\big)\Big)\,\mathrm{d}t\Big]\in\mathbb{R}.
		\]
    Moreover, as both summands inside of the integral are bounded above under the expectation and integral - see Lemma \ref{lem:quasi-integrable} and Lemma \ref{lem:$sup=sup$} - we find that
    \begin{equation}\label{eq:$cXinR$}\mathbb{E}\Big[\int_{0}^\infty\!e^{-\rho t}S(t)\log\big(c(t)X(t)\big)\,\mathrm{d}t\Big]\in\mathbb{R}.\end{equation}
    In view of \citet[p.$84$, Theorem $37$]{Protter2005}, equation (\ref{eq:$mathrmdX$}) has a unique solution; see (\ref{eq:$logX=$}).  In particular, we can rewrite (\ref{eq:$cXinR$}) as
    \begin{align}\nonumber&\mathbb{E}\Big[\int_{0}^\infty\!e^{-\rho t}S(t)\log\big(c(t)X(t)\big)\,\mathrm{d}t\Big]
    \\\label{eq:$-1/8-1/8-1/4$}&\begin{aligned}
        =\mathbb{E}\Big[\int_{0}^\infty\!e^{-\rho t}S(t)\Big(&\log X(0)+\int_0^t\!\alpha\lambda(s)\,\mathrm{d}s+\log c(t)-\frac{1}{2}\int_0^t\!c(s)\,\mathrm{d}s-\frac{1}{2}\int_0^t\!c(s)\,\mathrm{d}s
        \\&+\int_0^t\!(1-\omega(s))r+\omega(s)\mu-\frac{1}{8}\sigma^2\omega(s)^2\,\mathrm{d}s-\frac{1}{8}\int_0^t\!\sigma^2\omega(s)^2\,\mathrm{d}s
        \\&+\int_0^t\!\sigma\omega(s)\,\mathrm{d}W(s)\,-\frac{1}{4}\int_0^t\!\sigma^2\omega(s)^2\,\mathrm{d}s\Big)\,\mathrm{d}t\Big].\end{aligned}
    \end{align}
    We use the same argument as before: as all of the above terms are bounded above in expectation and integral - see (\ref{eq:estimate1a}),(\ref{eq:estimate2a}),(\ref{eq:estimate4a}),(\ref{eq:estimate3a}),(\ref{eq:estimate5}) - we can show that the finiteness of (\ref{eq:$cXinR$}) implies that
    \begin{gather}
        \label{eq:$alphainR$}\mathbb{E}\Big[\int_{0}^\infty\!e^{-\rho t}S(t)\Big(\log X(0)+\int_0^t\!\alpha\lambda(s)\,\mathrm{d}s\Big)\mathrm{d}t\Big]\in\mathbb{R},
        \\\label{eq:$intcinR$}\mathbb{E}\Big[\int_{0}^\infty\!e^{-\rho t}S(t)\Big(\frac{1}{2}\int_0^t\!c(s)\,\mathrm{d}s\Big)\,\mathrm{d}t\Big]\in\mathbb{R},
        \\\label{eq:$omega^2inR$}\mathbb{E}\Big[\int_{0}^\infty\!e^{-\rho t}S(t)\Big(\frac{1}{8}\int_0^t\!\sigma^2\omega(s)^2\,\mathrm{d}s\Big)\,\mathrm{d}t\Big]\in\mathbb{R}.
        \end{gather}
    From (\ref{eq:$omega^2inR$}), we deduce that $\,\mathbb{E}[\int_0^t\!\sigma^2\omega(s)^2\,\mathrm{d}s]<\infty\,$ for all $\,t\geq0$, and in turn that $\,t\mapsto\int_0^t\!\sigma\omega(s)\,\mathrm{d}W(s)\,$ is a true martingale.  Hence, using (\ref{eq:$-1/8-1/8-1/4$}) again,
    \begin{equation}\label{eq:$w-w^2inR$}\mathbb{E}\Big[\int_{0}^\infty\!e^{-\rho t}S(t)\Big(\int_0^t\!(1-\omega(s))r+\omega(s)\mu-\frac{1}{2}\sigma^2\omega(s)^2\,\mathrm{d}s\Big)\,\mathrm{d}t\Big]\in\mathbb{R}.\end{equation}
    The results (\ref{eq:$alphainR$}),(\ref{eq:$intcinR$}),(\ref{eq:$w-w^2inR$}) and the martingale property of $\,t\mapsto\int_0^t\!\sigma\omega(s)\,\mathrm{d}W(s)\,$ imply that
    \[ \mathbb{E}\Big[\int_{0}^\infty\!e^{-\rho t}S(t)\log X(t)\,\mathrm{d}t\Big]\in\mathbb{R}.\]
    Thus there exists a sequence of real numbers $\,t_n\uparrow\infty\,$ such that
    \[\lim_{n\rightarrow\infty}e^{-\rho t_n}S(t_n)\mathbb{E}[\log X(t_n)]=0.\]
    Finally, comparing this with (\ref{eq:$psi=$}),(\ref{eq:$V=$}),(\ref{eq:$varphi<=$}) establishes the result. Notice that $\,\psi(t)\rightarrow0\,$ as $\,t\uparrow\infty$.
\end{proof}
\end{lem}

\begin{pro}[Verification]\label{pro:Verification}
    Let $\,\alpha\leq1\,$ and $X$ be the corresponding wealth process from (\ref{eq:$mathrmdX$}). Furthermore let $\,c^*,\omega^*$ be defined by (\ref{eq:$omega^*=$}),(\ref{eq:$c^*=$}), respectively, with $X^*$ be the corresponding wealth process. Then 
    \begin{equation}\label{eq:$V(0,X_0)geq$}
		V(0,X_0)\geq\mathbb{E}\Big[\int_{0}^\infty\!S(t)e^{-\rho t}\Big(\log\big(c(t)X(t)\big)+b\log\big((1-\alpha)X(t)\big)\lambda(t)\Big)\,\mathrm{d}t\Big]
		\end{equation}
		and
		\[
		V(0,X_0)=\mathbb{E}\Big[\int_{0}^\infty\!S(t)e^{-\rho t}\Big(\log\big(c^*(t)X^*(t)\big)+b\log\big((1-\alpha)X^*(t)\big)\lambda(t)\Big)\,\mathrm{d}t\Big].
		\]
\begin{proof}
    For $\alpha=1$ the right-hand side of (\ref{eq:$V(0,X_0)geq$}) equals $-\infty$. Thus, we can assume that $\alpha<1$. If the expectation in (\ref{eq:$>-infty$}) equals $-\infty$, then inequality (\ref{eq:$V(0,X_0)geq$}) is trivial. Thus, we can assume that the expectation in (\ref{eq:$>-infty$}) is finite.
    Moreover, in this case, (\ref{eq:$omega^2inR$}) holds true and therefore $\,\mathbb{E}[\int_0^t\!\sigma^2\omega(s)^2\varphi(s)^2\,\mathrm{d}s]<\infty\,$ for all $\,t\geq0$, as $\varphi$ is a continuous and deterministic function.  Hence $\,t\mapsto$ $\int_0^t\!\sigma\omega(s)\varphi(s)\,\mathrm{d}W(s)\,$ is a martingale. 
    
    Now, the statement follows from an application of the It\^{o}-formula to the $C^{1,1,2}$-function $\,(\varphi,\log(x),$ $\psi)\mapsto\varphi\log(x)+\psi\,$ evaluated at the special sequence from Lemma \ref{lem:Transversality} for a given $c,\omega$ corresponding to $X$ together with Lemma \ref{lem:HJB}.
\end{proof}
\end{pro}
 
 In particular, combining Lemma \ref{lem:$sup=sup$} and Proposition \ref{pro:Verification} shows that $w^*,c^*,\alpha^*$ from (\ref{eq:$omega^*=$})--(\ref{eq:$alpha^*=$}) is the optimal solution to the logarithmic Problem (\ref{eq:$sup_omega,c,alpha log$}).

\newpage
\section{Tables, diagrams and figures}
\subsection{Diagrams}

\newpage
\subsection{Figures}

\newpage

\end{document}